{}%disable hyper at arxiv
{}%use pdflatex at arxiv

\documentclass[11pt,a4paper]{article}

\newif\ifpublic\publictrue

\ifpublic\else\usepackage{showkeys}\fi
\usepackage[bookmarks=true,hyperfigures=true,colorlinks=true,linkcolor=black,citecolor=black,urlcolor=black,bookmarksnumbered]{hyperref}
\usepackage[nosort]{cite}
\usepackage[parsep]{collref}
\usepackage[english]{babel}
\usepackage[T1]{fontenc}
\usepackage[latin1]{inputenc}
\usepackage{amsfonts,amsbsy,dsfont,euscript,mathrsfs,fixmath}
\usepackage{amsmath,amssymb}
\usepackage{enumerate}
\def\showkeysrefformat#1{{\normalfont\tiny\ttfamily#1}}
\makeatletter
\def\SK@@ref#1>#2\SK@{{\@inlabelfalse\leavevmode\vbox to\z@{\vss\SK@refcolor\rlap{\vrule\raise .75em \hbox{\showkeysrefformat{#2}}}}}}
\makeatother

\usepackage[a4paper,text={160mm,247mm},centering]{geometry}
\allowdisplaybreaks[3]
\numberwithin{equation}{section}
\def\[{\begin{equation}\begin{aligned}}
\def\]{\end{aligned}\end{equation}}

\usepackage[font=small,labelfont=bf,width=0.85\textwidth]{caption}

\expandafter\def\expandafter\bfseries\expandafter{\bfseries\ifmmode\else\boldmath\fi}
\expandafter\def\expandafter\mdseries\expandafter{\mdseries\ifmmode\else\unboldmath\fi}
\expandafter\def\expandafter\normalfont\expandafter{\normalfont\ifmmode\else\unboldmath\fi}

\RequirePackage{verbatim}

\makeatletter
\newwrite\bibinl@out
\newenvironment{bibtex}[1][\jobname]{%
\immediate\openout\bibinl@out #1.bib%
\immediate\write\bibinl@out{\@percentchar generated from `\jobname' starting line \the\inputlineno^^J}%
\def\verbatim@processline{\immediate\write\bibinl@out{\the\verbatim@line}}%
\@bsphack\let\do\@makeother\dospecials\catcode`\^^M\active\verbatim@start%
}
{\immediate\closeout\bibinl@out\@esphack}
\makeatother

\let\barefrac=\frac
\renewcommand{\frac}[2]{\mathinner{\barefrac{#1}{#2}}}

\let\baresqrt=\sqrt
\makeatletter
\renewcommand{\sqrt}{\@ifnextchar[\@sqrt@space@a\@sqrt@space@b}
\def\@sqrt@space@a[#1]#2{\mathinner{\mathchoice{\mkern-3mu}{\mkern-3mu}{}{}\baresqrt[#1]{#2}}}
\def\@sqrt@space@b#1{\mathinner{\mathchoice{\mkern-3mu}{\mkern-3mu}{}{}\baresqrt{#1}}}
\makeatother

\makeatletter
\let\per@dot@old=\.
\def\.{\ifmmode\def\per@dot@sel{\mkern3mu}\else\def\per@dot@sel{\per@dot@old}\fi\per@dot@sel}
\makeatother

\let\barefootnote=\footnote
\renewcommand{\footnote}[1]{\barefootnote{#1\vspace{3pt}}}

\newcommand{\sfrac}[2]{{\textstyle\frac{#1}{#2}}}
\newcommand{\half}{\sfrac{1}{2}}
\newcommand{\ihalf}{\sfrac{i}{2}}
\newcommand{\vfrac}[2]{\ifmmode\mathinner{\textstyle^{#1}\!/\!_{#2}}\else$^{#1}\!/\!_{#2}$\fi}

\newcommand{\identity}{\mathds{1}}
\DeclareMathOperator{\Mat}{Mat}

\DeclareMathOperator{\Tr}{Tr}

\newcommand{\transpose}{t}

\newcommand{\Order}{\mathcal{O}}

\newcommand{\set}[1]{\{#1\}}
\newcommand{\vectorspan}[1]{\operatorname{span}\{#1\}}
\newcommand{\Real}{\mathds{R}}
\newcommand{\Complex}{\mathds{C}}
\newcommand{\Integer}{\mathds{Z}}

\let\Im\relax\DeclareMathOperator{\Im}{Im}

\newcommand{\ind}[1]{{\scriptscriptstyle{#1}}}

\newcommand{\alg}[1]{\mathfrak{#1}}
\newcommand{\grp}[1]{\mathrm{#1}}
\DeclareMathOperator{\rank}{rank}
\DeclareMathOperator{\Lie}{Lie}
\DeclareMathOperator{\ad}{ad}
\DeclareMathOperator{\Ad}{Ad}
\newcommand{\killing}{\kappa}
\newcommand{\dsum}{+}

\newcommand{\com}[2]{[#1,#2]}

\newcommand{\Proj}[1]{\text{P}_{#1}}
\newcommand{\proj}{\text{P}}

\def\<{\big\langle}
\def\>{\big\rangle}

\newcommand{\geom}[1]{\mathrm{#1}}

\newcommand{\AdS}{\geom{AdS}}

\newcommand{\Sp}{\geom{S}}

\newcommand{\extder}{\mathrm{d}}

\newcommand{\lieder}{\mathcal{L}}

\newcommand{\Act}{\mathcal{S}}

\def\En{E_0}
\def\Gn{G_0}
\def\Bn{B_0}
\def\Ent{{\En^\prime}}
\def\Gnt{{\Gn^\prime}}
\def\Bnt{{\Bn^\prime}}

\def\lambdastar{\lambda^*}
\def\rot{\mathrm{rot}}
\def\WZ{\mathrm{WZ}}

\providecommand{\href}[2]{#2}

\makeatletter
\def\mr@ignsp#1 {\ifx\:#1\@empty\else #1\expandafter\mr@ignsp\fi}
\newcommand{\multiref}[1]{\begingroup%
\xdef\mr@no@sparg{\expandafter\mr@ignsp#1 \: }%
\def\mr@comma{}\def\mr@dash{-}%
\@for\mr@refs:=\mr@no@sparg\do{%
\ifx\mr@refs\mr@dash\def\mr@comma{}--\else%
\mr@comma\def\mr@comma{,}\ref{\mr@refs}\fi}%
\endgroup}
\renewcommand{\eqref}[1]{(\multiref{#1})}
\makeatother

\makeatletter
\newcommand{\namedref}[2]{\hyperref[#2]{#1~\ref*{#2}}}
\newcommand{\secref}{\@ifstar{\namedref{Section}}{\namedref{sec.}}}
\newcommand{\appref}{\@ifstar{\namedref{Appendix}}{\namedref{app.}}}
\newcommand{\tabref}{\@ifstar{\namedref{Table}}{\namedref{tab.}}}
\newcommand{\figref}{\@ifstar{\namedref{Figure}}{\namedref{fig.}}}
\makeatother

\let\oldbib=\thebibliography
\def\thebibliography{\phantomsection\addcontentsline{toc}{section}{\refname}\oldbib}

\let\oldtoc=\tableofcontents
\def\tableofcontents{\phantomsection\addcontentsline{toc}{section}{\contentsname}\oldtoc}

\newcommand{\foreign}[1]{\textit{#1}}

\providecommand{\hypersetup}[1]{}
\providecommand{\texorpdfstring}[2]{#1}
\hypersetup{plainpages=false}
\hypersetup{pdfpagemode=UseNone}
\hypersetup{bookmarksnumbered=true}
\hypersetup{pdfstartview=FitH}
\hypersetup{colorlinks=false}
\hypersetup{citebordercolor={1 1 1}}
\hypersetup{urlbordercolor={1 1 1}}
\hypersetup{linkbordercolor={1 1 1}}
\makeatletter
\let\@keywords\@empty
\let\@subject\@empty
\providecommand{\keywords}[1]{\gdef\@keywords{#1}}
\providecommand{\subject}[1]{\gdef\@subject{#1}}
\def\thetitle{\@title}
\def\theauthor{\@author}
\def\thesubject{\@subject}
\def\thedate{\@date}
\def\thekeywords{\@keywords}
\makeatother
\AtBeginDocument{
\hypersetup{pdftitle={\thetitle}}
\hypersetup{pdfauthor={\theauthor}}
\hypersetup{pdfsubject={\thesubject}}
\hypersetup{pdfkeywords={\thekeywords}}}

\newif\ifshownote
\shownotetrue

\ifpublic\shownotefalse\fi

\ifshownote

\ifpdf\else\RequirePackage[active]{srcltx}\fi
\RequirePackage{xcolor}
\RequirePackage{ulem}

\newcommand{\remark}[2][]{{\normalfont\sffamily\hspace{1ex}
\def\emph{\textsl}\def\textbullet{$\bullet$}
\def\tmparga{#1}
\def\tmpargb{BH}\ifx\tmparga\tmpargb\color[rgb]{0.5,0,0}\fi
\def\tmpargb{FS}\ifx\tmparga\tmpargb\color[rgb]{0,0.5,0}\fi
\def\tmpargb{}\ifx\tmparga\tmpargb\color{red}\fi
\def\tmpargb{}\ifx\tmparga\tmpargb\else \textbf{#1:}\fi
#2\hspace{1ex}}}

\else
\newcommand{\remark}[2][]{\ignorespaces}

\fi

\title{Poisson--Lie duals of the \texorpdfstring{$\eta$}{eta}~deformed \texorpdfstring{\unskip\\}{} symmetric space sigma model}
\author{Ben~Hoare and Fiona~K.~Seibold}

\begin{document}

\pdfbookmark[1]{Title Page}{title}
\thispagestyle{empty}

%\begingroup\raggedleft\footnotesize\ttfamily
%\arxivlink{yymm.nnnnnn}
%\par\endgroup

\vspace*{2cm}
\begin{center}
\begingroup\Large\bfseries\thetitle\par\endgroup
\vspace{1cm}

\begingroup\theauthor\par\endgroup
\vspace{1cm}

\textit{
Institut f\"ur Theoretische Physik,\\
Eidgen\"ossische Technische Hochschule Z\"urich,\\
Wolfgang-Pauli-Strasse 27, 8093 Z\"urich, Switzerland}
\vspace{5mm}

\begingroup\ttfamily\small
\verb+{+bhoare,fseibold\verb+}+@itp.phys.ethz.ch\par
\endgroup
\vspace{5mm}

\vfill

\textbf{Abstract}\vspace{5mm}

\begin{minipage}{12.5cm}\small
Poisson--Lie dualising the $\eta$~deformation of the $\grp{G}/\grp{H}$ symmetric space sigma model with respect to the simple Lie group $\grp{G}$ is conjectured to give an analytic continuation of the associated $\lambda$~deformed model.
In this paper we investigate when the $\eta$~deformed model can be dualised with respect to a subgroup $\grp{G}_0$ of $\grp{G}$.
Starting from the first-order action on the complexified group and integrating out the degrees of freedom associated to different subalgebras, we find it is possible to dualise when $\grp{G}_0$ is associated to a sub-Dynkin diagram.
Additional $\grp{U}_1$ factors built from the remaining Cartan generators can also be included.
The resulting construction unifies both the Poisson--Lie dual with respect to $\grp{G}$ and the complete abelian dual of the $\eta$~deformation in a single framework, with the integrated algebras unimodular in both cases.
We speculate that extending these results to the path integral formalism may provide an explanation for why the $\eta$~deformed $\AdS_5 \times \Sp^5$ superstring is not one-loop Weyl invariant, that is the couplings do not solve the equations of type IIB supergravity, yet its complete abelian dual and the $\lambda$~deformed model are.
\end{minipage}

\vspace*{2cm}

\end{center}

\newpage

\tableofcontents

%%%%%%%%%%%%%%%%%%%%%%%%%%%%%%%%%%%%%%%%%%%%%%%%%%%%%%%%%%%%%%%%%%%%%%%%%%%%%%%%
\section{Introduction}\label{sec:intro}

Recent progress in understanding integrable deformations of the $\AdS_5 \times \Sp^5$ superstring, along with its various integrable cousins and bosonic truncations, has led to significant advances relating the underlying algebraic and geometric structures.
Two of the most well-studied deformations are associated to $q$~deformations of the isometry algebra $\alg{psu}_{2,2|4}$.
These are the $\eta$~deformation \cite{Delduc:2013qra,Delduc:2014kha} and the $\lambda$~deformation \cite{Hollowood:2014qma}.
The former is a deformation of the Green-Schwarz type IIB superstring \cite{Green:1983wt,Green:1983sg,Witten:1985nt,Grisaru:1985fv} written as a sigma model on the supercoset \cite{Metsaev:1998it,Berkovits:1999zq}
\[
\frac{\grp{PSU}_{2,2|4}}{\grp{SO}_{1,4} \times \grp{SO}_5} ,
\]
while the latter is a deformation of its non-abelian dual with respect to the full isometry group $\grp{PSU}_{2,2|4}$.

\medskip

The $\eta$~deformation of the $\AdS_5 \times \Sp^5$ superstring \cite{Delduc:2013qra,Delduc:2014kha} generalises a certain integrable deformation of the principal chiral model \cite{Klimcik:2002zj,Klimcik:2008eq} and of the symmetric space sigma model \cite{Delduc:2013fga}, often referred to as Yang--Baxter sigma models due to their explicit dependence on a non-split solution of the modified classical Yang--Baxter equation.
The model depends on two parameters: an overall coupling $h$, which plays the role of the effective string tension, and the deformation parameter $\eta$.
After the deformation only the Cartan subalgebra of the original isometry algebra remains a symmetry.
However, the other charges are not lost; rather they are hidden, satisfying the relations of the $q$~deformed isometry algebra, where $q \in \Real$ is a function of the string tension and $\eta$ \cite{Arutyunov:2013ega,Delduc:2014kha,Delduc:2016ihq}.

From the deformed superstring sigma model one can determine the corresponding deformation of the maximally supersymmetric $\AdS_5 \times \Sp^5$ type IIB supergravity background \cite{Arutyunov:2013ega,Arutyunov:2015qva}.
The resulting set of fields do not satisfy the type IIB supergravity equations of motion.
However, one can show that the maximal abelian dual is a supergravity solution with a dilaton that is linear in the dualised isometric coordinates \cite{Hoare:2015gda,Hoare:2015wia}.
As a consequence the deformed background solves a set of generalised type IIB supergravity equations \cite{Arutyunov:2015mqj}, which are compatible with the $\kappa$~symmetry of the worldsheet sigma model \cite{Wulff:2016tju}.
For further progress in ascertaining and relating the background fields of the generalised theory for $\eta$ deformed models see, for example, \cite{Hoare:2016ibq,Borsato:2016ose,Araujo:2017enj}.
Various interpretations of these results and the consequences for Weyl invariance have been given in \cite{Hoare:2015gda,Hoare:2015wia,Arutyunov:2015mqj,Wulff:2016tju,Araujo:2017enj}, and, in the context of double field theory, in \cite{Sakatani:2016fvh,Baguet:2016prz,Sakamoto:2017wor}.
However, currently there is no final word on how to (or if it is even possible to) interpret the $\eta$~deformation as a string theory.

The $\lambda$~deformation of the non-abelian dual of the $\AdS_5 \times \Sp^5$ superstring \cite{Hollowood:2014qma} generalises the bosonic models of \cite{Sfetsos:2013wia,Hollowood:2014rla}.
This model also depends on two parameters: an overall level $k$ of the underlying gauged Wess--Zumino--Witten action, playing the role of the effective string tension, and the parameter $\lambda$.
In this case the corresponding deformed type IIB supergravity background does solve the supergravity equations of motion \cite{Borsato:2016zcf,Chervonyi:2016bfl,Borsato:2016ose}.
This can be understood as a consequence of the $\kappa$~symmetry of the model, together with the lack of any Killing vectors admitted by the geometry \cite{Wulff:2016tju}.
However, it should be noted that there are conserved charges, which do not manifest as Killing vectors, that can be used to characterise the spectrum of the model \cite{Appadu:2017xku}.
The hidden symmetry is again a $q$~deformation, but now with $|q| = 1$, where $q$ is a function of the level $k$ \cite{Appadu:2017xku}.
Furthermore, the level $k$ is quantised such that the deformation parameter $q$ is a root of unity and hence discrete.

\medskip

The $\eta$~deformation is conjectured to be related by Poisson--Lie duality, a generalisation of non-abelian duality to sigma models with Poisson--Lie symmetry \cite{Klimcik:1995ux,Klimcik:1995jn}, to an analytic continuation of the corresponding $\lambda$~deformation \cite{Hoare:2015gda}.
A duality of this type was considered in \cite{Vicedo:2015pna}, which relates an $\eta^s$~deformation (the superscript denoting it is based on a split solution of the modified classical Yang--Baxter equation) to the corresponding $\lambda$~deformation.
In both cases the duality is with respect to the full symmetry group of the undeformed model.
Hints of such a relation, in particular the need for analytic continuation, were first seen in the analysis of low-dimensional models in \cite{Hoare:2014pna}.

Our eventual goal is to explore the deformations of the $\AdS_5 \times \Sp^5$ superstring.
The associated symmetry algebra $\alg{psu}_{2,2|4}$ does not admit split solutions of the modified classical Yang--Baxter equation, and hence we will focus on the duality between the $\eta$~deformation and an analytic continuation of the corresponding $\lambda$~deformation.
The required analytic continuation of the $\lambda$~deformation acts on both the fields and parameters of the model, and is such that $|q| = 1$ is mapped to $q \in \Real$: how $k$ and $\lambda$ are related to $h$ and $\eta$ is given in \cite{Hoare:2015gda}.
We will refer to this analytic continuation as the $\lambdastar$~deformation.

The claim of duality has been demonstrated for various low-dimensional examples in \cite{Hoare:2015gda,Sfetsos:2015nya} using results of \cite{Sfetsos:1999zm}.
Furthermore, starting from a certain first-order action on a Drinfel'd double \cite{Klimcik:1995dy,Klimcik:1996nq}, which generalises the duality-invariant action of \cite{Tseytlin:1990nb,Tseytlin:1990va} underlying abelian duality, it has been proven for the deformation of the principal chiral model \cite{Klimcik:2015gba}.
In this paper we do not provide new evidence for this conjecture.
We instead \foreign{assume} it to be true with the aim of further exploring the underlying structures.

Thus far we have discussed two alternative maps of the $\eta$~deformation of $\AdS_5 \times \Sp^5$ to a solution of type IIB supergravity: the complete abelian dual and the Poisson--Lie dual with respect to $\grp{PSU}_{2,2|4}$.
This motivates a number of interesting questions.
First, can we incorporate the two transformations in a unified framework?
Second, is it possible to Poisson--Lie dualise with respect to subgroups of $\grp{PSU}_{2,2|4}$?
And third, what are the consequences for Weyl invariance?
In this paper we focus on the first two of these three questions for the $\eta$~deformation of the bosonic symmetric space sigma model.
In \secref{sec:conc} we briefly comment on the extension to the superstring and speculate on the implications for Weyl invariance.

\medskip

The symmetric space sigma model for the coset $\grp{G}/\grp{H}$ is invariant under the action of the isometry group $\grp{G}$.
In particular, one can dualise with respect to any subgroup of $\grp{G}$, including the Cartan subgroup $\grp{U}_1{}^{\rank\grp{G}}$ and $\grp{G}$ itself.
These two dualities both survive the $\eta$~deformation: the first as the complete abelian dual and the second as the Poisson--Lie dual with respect to $\grp{G}$, the $\lambdastar$~deformation.
Our aim is to understand how these dualities fit into a single framework, and to investigate which other dualities also extend to the deformed model.

In order to answer these questions we work with the first-order action on a Drinfel'd double $\grp{D}$ \cite{Klimcik:1995dy,Klimcik:1996nq} and its generalisation to coset spaces \cite{Klimcik:1996np,Squellari:2011dg}.
As a vector space, the Lie algebra of the Drinfel'd double $\alg{d} = \Lie(\grp{D})$ can be decomposed as
\[
\alg{d} = \alg{g} \dsum \tilde{\alg{g}} ~,
\]
where $\alg{g}$ and $\tilde{\alg{g}}$ are two subalgebras, maximally isotropic with respect to a non-degenerate ad-invariant inner product on $\alg{d}$.
Integrating out the degrees of freedom associated to $\tilde{\alg{g}}$ or $\alg{g}$ gives a second-order Poisson--Lie symmetric model on $\tilde{\grp{G}}\backslash\grp{D} \sim \grp{G}$ or $\grp{G}\backslash\grp{D} \sim \tilde{\grp{G}}$ respectively.
These are then said to be Poisson--Lie dual.

After reviewing this construction we turn our attention to alternative maximally isotropic decompositions of the same Drinfel'd double
\[
\alg{d} = \alg{k} \dsum \tilde{\alg{k}} ~,
\]
where $\tilde{\alg{k}}$ is a subalgebra of $\alg{d}$ with associated Lie group $\tilde{\grp{K}}$, but $\alg{k}$ need not be.
In this case, the degrees of freedom associated to $\tilde{\alg{k}}$ can be integrated out to define a model on $\tilde{\grp{K}}\backslash\grp{D}$ \cite{Klimcik:1996nq,Klimcik:2002zj,Klimcik:2015gba}.
We consider, in particular, Lie algebras $\tilde{\alg{k}}$ whose intersection with $\alg{g}$ defines a common subalgebra $\alg{g}_0$.
Then, given that the four spaces $\alg{g}$, $\tilde{\alg{g}}$, $\alg{k}$ and $\tilde{\alg{k}}$ admit the following decompositions
\[
\alg{g} & = \alg{g}_0 \dsum \alg{m} ~, & \qquad \tilde{\alg{g}} & = \tilde{\alg{g}}_0 \dsum \tilde{\alg{m}} ~,
\\
\alg{k} & = \tilde{\alg{g}}_0 \dsum \alg{m} ~, & \qquad \tilde{\alg{k}} & = \alg{g}_0 \dsum \tilde{\alg{m}} ~,
\]
we say that the model on $\tilde{\grp{K}}\backslash\grp{D}$ is the Poisson--Lie dual of the model on $\tilde{\grp{G}}\backslash\grp{D} \sim \grp{G}$ with respect to $\grp{G}_0$.
The requirement that $\tilde{\alg{k}}$ forms a subalgebra places restrictions on the Lie groups $\grp{G}_0$ with respect to which we can dualise.

The $\eta$~deformation of the symmetric space sigma model can be formulated in the above framework, with the Drinfel'd double given by the complexified group $\grp{D} \equiv \grp{G}^\Complex$.
The relevant maximally isotropic decomposition is then
\[
\alg{d} = \alg{g} \dsum \alg{b} ~,
\]
where $\alg{b}$ is the Borel subalgebra generated by the Cartan generators and positive roots.
The $\eta$~deformed model is found by integrating out the degrees of freedom associated to $\alg{b}$.
We construct a class of subalgebras $\alg{g}_0$ with respect to which we can dualise.
This includes both the Poisson--Lie dual with respect to $\grp{G}$ as well as the complete abelian dual.
We illustrate our results on a number of examples for $\Sp^2$ and $\Sp^5$, both spaces of interest in the context of integrable superstring sigma models \cite{Zarembo:2010sg,Wulff:2014kja,Wulff:2015mwa}.
For recent progress towards classifying such models, in particular incorporating those without supersymmetry, see \cite{Wulff:2017hzy,Wulff:2017zbl,Wulff:2017lxh}.

\medskip

The layout of this paper is as follows.
We start in \secref{sec:general} with a review of the model on the Drinfel'd double and the general formalism for constructing Poisson--Lie dual sigma models.
Then in \secref{sec:application} we apply the formalism to the $\eta$~deformation of the symmetric space sigma model.
We conclude with comments and open questions in \secref{sec:conc}.

%%%%%%%%%%%%%%%%%%%%%%%%%%%%%%%%%%%%%%%%%%%%%%%%%%%%%%%%%%%%%%%%%%%%%%%%%%%%%%%%
\section{Poisson--Lie duality and the Drinfel'd double}\label{sec:general}

Two sigma models are said to be Poisson--Lie dual if they are described by the same set of equations after appropriate non-local field and parameter redefinitions \cite{Klimcik:1995ux,Klimcik:1995jn}.
Quantities computed within the framework of one sigma model thus have an equivalent in the dual theory.
Both abelian and non-abelian duality are special cases of Poisson--Lie duality.
The underlying algebraic structure of Poisson--Lie duality is the Drinfel'd double.
A Drinfel'd double is defined as a $2n$-dimensional connected Lie group $\grp{D}$ such that its Lie algebra $\alg{d}$ can be decomposed into a pair of $n$-dimensional subalgebras $\alg{g}$ and $\tilde{\alg{g}}$, maximally isotropic with respect to a non-degenerate ad-invariant inner product $\< \cdot , \cdot \>$ on $\alg{d}$.
In this paper we will use ``Drinfel'd double'' to refer to both the group and its algebra.
The two sigma models can be obtained from a first-order action on the Drinfel'd double by integrating out the dual degrees of freedom \cite{Klimcik:1995dy,Klimcik:1996nq,Klimcik:2002zj,Klimcik:2015gba}.

In this section, after introducing the actions for the two sigma models related by Poisson--Lie duality, we will investigate when it is possible to dualise with respect to a subgroup $\grp{G}_0 \subset \grp{G}$.
In order to dualise it transpires that the corresponding Lie algebra, $\alg{g}_0 = \Lie(\grp{G}_0)$, should satisfy certain commutation relations within the Drinfel'd double.
These restrictions have a natural algebraic interpretation: the new Poisson--Lie duals are associated to more general decompositions of the Drinfel'd double of the type $\alg{d}=\alg{k} \dsum \tilde{\alg{k}}$, where $\tilde{\alg{k}}$ is a subalgebra, but $\alg{k}$ need not be.

%%%%%%%%%%%%%%%%%%%%%%%%%%%%%%%%%%%%%%%%%%%%%%%%%%%%%%%%%%%%%%%%%%%%%%%%%%%%%%%%
\subsection{Poisson--Lie dual sigma models}\label{ssec:pldsigma}

Let us begin by reviewing the construction of the Poisson--Lie symmetric sigma model and its dual \cite{Klimcik:1995ux,Klimcik:1995jn} (see also \cite{Tyurin:1995bu}).
Our starting point is the two-dimensional non-linear sigma model with target-space metric and $B$-field
\[
\Act = \int \extder^2 \sigma \, F_\ind{IJ}(X) \partial_+ X^\ind{I} \partial_- X^\ind{J} ~,
\qquad F_\ind{IJ}(X) = G_\ind{IJ}(X) + B_\ind{IJ}(X) ~.
\]
The light-cone coordinates $(\sigma^+, \sigma^-)$ are defined in terms of the usual two-dimensional Minkowski coordinates $(\tau,\sigma)$ as $\sigma^{\pm} = (\tau \pm \sigma)/2$, and we use the shorthand notation $\extder^2\sigma=\extder \sigma^+ \extder \sigma^- = \half \extder\tau\extder\sigma$.
The scalar fields $X^\ind{I} (\tau,\sigma)$ parametrise an element $g$ of the Lie group $\grp{G}$, such that under the left action of $\grp{G}$, they transform linearly, $X^\ind{I} \to X^\ind{I} + \epsilon^\ind{A} v_\ind{A}^\ind{I}$, where $v_\ind{A}^\ind{I}$ are the right-invariant frames.
The Noetherian forms can be defined by considering the variation of the action with respect to this transformation
\[
\delta_\epsilon \Act = \int \epsilon^\ind{A} \lieder_{v_\ind{A}} L + \int \extder \epsilon^\ind{A} \wedge \star K_\ind{A} ~,
\]
where $\star K_\ind{A} = v_\ind{A}^\ind{I}(X) F_\ind{IJ}(X) \partial_- X^\ind{J} \, \extder \sigma^- - v_\ind{A}^\ind{J}(X) F_\ind{IJ}(X) \partial_+ X^\ind{I} \, \extder \sigma^+$.
If the left action of $\grp{G}$ corresponds to an isometry of the target space, then the Lie derivative of the Lagrangian, $\lieder_{v_\ind{A}} L$, vanishes and $\star K_\ind{A}$ are closed one-forms on-shell.
We can relax the isometry condition and consider the case in which the currents obey the Maurer--Cartan equation \cite{Klimcik:1995ux,Klimcik:1995jn}
\[
\extder \star K_\ind{C} = \half \tilde{f}^{\ind{AB}}{}_\ind{C} \star K_\ind{A} \wedge \star K_\ind{B} ~,
\]
where $\tilde{f}^{\ind{AB}}{}_{\ind{C}}$ are the structure constants of some Lie algebra $\tilde{\alg{g}}$.
The sigma model exhibits a Poisson--Lie symmetry with respect to the coduality group $\tilde{\grp{G}}$.
The background is said to have generalised isometries, and should satisfy
\[\label{eq:LieF}
\lieder_{v_\ind{C}} F_\ind{IJ} = \tilde{f}^{\ind{AB}}{}_\ind{C} \, F_\ind{IL} F_\ind{KJ} \, v_\ind{A}^\ind{K} v_\ind{B}^\ind{L} ~.
\]
The dual sigma model leads to the same equation of motion, but with the roles of $\grp{G}$ and $\tilde{\grp{G}}$ interchanged.
The Lie derivative acting on differential forms should obey $\com{\lieder_{v_\ind{A}}}{\lieder_{v_\ind{B}}} = \lieder_{\com{v_\ind{A}}{v_\ind{B}}}$, which, combined with \eqref{eq:LieF}, imposes the following compatibility requirement on the structure constants of the original and dual Lie algebras
\[\label{eq:jacobi}
f_{\ind{AB}}{}^\ind{E} \tilde{f}^{\ind{CD}}{}_\ind{E} = f_{\ind{EA}}{}^\ind{C} \tilde{f}^{\ind{DE}}{}_\ind{B} - f_{\ind{EA}}{}^\ind{D} \tilde{f}^{\ind{CE}}{}_\ind{B} - f_{\ind{EB}}{}^\ind{C} \tilde{f}^{\ind{DE}}{}_\ind{A} + f_{\ind{EB}}{}^\ind{D} \tilde{f}^{\ind{CE}}{}_\ind{A} ~.
\]
This condition is manifestly dual in the sense that it is invariant under interchanging the roles of $\grp{G}$ and $\tilde{\grp{G}}$.
Furthermore, it is nothing other than the Jacobi identity on the Drinfel'd double $\alg{d} = \alg{g} \dsum \tilde{\alg{g}}$, turning the pair $(\alg{g},\tilde{\alg{g}})$ into a Lie bialgebra.
By definition the Drinfel'd double has an inner product with respect to which the decomposition $\alg{d} = \alg{g} \dsum \tilde{\alg{g}}$ is maximally isotropic
\[\label{eq:isotropyggt}
\< \alg{g}, \alg{g} \> = \< \tilde{\alg{g}}, \tilde{\alg{g}} \> = 0 ~.
\]

In order to gain more insight into the structure of this Drinfel'd double let us work in a specific basis: we denote the generators of $\alg{g}$ (respectively $\tilde{\alg{g}}$) by $T_\ind{A}$, $A = 1,2,\ldots,\dim\grp{G}$ (respectively $\tilde{T}^\ind{A}$).
With respect to this basis we assume the inner product \eqref{eq:isotropyggt} takes the form
\[\label{eq:bilinear2}
\< T_\ind{A}, T_\ind{B} \> = \< \tilde{T}^\ind{A}, \tilde{T}^\ind{B} \> = 0 ~, \qquad \< T_\ind{A}, \tilde{T}^\ind{B} \> = \delta_\ind{A}^\ind{B} ~.
\]
Consequently, there is a canonical pairing between the generators of the original and dual Lie algebras.
The commutation relations of the Drinfel'd double read
\[
& \com{T_\ind{A}}{T_\ind{B}} = f_\ind{AB}{}^\ind{C} T_\ind{C} ~, \qquad
\com{\tilde{T}^\ind{A}}{\tilde{T}^\ind{B}} = \tilde{f}^\ind{AB}{}_\ind{C} \tilde{T}^\ind{C} ~,
\\
& \com{T_\ind{A}}{\tilde{T}^\ind{B}} = f_\ind{CA}{}^\ind{B} \tilde{T}^\ind{C} + \tilde{f}^\ind{BC}{}_\ind{A} T_\ind{C} ~,
\]
where the third commutation relation follows from the ad-invariance of the inner product \eqref{eq:bilinear2} and is consistent with the Jacobi identity \eqref{eq:jacobi}.

Before giving the actions of the Poisson--Lie dual sigma models we introduce some notation.
Let $g$ (respectively $\tilde{g}$) be an arbitrary element of the Lie group $\grp{G}$ (respectively $\tilde{\grp{G}}$), and define
\[\label{eq:matricesabpi}
g^{-1} T_\ind{A} g & = a_\ind{A}{}^\ind{B} T_\ind{B} ~, \qquad
& g^{-1} \tilde{T}^\ind{A} g & = b^\ind{AB} T_\ind{B} + (a^{-1})_\ind{B}{}^\ind{A} \tilde{T}^\ind{B} ~, \qquad
& \Pi^\ind{AB} & = b^\ind{CA} a_\ind{C}{}^\ind{B} ~,
\\
\tilde{g}^{-1} \tilde{T}^\ind{A} \tilde{g} & = \tilde{a}^\ind{A}{}_\ind{B} \tilde{T}^\ind{B} ~, \qquad
& \tilde{g}^{-1} T_\ind{A} \tilde{g} & = \tilde{b}_\ind{AB} \tilde{T}^\ind{B} + (\tilde{a}^{-1})^\ind{B}{}_\ind{A} T_\ind{B} ~, \qquad
& \tilde{\Pi}_\ind{AB} & = \tilde{b}_\ind{CA} \tilde{a}^\ind{C}{}_\ind{B} ~.
\]
The matrices $a$, $b$ and $\Pi$ obey
\[
a(g^{-1}) = a^{-1}(g) ~, \qquad b(g^{-1}) = b^\transpose(g) ~, \qquad \Pi^\transpose(g) = -\Pi(g) ~,
\]
with similar relations holding for $\tilde{a}$, $\tilde{b}$ and $\tilde{\Pi}$.
In the absence of spectator fields, the sigma model whose background satisfies \eqref{eq:LieF} and its dual are then given by
\[\label{eq:dual_actions}
\Act_{\grp{G}}(g) & = \int \extder^2 \sigma \, (g^{-1} \partial_+ g)^\ind{A} E_\ind{AB} (g^{-1} \partial_- g)^\ind{B} ~, \qquad & E & = (\En + \Pi(g))^{-1} ~, \\
\Act_{\tilde{\grp{G}}}(\tilde{g}) & = \int \extder^2 \sigma \, (\tilde{g}^{-1} \partial_+ \tilde{g})_\ind{A} \tilde{E}^\ind{AB} (\tilde{g}^{-1} \partial_- \tilde{g} )_\ind{B} ~, \qquad & \tilde{E} & = (\En^{-1} + \tilde{\Pi}(\tilde{g}))^{-1} ~.
\]

\medskip

Let us now analyse what happens if we attempt to dualise in a subgroup $\grp{G}_0 \subset \grp{G}$ with the corresponding Lie algebra $\alg{g}_0 = \Lie(\grp{G}_0)$.
The Lie algebra $\alg{g}$ can be decomposed as $\alg{g} = \alg{g}_0 \dsum \alg{m}$.
We denote the generators of $\alg{g}_0$ by $T_a$, $a = 1,2,\ldots,\dim\grp{G}_0$, with the remaining generators of $\alg{g}$ forming a basis of $\alg{m}$ and labelled by a Greek index, $T_\alpha$, $\alpha = 1,2,\ldots,\dim\grp{G} - \dim\grp{G}_0$.
Starting from the sigma model on the Lie group $\grp{G}$ \eqref{eq:dual_actions}, which we write in the form
\[\label{eq:Lag_sigma}
\Act_{\grp{G}}(g) = \int \extder^2 \sigma \, \< g^{-1} \partial_+ g, (\En + \Pi(g) )^{-1} g^{-1} \partial_- g \> ~,
\]
the currents associated to the left action of $\grp{G}$ obey the Maurer--Cartan equation
\[\label{eq:eomres}
\extder \star K_\ind{A} = \half \tilde{f}^\ind{BC}{}_\ind{A} \star K_\ind{B} \wedge \star K_\ind{C} ~,
\]
where $\tilde{f}^\ind{BC}{}_\ind{A}$ are the structure constants of the dual algebra $\tilde{\alg{g}} = \Lie(\tilde{\grp{G}})$.
This Lie algebra has the same dimensionality as $\alg{g}$ and hence, using the canonical pairing \eqref{eq:bilinear2}, it is also possible to split its generators into $\tilde{T}^a$ and $\tilde{T}^\alpha$, defining bases of $\tilde{\alg{g}}_0$ and $\tilde{\alg{m}}$ respectively.

Restricting the equation of motion \eqref{eq:eomres} to the subalgebra $\alg{g}_0$
\[
\extder \star K_a = \half (\tilde{f}^{bc}{}_{a} \star K_b \wedge \star K_c + \tilde{f}^{\beta c}{}_a \star K_\beta \wedge \star K_c + \tilde{f}^{b \gamma}{}_a \star K_b \wedge \star K_\gamma + \tilde{f}^{\beta \gamma}{}_a \star K_\beta \wedge \star K_\gamma) ~,
\]
we see that in order to be able to dualise with respect to the subgroup $\grp{G}_0$, the dual algebra $\tilde{\alg{g}} = \tilde{\alg{g}}_0 \dsum \tilde{\alg{m}}$ should have the following structure
\[\label{eq:commutdual}
\com{\tilde{\alg{g}}_0}{\tilde{\alg{m}}} \subset \tilde{\alg{m}} ~, \qquad
\com{\tilde{\alg{m}}}{\tilde{\alg{m}}} \subset \tilde{\alg{m}} ~.
\]
This in turn imposes conditions on the subgroups $\grp{G}_0$ with respect to which we can Poisson--Lie dualise.

Before we proceed to investigate these conditions further, let us briefly comment on the relation to non-abelian duality.
An action that is invariant under the left action of $\grp{G}$ is also invariant under the left action of any subgroup $\grp{G}_0$ of $\grp{G}$, and hence it is always possible to dualise.
Indeed, when the action of $\grp{G}$ is a symmetry, the structure constants of the dual algebra $\tilde{\alg{g}}$ vanish by definition and \eqref{eq:commutdual} is trivially satisfied for any choice of $\alg{g}_0$.
In this case the associated Drinfel'd double is the semi-abelian double \cite{Klimcik:1995ux,Klimcik:1995jn}.
Thus, while it is possible to non-abelian dualise in any subgroup of $\grp{G}$, this is not the case for the Poisson--Lie duality.

The ad-invariance of the inner product \eqref{eq:bilinear2} imposes further structure on the commutation relations between $\alg{g}$ and $\tilde{\alg{g}}$
\[\label{eq:commutdualcons}
\com{\alg{g}_0}{\tilde{\alg{m}}} \subset \tilde{\alg{m}} ~.
\]
It follows that the space $\tilde{\alg{k}} = \alg{g}_0 \dsum \tilde{\alg{m}}$ also forms an algebra.
By construction $\tilde{\alg{k}}$ is an isotropic subalgebra of $\alg{d}$ with respect to the inner product \eqref{eq:isotropyggt}.
The Drinfel'd double can then always be decomposed as $\alg{d} = \alg{k} \dsum \tilde{\alg{k}}$, where $\alg{k}$ denotes the complement of $\tilde{\alg{k}}$ in $\alg{d}$, such that we again have the isotropy of both spaces
\[\label{eq:isotropykkt}
\< \alg{k}, \alg{k} \> = \< \tilde{\alg{k}}, \tilde{\alg{k}} \> = 0 ~.
\]
It is important to note that $\alg{k}$ does not necessarily form an algebra.

To summarise, we have two maximally isotropic decompositions, \eqref{eq:isotropyggt} and \eqref{eq:isotropykkt}, of the Drinfel'd double, $\alg{d} = \alg{g} \dsum \tilde{\alg{g}} = \alg{k} \dsum \tilde{\alg{k}}$, where these spaces have the additional structure
\[\label{eq:sumstruct}
\alg{g} & = \alg{g}_0 \dsum \alg{m} ~, & \qquad \tilde{\alg{g}} & = \tilde{\alg{g}}_0 \dsum \tilde{\alg{m}} ~,
\\
\alg{k} & = \tilde{\alg{g}}_0 \dsum \alg{m} ~, & \qquad \tilde{\alg{k}} & = \alg{g}_0 \dsum \tilde{\alg{m}} ~.
\]
The spaces $\alg{g}$ and $\tilde{\alg{g}}$ are Lie algebras by definition, while $\alg{g}_0$ forms an algebra by assumption.
To be able to dualise the Poisson--Lie symmetric model on $\grp{G}$ with respect to the subgroup $\grp{G}_0$ we require that \eqref{eq:commutdual} and \eqref{eq:commutdualcons} are satisfied, which in particular implies that $\tilde{\alg{k}}$ also forms an algebra.

One approach to obtaining the backgrounds of the Poisson--Lie duals of \eqref{eq:Lag_sigma} is to rewrite the original sigma model in terms of a field taking values in the subgroup $\grp{G}_0$ together with spectator fields
\[
\< g^{-1} \partial_+ g, (\En + \Pi(g) )^{-1} g^{-1} \partial_- g \> & =
P^{(0)}_{a b} (g_0^{-1} \partial_+ g_0)^a (g_0^{-1} \partial_- g_0)^b + P^{(1)}_{a\nu} (g_0^{-1} \partial_+ g_0 )^a \partial_-x^\nu
\\ & \quad + P^{(2)}_{\mu b} \partial_+ x^\mu (g_0^{-1} \partial_- g_0 )^b + P^{(3)}_{\mu \nu} \partial_+ x^{\mu} \partial_- x^\nu ~.
\]
In order to dualise, for example using the Poisson--Lie duality relations in the presence of spectators presented in \cite{Bossard:2001au}, the operator $Q$ defined through $P^{(0)} = (Q + \Pi(g_0))^{-1}$ should be a function of the spectator fields $x^\mu$ only.
This is satisfied precisely when the dual algebra $\tilde{\alg{g}}$ has the structure \eqref{eq:commutdual}.

%%%%%%%%%%%%%%%%%%%%%%%%%%%%%%%%%%%%%%%%%%%%%%%%%%%%%%%%%%%%%%%%%%%%%%%%%%%%%%%%
\subsection{First-order action on the Drinfel'd double and Poisson--Lie duality}\label{ssec:firstorder}

In this section we will take an alternative path to finding the dual model and computing its background.
We start from the first-order action on the Drinfel'd double \cite{Klimcik:1995dy,Klimcik:1996nq,Klimcik:2002zj,Klimcik:2015gba}.
In the standard decomposition of the Drinfel'd double $\alg{d} = \alg{g} \dsum \tilde{\alg{g}}$, both $\alg{g}$ and $\tilde{\alg{g}}$ are Lie algebras.
To recover the Poisson--Lie dual models on $\grp{G}$ and $\tilde{\grp{G}}$ \eqref{eq:dual_actions} we integrate out the degrees of freedom associated to $\tilde{\alg{g}}$ and $\alg{g}$ respectively.
As discussed in \secref{ssec:pldsigma}, dualising a Poisson--Lie symmetric model with respect to a subgroup $\grp{G}_0 \subset \grp{G}$ corresponds to considering alternative decompositions of the Drinfel'd double $\alg{d} = \alg{k} \dsum \tilde{\alg{k}}$, where $\tilde{\alg{k}}$ is a Lie algebra, but $\alg{k}$ need not be.
The Poisson--Lie dual is then given by integrating out the degrees of freedom associated to $\tilde{\alg{k}}$.
Such configurations have been considered previously in \cite{Klimcik:1996nq,Klimcik:2002zj,Klimcik:2015gba}.
Although given in a different form, which is of use in the application to the $\eta$~deformation in \secref{sec:application}, the presented results are equivalent.

%%%%%%%%%%%%%%%%%%%%%%%%%%%%%%%%%%%%%%%%
\paragraph{First-order action on the Drinfel'd double.}
The first-order action on the Drinfel'd double that underlies the Poisson--Lie duality is \cite{Klimcik:1995dy,Klimcik:1996nq}
\[\label{eq:action_double}
\Act_\grp{D}(l) = \int \extder \tau \extder \sigma \, \Big[ \half \< l^{-1} \partial_\sigma l , l^{-1} \partial_\tau l \> - \half K(l^{-1} \partial_\sigma l ) \Big] +\WZ(l)~,
\]
where the dynamical field $l \in \grp{D}$ and
\[
\WZ(l) = - \sfrac{1}{12} \int \, \extder^{-1} \< l^{-1} \extder l , \com{l^{-1} \extder l}{l^{-1} \extder l} \>~,
\]
is the standard Wess--Zumino term.
The first and third terms together form a Wess--Zumino--Witten model on the Drinfel'd double, albeit with the light-cone coordinates replaced by $\tau$ and $\sigma$.
The quadratic form $K$ encodes the details of the particular model.
Let us note that this action is not Lorentz invariant and generalises the duality-invariant action of \cite{Tseytlin:1990nb,Tseytlin:1990va} underlying abelian duality.

The field $l \in \grp{D}$ can be parametrised as $l = \tilde{k} k$, where $\tilde{k} \in \tilde{\grp{K}} = \exp[\tilde{\alg{k}}]$ and $k \in \tilde{\grp{K}}\backslash\grp{D}$.
(Note that in this paper we will not be concerned with any global issues of such parametrisations.)
Using the fact that $\tilde{k}$ is a group-valued field, the action \eqref{eq:action_double} simplifies to
\[\label{eq:action_double_simp}
\Act_\grp{D}(\tilde{k} k) & = \int \extder \tau \extder \sigma \, \Big[\< \tilde{k}^{-1} \partial_\sigma \tilde{k}, \partial_\tau k k^{-1} \> - \half K ( k^{-1} \partial_\sigma k + \Ad_k^{-1} \tilde{k}^{-1} \partial_\sigma \tilde{k} )\Big] \\
& \quad + \half \int \extder \tau \extder \sigma \, \< k^{-1}\partial_\sigma k, k^{-1}\partial_\tau k \> + \WZ(k) ~.
\]
Note that if $\alg{k}$ is a Lie algebra we have that $\tilde{\grp{K}}\backslash\grp{D} \sim \grp{K}$.
In this case we can take $k \in \grp{K}$ and the second line of \eqref{eq:action_double_simp} vanishes.

%%%%%%%%%%%%%%%%%%%%%%%%%%%%%%%%%%%%%%%%
\paragraph{Integrating out $\tilde{\alg{k}}$.}
In addition to the basis $\set{T_\ind{A}, \tilde{T}^\ind{A}}$ of $\alg{d} = \alg{g} \dsum \tilde{\alg{g}}$, we introduce bases of $\alg{k}$ and $\tilde{\alg{k}}$, denoted $S_\ind{A}$ and $\tilde{S}^\ind{A}$ respectively, such that the inner product \eqref{eq:bilinear2} again has the form
\[\label{eq:bilinears}
\< S_\ind{A}, S_\ind{B} \> = \< \tilde{S}^\ind{A}, \tilde{S}^\ind{B} \> = 0 ~, \qquad \< S_\ind{A}, \tilde{S}^\ind{B} \> = \delta_\ind{A}^\ind{B} ~.
\]
In the spirit of \secref{ssec:pldsigma} we define the matrices
\[\label{eq:kmatrices}
k^{-1} S_\ind{A} k = a(k)_\ind{A}{}^\ind{B} S_\ind{B} + c(k)_\ind{AB} \tilde{S}^\ind{B} ~, & \qquad
k^{-1} \tilde{S}^\ind{A} k = b(k)^\ind{AB} S_\ind{B} + a(k^{-1})_\ind{B}{}^\ind{A} \tilde{S}^\ind{B} ~, \\
\Pi(k)^\ind{AB} & = b^\ind{CA}(k)(a(k^{-1})^{-1})_\ind{C}{}^\ind{B} ~,
\]
where, in contrast to \eqref{eq:matricesabpi}, the new matrix $c(k)$ appears as $\alg{k}$ is no longer assumed to form an algebra.
Using the ad-invariance of the inner product \eqref{eq:bilinears}, it is possible to show that $\Pi(k)$ is antisymmetric, and reduces to the standard form when $\alg{k}$ forms an algebra and we take $k \in \grp{K}$.
Indeed, in this case $c(k)$ vanishes and the identity $a(k^{-1}) = a(k)^{-1}$ follows.
We also introduce the projectors $\Proj{\alg{k}}$ and $\Proj{\tilde{\alg{k}}}$ onto the spaces $\alg{k}$ and $\tilde{\alg{k}}$ respectively.

In order to integrate out the degrees of freedom contained in $\tilde{k}$, we need to specify the action of the bilinear form $K$.
An arbitrary element $x \in \alg{d}$ can be uniquely decomposed as $x = y + z$ where $y \in \alg{k}$ and $z \in \tilde{\alg{k}}$.
Without loss of generality we may take
\[
K(x) = \< z, \Gnt z \> + \< (y + \Bnt z), \Gnt^{-1} (y + \Bnt z) \> ~,
\]
where $\Gnt$ and $\Bnt$ are the symmetric and antisymmetric parts of a general operator $\Ent: \tilde{\alg{k}} \to \alg{k}$.
The prime symbol highlights that these operators act on the space $\tilde{\alg{k}}$.
This is in contrast to $\En$, which we reserve for the corresponding operator acting on $\tilde{\alg{g}}$.
Taking $x = k^{-1} \partial_\sigma k + \Ad_k^{-1} \tilde{k}^{-1} \partial_\sigma \tilde{k}$ we have
\[
y & = \Proj{\alg{k}} k^{-1} \partial_\sigma k + \Proj{\alg{k}} \Ad_k^{-1} \Proj{\tilde{\alg{k}}} \tilde{k}^{-1} \partial_\sigma \tilde{k} ~, \\
z & = \Proj{\tilde{\alg{k}}} k^{-1} \partial_\sigma k + \Proj{\tilde{\alg{k}}} \Ad_k^{-1} \Proj{\tilde{\alg{k}}} \tilde{k}^{-1} \partial_\sigma \tilde{k} ~.
\]
The operator $\Proj{\tilde{\alg{k}}} \Ad_k^{-1} \Proj{\tilde{\alg{k}}}$ is invertible on $\tilde{\alg{k}}$ and hence we can solve for $\tilde{k}^{-1} \partial_\sigma \tilde{k}$
\[
\tilde{k}^{-1} \partial_\sigma \tilde{k} = (\Proj{\tilde{\alg{k}}} \Ad_k^{-1} \Proj{\tilde{\alg{k}}})^{-1} [z - \Proj{\tilde{\alg{k}}} k^{-1} \partial_\sigma k] ~.
\]
This in turn allows us to write the following expression for $y$ in terms of $k$ and $z$
\[
y & = \Proj{\alg{k}} k^{-1} \partial_\sigma k + \Proj{\alg{k}} \Ad_k^{-1} \Proj{\tilde{\alg{k}}} (\Proj{\tilde{\alg{k}}} \Ad_k^{-1} \Proj{\tilde{\alg{k}}})^{-1} [z - \Proj{\tilde{\alg{k}}} k^{-1} \partial_\sigma k] \\
& = \Proj{\alg{k}} k^{-1} \partial_\sigma k + \Pi(k) [z - \Proj{\tilde{\alg{k}}} k^{-1} \partial_\sigma k] ~,
\]
where
\[
\Pi(k) = \Proj{\alg{k}} \Ad_k^{-1} \Proj{\tilde{\alg{k}}} (\Proj{\tilde{\alg{k}}} \Ad_k^{-1} \Proj{\tilde{\alg{k}}})^{-1},
\]
is the operator form of the matrix introduced in \eqref{eq:kmatrices}.
Using these relations we can eliminate $\tilde{k}$ in the action \eqref{eq:action_double_simp} in favour of $k$ and $z$.
The resulting action is quadratic in $z$, which can then be integrated out.
Doing so, we find the following Lorentz-invariant action for the field $k$
\[\label{eq:actionS1}
\Act_{\tilde{\grp{K}}\backslash\grp{D}}(k) = & \int \extder^2 \sigma \, \< k^{-1} \partial_+ k, (\half \Proj{\tilde{\alg{k}}}^\rot - \half \Proj{\alg{k}}^\rot + \Proj{\tilde{\alg{k}}}^\rot (\Ent+\Pi(k))^{-1} \, \Proj{\alg{k}}^\rot) k^{-1} \partial_- k \> + \WZ(k) ~,
\]
where $\Proj{\alg{k}}^\rot = \Proj{\alg{k}} - \Pi(k) \Proj{\tilde{\alg{k}}}$ and $\Proj{\tilde{\alg{k}}}^\rot = \Proj{\tilde{\alg{k}}} + \Pi(k) \Proj{\tilde{\alg{k}}}$.
The combination $\half\Proj{\tilde{\alg{k}}}^\rot-\half\Proj{\alg{k}}^\rot$ is an antisymmetric operator under the inner product and thus, along with the Wess--Zumino term, only contributes to the $B$-field of the background.
As $k \in \tilde{\grp{K}} \backslash \grp{D}$ this action should be invariant under the gauge transformation $k \to \tilde{k} k$ where $\tilde{k} \in \tilde{\grp{K}}$.
This can be readily seen from the identities $\Proj{\alg{k}}^\rot \Ad_k^{-1} \tilde{k}^{-1} \partial_{\pm} \tilde{k} = 0$ and $\Pi(\tilde{k} k) = \Pi(k)$, and the Polyakov--Wiegmann identity for the Wess--Zumino term.
Finally, let us note that if $\alg{k}$ forms an algebra and we take $k \in \grp{K}$, then the action reduces to
\[
\Act_{\grp{K}}(k) = \int \extder^2 \sigma \, \< k^{-1} \partial_+ k, (\Ent+\Pi(k))^{-1} k^{-1} \partial_- k \> ~,
\]
due to the isotropy of $\alg{k}$ with respect to the inner product.
As expected we recover an action that takes the form of the Poisson--Lie dual models \eqref{eq:dual_actions} considered in \secref{ssec:pldsigma}.

%%%%%%%%%%%%%%%%%%%%%%%%%%%%%%%%%%%%%%%%
\paragraph{Wess--Zumino term.}
The topological term
\[
\WZ(k) = - \sfrac{1}{12} \int \, \extder^{-1} \< k^{-1} \extder k, \com{k^{-1} \extder k}{k^{-1} \extder k} \> ~,
\]
is only defined up to a total derivative.
To fix its contribution to the $B$-field uniquely we may assume that we can fix a gauge such that $k = \tilde{g} g$, where $\tilde{g} \in \tilde{\grp{G}}$ and $g \in \grp{G}$.
Using the Polyakov--Wiegmann identity we may write the Wess--Zumino term as the following two-dimensional integral
\[
\WZ(\tilde{g} g) = - \half \int \extder^2 \sigma \, \Big[ \< \partial_+ g g^{-1}, \tilde{g}^{-1} \partial_- \tilde{g} \> - \< \partial_- g g^{-1}, \tilde{g}^{-1} \partial_+ \tilde{g} \> \Big] ~.
\]

%%%%%%%%%%%%%%%%%%%%%%%%%%%%%%%%%%%%%%%%
\paragraph{Equations of motion.}
Defining the gauge invariant currents
\[
\label{eq:currents}
K_- = (E-\half)J_- ~, \quad K_+ = (E^\transpose+\half) J_+ ~,
\]
where $J_\pm = k^{-1} \partial_\pm k$ and
\[
E \equiv \half \Proj{\tilde{\alg{k}}}^\rot - \half \Proj{\alg{k}}^\rot + \Proj{\tilde{\alg{k}}}^\rot (\Ent+\Pi(k))^{-1} \, \Proj{\alg{k}}^\rot~,
\]
varying the action \eqref{eq:actionS1} leads to the following equations of motion
\[
\label{eq:eom1}
\< \Delta, \partial_+ K_- + \partial_- K_+ + \com{K_+}{K_-} \> =0 ~, \qquad \Delta \equiv k^{-1}\delta k ~.
\]

%%%%%%%%%%%%%%%%%%%%%%%%%%%%%%%%%%%%%%%%
\paragraph{$\Ent$ operator.}
The operator $\Ent$ acts on the space $\tilde{\alg{k}}$, and is related in a non-trivial way to its counterpart $\En$, reserved for the corresponding operator acting on $\tilde{\alg{g}}$.
Let us now present the precise relationship between $\Ent$ and $\En$.
The two pairs of dual bases $\set{T,\tilde{T}}$ and $\set{S,\tilde{S}}$ are related by an $\grp{O}_{n,n}(\Real)$ transformation preserving the inner product \eqref{eq:bilinear2}, that is
\[
& \begin{pmatrix} S \\ \tilde{S} \end{pmatrix}
=
\begin{pmatrix} W & X \\ Y & Z \end{pmatrix}
\begin{pmatrix} T \\ \tilde{T} \end{pmatrix} ~,
\qquad
\begin{pmatrix} W & X \\ Y & Z \end{pmatrix}
\begin{pmatrix} 0 & \identity \\ \identity & 0 \end{pmatrix}
\begin{pmatrix} W^\transpose & Y^\transpose \\ X^\transpose & Z^\transpose \end{pmatrix}
=
\begin{pmatrix} 0 & \identity \\ \identity & 0 \end{pmatrix} ~.
\]
Taking the quadratic form in the basis $\set{T,\tilde{T}}$ to be given by
\[
K_{T} = \begin{pmatrix} \Gn^{-1} & \Gn^{-1} \Bn \\ - \Bn \Gn^{-1} & \Gn-\Bn \Gn^{-1} \Bn \end{pmatrix} ~,
\]
the quadratic form in the basis $\set{S,\tilde{S}}$ becomes
\[
K_{S} & = \begin{pmatrix} W & X \\ Y & Z \end{pmatrix}
\begin{pmatrix} \Gn^{-1} & \Gn^{-1} \Bn \\ - \Bn \Gn^{-1} & \Gn-\Bn \Gn^{-1} \Bn \end{pmatrix}
\begin{pmatrix} W^\transpose & Y^\transpose \\ X^\transpose & Z^\transpose \end{pmatrix}
\\ & \equiv \begin{pmatrix} \Gnt^{-1} & \Gnt^{-1} \Bnt \\ - \Bnt \Gnt^{-1} & \Gnt-\Bnt \Gnt^{-1} \Bnt \end{pmatrix} ~,
\]
corresponding to the non-linear transformation rule
\[\label{eq:E0t}
\Ent = (X^\transpose + \En^{-1} W^\transpose)^{-1} (Z^\transpose + \En^{-1} Y^\transpose) ~.
\]

Let us check the validity of this expression for certain special cases.
For the Poisson--Lie duality presented in \secref{ssec:pldsigma}, the first model is obtained by integrating out the $\tilde{\grp{G}}$-valued field.
In this case, $W = Z = \identity$ and $X = Y = 0$, and hence $\Ent = \En$, as expected.
The dual sigma model on the other hand corresponds to the integrating out the $\grp{G}$-valued field.
In this second case, $X = Y = \identity$ and $W = Z = 0$, and the relation $\Ent = \En^{-1}$ follows.
This is nothing other than the relation between the Poisson--Lie dual models \eqref{eq:dual_actions}.

We introduced the spaces $\alg{k}$ and $\tilde{\alg{k}}$ to study Poisson--Lie dual models where we dualise with respect to a subgroup $\grp{G}_0 \subset \grp{G}$.
Recalling the associated structure \eqref{eq:sumstruct}, along with the splitting of the generators for $\alg{g}$ and $\tilde{\alg{g}}$ as $\set{T_\ind{A}} = \set{T_a,T_\alpha}$ and $\set{\tilde{T}^\ind{A}} = \set{\tilde{T}^a,\tilde{T}^\alpha}$ respectively, it follows that the bases of $\alg{k}$ and $\tilde{\alg{k}}$ are given by $\set{S_\ind{A}} = \set{\tilde{T}^a,T_\alpha}$ and $\set{S^\ind{A}} = \set{T_a,\tilde{T}^\alpha}$ respectively.
It is therefore natural to introduce the additional block structure
\[
\label{eq:blockT}
T_\ind{A} = \begin{pmatrix} T_a \\ T_\alpha \end{pmatrix} ~, \qquad
\tilde{T}^\ind{A} = \begin{pmatrix} \tilde{T}^a \\ \tilde{T}^\alpha \end{pmatrix} ~.
\]
For the case of interest, that is the Poisson--Lie dual with respect to $\grp{G}_0$, we therefore have
\[
X = Y = \begin{pmatrix} \identity & 0 \\ 0 & 0 \end{pmatrix} ~, \qquad
W = Z = \begin{pmatrix} 0 & 0 \\ 0 & \identity \end{pmatrix} ~.
\]
Substituting these expressions into the right-hand side of \eqref{eq:E0t} and defining
\[
\En = \begin{pmatrix} A & B \\ C & D \end{pmatrix} ~,
\]
we find
\[\label{eq:E0tmatrix}
\Ent = \begin{pmatrix} A^{-1} & A^{-1} B \\ - C A^{-1} & D - C A^{-1} B \end{pmatrix} ~.
\]
assuming that $A$ is invertible.

%%%%%%%%%%%%%%%%%%%%%%%%%%%%%%%%%%%%%%%%
\paragraph{Generalisation to coset spaces.}
We conclude this section with a discussion of how the above construction generalises to coset spaces \cite{Klimcik:1996np,Sfetsos:1999zm,Squellari:2011dg}.
Let us introduce a subgroup $\grp{H} \subset \grp{G}$ and the corresponding Lie algebra $\alg{h} = \Lie(\grp{H})$.
Note that the dual space $\tilde{\alg{h}}$ identified by the canonical pairing \eqref{eq:bilinear2} does not necessarily form an algebra.
Moreover, in the most general case there is no basis of generators for $\alg{g}$ such that a subset generates $\alg{g}_0$ and another subset generates $\alg{h}$.
Consequently, in addition to the two pairs of dual bases $\set{T_\ind{A},\tilde{T}^\ind{A}}$ and $\set{S_\ind{A},\tilde{S}^\ind{A}}$ introduced above, we also introduce a new basis of $\alg{d} = \alg{g} \dsum \tilde{\alg{g}}$ denoted $\set{U_\ind{A},\tilde{U}^\ind{A}}$ such that a subset of $\set{U_\ind{A}}$ generates the gauge group $H$.
More explicitly,
\[
U_\ind{A} = \begin{pmatrix} U_{\bar a} \\ U_{\bar\alpha} \end{pmatrix} ~,
\]
where $\vectorspan{U_{\bar a}} = \alg{h}$, $\bar a = 1,\ldots,\dim\grp{H}$.
The remaining generators are labelled by a Greek index $\bar\alpha = 1,\ldots,\dim\grp{G}-\dim\grp{H}$.
The operator $\En: \tilde{\alg{g}} \to \alg{g}$ written in the $\set{U_\ind{A},\tilde{U}^\ind{A}}$ basis can be written in block form
\[
\En = \left(\begin{array}{cc}
\bar A & \bar B \\
\bar C & \bar D
\end{array}\right) ~,
\]
where $\bar A: \tilde{\alg{h}}\to \alg{h}$.
Following the construction of \cite{Sfetsos:1999zm}, we are interested in the limit where the number of fields in the original model is reduced by $\dim\grp{H}$.
For this purpose a parameter $\epsilon$ is introduced
\[
\En = \left(\begin{array}{cc}
\epsilon^{-1} \bar A & \bar B \\
\bar C & \bar D
\end{array}\right) ~,
\]
and as $\epsilon \to 0$, under some mild assumptions outlined in \cite{Sfetsos:1999zm}, the model develops a local gauge invariance.
This gauge invariance also appears in the first-order action on the Drinfel'd double and hence we would expect this to also be the case for the various Poisson--Lie dual models.
Furthermore, this has been explicitly demonstrated for the Poisson--Lie dual with respect to $\grp{G}$ in \cite{Squellari:2011dg} and the proof should go in a similar way for the more general Poisson--Lie duals considered here.

%%%%%%%%%%%%%%%%%%%%%%%%%%%%%%%%%%%%%%%%%%%%%%%%%%%%%%%%%%%%%%%%%%%%%%%%%%%%%%%%
\section{Application to the \texorpdfstring{$\eta$}{eta} and \texorpdfstring{$\lambdastar$}{lambda-star} deformations}\label{sec:application}

In this section we explore the application of the methods of \secref{sec:general} to the $\eta$~deformation of the symmetric space sigma model for the coset $\grp{G}/\grp{H}$ \cite{Delduc:2013fga}.
The action for this model can be written in the form \eqref{eq:Lag_sigma} for a certain choice of the operator $\En$ that preserves the integrability of the undeformed model.

%%%%%%%%%%%%%%%%%%%%%%%%%%%%%%%%%%%%%%%%%%%%%%%%%%%%%%%%%%%%%%%%%%%%%%%%%%%%%%%%
\subsection{The \texorpdfstring{$\eta$}{eta} and \texorpdfstring{$\lambdastar$}{lambda-star}~deformed models}\label{ssec:etalambda}

We begin by reviewing the construction of the $\eta$~deformation of the symmetric space sigma model for the coset $\grp{G}/\grp{H}$ \cite{Delduc:2013fga} (with $\grp{G}$ a compact simple Lie group).
In particular, we write the action in the form \eqref{eq:Lag_sigma}.
Starting from the corresponding model on the Drinfel'd double \eqref{eq:action_double} this allows us to investigate in which subgroups of $\grp{G}$ the $\eta$~deformed model can be Poisson--Lie dualised and construct the corresponding backgrounds.

Our starting point for the construction of the $\eta$~deformed model will be the compact simple Lie group $\grp{G}$, with Lie algebra $\alg{g} = \Lie(\grp{G})$, and the complex Drinfel'd double $\grp{D} \equiv \grp{G}^\Complex$, with Lie algebra $\alg{d} \equiv \alg{g}^\Complex = \Lie(\grp{D})$ \cite{Klimcik:2002zj,Delduc:2013fga,Vicedo:2015pna,Klimcik:2015gba}.

%%%%%%%%%%%%%%%%%%%%%%%%%%%%%%%%%%%%%%%%
\paragraph{The complex Drinfel'd double.}
Let us introduce a Cartan-Weyl basis for the \foreign{complex} Lie algebra $\alg{d} \equiv \alg{g}^\Complex$, composed of the Cartan generators $\set{h_i}$ and positive $\set{e_\ind{M}}$ and negative $\set{f_\ind{M}}$ roots.
The Cartan generators and the simple roots satisfy the defining relations
\[
\com{h_i}{e_j} = a_{ij} e_j ~, \qquad \com{h_i}{f_j} = - a_{ij} f_j ~, \qquad \com{e_i}{f_j} = \delta_{ij} h_j ~,
\]
where $a_{ij}$ is the symmetrised Cartan matrix.
We furthermore define the non-simple roots such that if $e_\ind{M}$ and $e_\ind{N}$ are two positive roots commuting to give a further positive root $e_\ind{P}$, then
\[
\com{e_\ind{M}}{e_\ind{N}} = \mathcal{N}_\ind{MNP} e_\ind{P} ~, \qquad \com{f_\ind{M}}{f_\ind{N}} = - \mathcal{N}_\ind{MNP} f_\ind{P} ~.
\]
This choice then allows us to normalise the non-vanishing components of the Killing form such that
\[
\killing(h_i,h_j) = a_{ij} ~, \qquad \killing(e_\ind{M}, f_\ind{N}) = \delta_\ind{MN} ~.
\]

The complex Lie algebra $\alg{d}$ can then be decomposed into two real subalgebras
\[\label{eq:doubdecomp}
\alg{d} = \alg{g} \dsum \alg{b} ~,
\]
where $\alg{g}$ is the compact real form generated by
\[\label{eq:compactreal}
\set{T_\ind{A}} \equiv \set{i c_{ij}h_j,i(e_\ind{M}+f_\ind{M}),-(e_\ind{M}-f_\ind{M})} ~,
\]
and $\alg{b}$, the coduality algebra, is the Borel subalgebra generated by
\[\label{eq:borel}
\set{\tilde{T}^\ind{A}} \equiv \set{b_{ij}h_j,e_\ind{M},i e_\ind{M}} ~.
\]
Here $c_{ij}$ and $b_{ij}$ are real constants such that the corresponding matrices $c$ and $b$ have non-vanishing determinant.
If the latter are further constrained in terms of the former through the relation
\[\label{eq:bca}
b = c^{-\transpose} a^{-1} ~,
\]
then the ad-invariant inner product on the double
\[
\< x , x' \> = \Im \killing(x , x') ~, \qquad x, x' \in \alg{d} ~,
\]
is normalised and maximally isotropic with respect to the decomposition \eqref{eq:doubdecomp}
\[
\<T_\ind{A}, \tilde{T}^\ind{B}\> = \delta_\ind{A}^\ind{B} ~, \qquad \<T_\ind{A},T_\ind{B}\> = \<\tilde{T}^\ind{A},\tilde{T}^\ind{B}\> = 0 ~.
\]

Introducing the structure constants for $\alg{g}$ and $\alg{b}$
\[
\com{T_\ind{A}}{T_\ind{B}} & = f_\ind{AB}{}^\ind{C} T_\ind{C} ~, \qquad \com{\tilde{T}^\ind{A}}{\tilde{T}^\ind{B}} = \tilde{f}^\ind{AB}{}_\ind{C} \tilde{T}^\ind{C} ~,
\]
we have
\[
\com{T_\ind{A}}{\tilde{T}^\ind{B}} & = f_\ind{CA}{}^\ind{B} \tilde{T}^\ind{C} + \tilde{f}^\ind{BC}{}_\ind{A} T_\ind{C} ~.
\]
We further introduce the components of the Killing form for the compact real form $\alg{g}$
\[\label{eq:killing}
\killing (T_\ind{A}, T_\ind{B}) = \killing_\ind{AB} ~,
\]
and its inverse, $\killing^\ind{AC} \killing_\ind{CB} = \delta^\ind{A}_\ind{B}$, with which we lower and raise indices.
Let us recall that lowering the upper index of $f_\ind{AB}{}^\ind{C}$, the structure constants of the simple Lie algebra $\alg{g}$, results in the completely antisymmetric tensor $f_\ind{ABC}$.

Finally let us note two useful relations for which we introduce the projectors $\Proj{\alg{g}}$ and $\Proj{\alg{b}}$ onto $\alg{g}$ and $\alg{b}$ respectively.
If we parametrise $g = \exp[\theta] \in \grp{G}$ and $\tilde{g} = \exp[\tilde{\theta}] \in \grp{B}$, with $\theta \in \alg{g}$ and $\tilde{\theta} \in \alg{b}$, then we have
\unskip\footnote{The final equalities in these relations follow from $\Proj{\alg{b}} \ad_\theta^n = (\Proj{\alg{b}} \ad_\theta)^n$ and $\Proj{\alg{g}} \ad_{\tilde{\theta}}^n = (\Proj{\alg{g}} \ad_{\tilde{\theta}})^n$, both of which can be proven by induction.
Here we give the proof for the first, while a similar proof can be easily constructed for the second.
The relation $\Proj{\alg{b}} \ad_\theta^n = (\Proj{\alg{b}} \ad_\theta)^n$ is clearly true for $n = 1$.
While if we assume it to be true for $n = k - 1$ we have $(\Proj{\alg{b}} \ad_\theta)^k = \Proj{\alg{b}} \ad_\theta (\Proj{\alg{b}} \ad_\theta)^{k-1} = \Proj{\alg{b}} \ad_\theta \Proj{\alg{b}} \ad_\theta^{k-1} = \Proj{\alg{b}} \ad_\theta^k - \Proj{\alg{b}} \ad_\theta \Proj{\alg{g}} \ad_\theta^{k-1} = \Proj{\alg{b}} \ad_\theta^k $, and hence it is true for $n = k$.
The final step follows as $\ad_\theta$ maps $\alg{g}$ to $\alg{g}$.}
\[\label{eq:rel1}
\Proj{\alg{b}} \Ad_g = \Proj{\alg{b}} \Ad_{\exp[\theta]} = \Proj{\alg{b}} \exp[\ad_\theta] = \exp[\Proj{\alg{b}} \ad_\theta] ~,
\\ \Proj{\alg{g}} \Ad_{\tilde{g}} = \Proj{\alg{g}} \Ad_{\exp[\tilde{\theta}]} = \Proj{\alg{g}} \exp[\ad_{\tilde{\theta}}] = \exp[\Proj{\alg{g}} \ad_{\tilde{\theta}}] ~.
\]
Therefore when $g$ and $\tilde{g}$ are exponentials of algebra elements, which we will henceforth always assume to be the case, we have that
\[\label{eq:rel2}
\Proj{\alg{b}} \Ad_g^{-1} \Proj{\alg{b}} \Ad_g & = \exp[-\Proj{\alg{b}} \ad_{\theta}] \exp[\Proj{\alg{b}} \ad_{\theta} ] = \Proj{\alg{b}} ~,
\\ \Proj{\alg{g}} \Ad_{\tilde{g}}^{-1} \Proj{\alg{g}} \Ad_{\tilde{g}} & = \exp[-\Proj{\alg{g}} \ad_{\tilde{\theta}}] \exp[\Proj{\alg{g}} \ad_{\tilde{\theta}} ] = \Proj{\alg{g}} ~.
\]

%%%%%%%%%%%%%%%%%%%%%%%%%%%%%%%%%%%%%%%%
\paragraph{Drinfel'd--Jimbo R-matrix.}
The Drinfel'd--Jimbo R-matrix will play an important role in the construction of the $\eta$~deformed models.
This R-matrix is defined by its action on the Cartan generators and the positive and negative roots
\[\label{eq:rmat}
R e_\ind{M} = i e_\ind{M} ~, \qquad R f_\ind{M} = - i f_\ind{M} ~, \qquad R h_i = 0 ~,
\]
where we have normalised such that the R-matrix satisfies the modified classical Yang--Baxter equation
\[
\com{R y}{R y'} - R(\com{y}{R y'} + \com{R y}{y'} ) = \com{y}{y'} ~, \qquad y,y' \in \alg{g} ~.
\]
We further define its action on the basis $\set{T_\ind{A}}$
\[
R T_\ind{A} = R_\ind{A}{}^\ind{B} T_\ind{B} ~,
\]
where we note that \eqref{eq:rmat} preserves the compact real form.
Furthermore, one can see that the R-matrix is an antisymmetric operator with respect to the Killing form \eqref{eq:killing}.
That is $\killing(T_\ind{A}, R T_\ind{B}) = - \killing(R T_\ind{A}, T_\ind{B})$ and hence $R_\ind{B}{}^\ind{C} \killing_\ind{CA} = - R_\ind{A}{}^\ind{C} \killing_\ind{CB}$.
Finally, the structure constants of the Borel algebra $\alg{b}$ are related to those of $\alg{g}$ through the R-bracket
\[\label{eq:strucrel}
R_\ind{A}{}^\ind{D} f_\ind{DB}{}^\ind{C} + R_\ind{B}{}^\ind{D} f_\ind{AD}{}^\ind{C} = \killing_\ind{AD} \killing_\ind{BE} \killing^\ind{CF} \tilde{f}^\ind{DE}{}_\ind{F} ~.
\]

%%%%%%%%%%%%%%%%%%%%%%%%%%%%%%%%%%%%%%%%
\paragraph{Symmetric space.}
To construct the $\eta$~deformed symmetric space sigma model for the coset $\grp{G}/\grp{H}$ we introduce the subgroup $\grp{H}$ with the corresponding Lie algebra $\alg{h} = \Lie(\grp{H})$.
As we have a symmetric space we can decompose $\alg{g}$ according to the underlying $\Integer_2$ automorphism
\[\label{eq:symspace}
\alg{g} = \alg{h} \dsum \alg{p} ~, \qquad \com{\alg{h}}{\alg{h}} \subset \alg{h} ~, \quad \com{\alg{h}}{\alg{p}} \subset \alg{p} ~, \quad \com{\alg{p}}{\alg{p}} \subset \alg{h} ~.
\]
Furthermore, the Killing form \eqref{eq:killing} is orthogonal under this decomposition.
Thus, recalling the bases $\{U_{\bar a}\}$ of $\alg{h}$ and $\{U_{\bar\alpha}\}$ of $\alg{p}$ introduced in \secref{ssec:firstorder}, we have
\[\label{eq:orthog}
\killing (U_{\bar a} , U_{\bar b}) = \killing_{\bar a \bar b} ~, \qquad \killing (U_{\bar\alpha}, U_{\bar\beta}) = \killing_{\bar\alpha\bar\beta} ~, \qquad \killing(U_{\bar a}, U_{\bar\beta}) = 0 ~.
\]
We can now use the Killing form of the Lie algebra $\alg{g}$ and its decomposition under the symmetric space structure to define a map from $\alg{g} \to \alg{b}$
\[\label{eq:projeps}
P_\epsilon : U_{\bar a} \to -\epsilon \killing_{\bar a \bar b} \tilde{U}^{\bar b} ~, \qquad P_\epsilon : U_{\bar\alpha} \to -\eta \killing_{\bar\alpha\bar\beta} \tilde{U}^{\bar\beta} ~,
\]
where $\epsilon$ and $\eta$ are free parameters.
In the following we will be interested in the limit $\epsilon \to 0$ and hence we define
\[\label{eq:projeps2}
P \equiv P_0 : \begin{cases} \alg{h} \to 0 ~, \\ \alg{p} \to \alg{b} ~. \end{cases}
\]

%%%%%%%%%%%%%%%%%%%%%%%%%%%%%%%%%%%%%%%%
\paragraph{Poisson--Lie symmetric model.}
An arbitrary element $x \in \alg{d}$ has a unique decomposition $x = y + z$, with $y \in \alg{g}$ and $z \in \alg{b}$.
Without loss of generality we may write the action of the bilinear form $K$ on an element $x \in \alg{d}$ (defining the model on the double \eqref{eq:action_double}) as
\[
K(x) = \<z, \Gn z\> + \<y + \Bn z, \Gn^{-1} (y + \Bn z) \> ~,
\]
where $\Gn$ and $\Bn$ are the symmetric and antisymmetric parts of a general operator $\En : \alg{b} \to \alg{g}$.

Starting from the model on the double \eqref{eq:action_double} and parametrising the group-valued field $l \in D$ as
\[
l = \tilde{g} g ~, \qquad \tilde{g} \in \grp{B} ~, \quad g \in \grp{G} ~,
\]
we integrate out the degrees of freedom contained in $b$ to find the following Lorentz-invariant action for $g$
\[\label{eq:actgroup}
\Act(g) = \int \extder^2 \sigma \, \< g^{-1} \partial_+ g, \frac{1}{\En + \Pi(g)} g^{-1} \partial_- g \> ~,
\]
where
\unskip\footnote{Inserting additional projectors in the first relation of \eqref{eq:rel2} we find
\begin{equation*}
(\Proj{\alg{b}} \Ad_g^{-1} \Proj{\alg{b}}) (\Proj{\alg{b}}\Ad_g \Proj{\alg{b}}) = \Proj{\alg{b}} \quad \Rightarrow \quad (\Proj{\alg{b}}\Ad_g \Proj{\alg{b}}) = (\Proj{\alg{b}} \Ad_g^{-1} \Proj{\alg{b}})^{-1} ~,
\end{equation*}
where we understand $(\Proj{\alg{b}} \Ad_g \Proj{\alg{b}})$ and $(\Proj{\alg{b}} \Ad_g^{-1} \Proj{\alg{b}})$ as invertible maps from $\alg{b} \to \alg{b}$.
Using this relation we can rewrite the operator $\Pi(g) = \Proj{\alg{g}} \Ad_g^{-1} \Proj{\alg{b}} (\Proj{\alg{b}} \Ad_g^{-1} \Proj{\alg{b}})^{-1}$ in the form given in \eqref{eq:pimap}.}
\[\label{eq:pimap}
\Pi(g) = \Proj{\alg{g}} \Ad_g^{-1} \Proj{\alg{b}} \Ad_g \Proj{\alg{b}} : \alg{b} \to \alg{g} ~.
\]

The inverse of the symmetric operator $\Gn$ is a map $\Gn^{-1} : \alg{g} \to \alg{b}$.
Therefore we may take $\Gn^{-1} = P$ as defined in \eqref{eq:projeps} and rewrite the action \eqref{eq:actgroup} as
\[\label{eq:actproj}
\Act(g) = \int \extder^2 \sigma \, \< g^{-1} \partial_+ g , P \frac{1}{1 + (\Bn + \Pi(g)) P } g^{-1} \partial_- g \> ~.
\]
We would now like to consider the limit $\epsilon \to 0$ and demonstrate that if we take $\Bn^\ind{AB} = \killing^\ind{AC} R_\ind{C}{}^\ind{B}$, where
\[
\Bn \tilde{T}^\ind{A} = \Bn^\ind{AB} T_\ind{B} ~,
\]
the action develops a $\grp{H}$ gauge symmetry.
Recall that $\killing^\ind{AB}$ and $R_\ind{A}{}^\ind{B}$ are the components of the inverse Killing form and the Drinfel'd--Jimbo R-matrix respectively.

%%%%%%%%%%%%%%%%%%%%%%%%%%%%%%%%%%%%%%%%
\paragraph{Gauge invariance.}
Under $g \to g h$ the left-invariant one-form transforms as
\[
g^{-1} d g \to \Ad_h^{-1} (g^{-1} d g) + h^{-1} d h ~.
\]
The non-covariant piece takes values in $\alg{h}$ and hence $P (h^{-1} d h) = 0$.
Furthermore the symmetric space commutation relations \eqref{eq:symspace} and the orthogonality of the Killing form \eqref{eq:orthog} imply that
\[\label{eq:projadcom}
& \Proj{\alg{b}} \com{\ad_\xi^n}{P} \Proj{\alg{g}} = 0 ~, & \qquad & \xi \in \alg{h} ~,
\\ & \Proj{\alg{b}} \com{\Ad_h}{P} \Proj{\alg{g}} = 0 ~, & \qquad & h \in \grp{H} ~.
\]

As a result the action transforms as
\[\label{eq:acttrans}
\Act(g h) = \int \extder^2 \sigma \, \< g^{-1} \partial_+ g , P \frac{1}{1 + \Ad_h \Proj{\alg{g}} ( \Bn + \Pi(gh)) \Proj{\alg{b}} \Ad_h^{-1} P } g^{-1} \partial_- g \> ~.
\]
For invariance we then require that
\[\label{eq:requirement}
P \left( \Ad_h \Proj{\alg{g}} (\Bn + \Pi(gh) ) \Proj{\alg{b}} \Ad_h^{-1} \right) P - P \left(\Bn + \Pi(g) \right) P = 0 ~,
\]
understood as an operator equation acting on $\alg{g}$.
In \appref{app:gauge} (see also \cite{Sfetsos:1999zm}) we show that this condition is satisfied if the components of $\Bn$ solve
\[\label{eq:gc8text}
\Bn^{\bar\alpha\ind{D}} f_{\bar c\ind{D}}{}^{\bar\beta} - f_{\ind{D}\bar c}{}^{\bar\alpha} \Bn^{\ind{D}\bar\beta} = \tilde{f}^{\bar\alpha\bar\beta}{}_{\bar c} ~.
\]

Let us consider stronger condition in which we allow the free indices to run over their full possible range.
Recalling that $\Bn^\ind{AB}$ and $f_\ind{ABC}$ are completely antisymmetric, we can rearrange \eqref{eq:gc8text} to take the form
\[\label{eq:gcfinal}
\killing_\ind{AE} \Bn^\ind{ED} f_\ind{DB}{}^\ind{C} + \killing_\ind{BE} \Bn^\ind{ED} f_\ind{AD}{}^\ind{C} = \killing_\ind{AD} \killing_\ind{BE} \killing^\ind{CF} \tilde{f}^\ind{DE}{}_\ind{F} ~.
\]
This equation relates the structure constants of $\alg{b}$ to those of $\alg{g}$ and furthermore takes exactly the same form as \eqref{eq:strucrel} if we identify
\[\label{eq:brrel}
\Bn^\ind{AB} \equiv \killing^\ind{AC} R_\ind{C}{}^\ind{B} ~.
\]
Hence with this choice in the limit $\epsilon \to 0$ we find that \eqref{eq:actproj} is invariant under gauge transformations $g \to g h$.

%%%%%%%%%%%%%%%%%%%%%%%%%%%%%%%%%%%%%%%%
\paragraph{Relation to the $\eta$~deformed model.}
With the identification \eqref{eq:brrel} the action \eqref{eq:actproj} becomes that of the $\eta$~deformed symmetric space sigma model.

Let us give a brief overview of the $\eta$~deformed principal chiral model \cite{Klimcik:2002zj,Klimcik:2008eq}.
The principal chiral model for the group-valued field $g \in \grp{G}$ is defined by the action
\[
\Act_{\text{PCM}} = -\int \extder^2 \sigma \, \killing \big( g^{-1} \partial_+ g, g^{-1} \partial_- g \big) ~,
\]
where, as before, $\killing$ denotes the Killing form on $\alg{g}$.
As the terminology suggests, this action is invariant under both the left and right action of $\grp{G}$.
Furthermore, it is possible to demonstrate that this model admits a Lax pair, and is therefore classically integrable.
The integrability preserving $\eta$~deformation of the principal chiral model (often referred to as the Yang--Baxter sigma model) is defined by the action \cite{Klimcik:2002zj,Klimcik:2008eq}
\[
\Act_{\text{PCM},\eta} = -\int \extder^2 \sigma \, \killing \big( g^{-1} \partial_+ g, \frac{1}{\identity-\eta R_g} g^{-1} \partial_- g \big) ~,
\]
where $R_g = \Ad_g^{-1} R \Ad_g$ and the antisymmetric operator $R$ acting on the Lie algebra $\alg{g}$ satisfies the non-split modified classical Yang--Baxter equation.
The deformation breaks the invariance of the action under the left action of $\grp{G}$.
The currents associated to this transformation
\[
K_{\pm, \ind{A}} = -\killing \big( T_\ind{A}, \frac{1}{\identity \pm \eta R} \partial_\pm g g^{-1} \big) ~,
\]
satisfy the equation
\[
\partial_+ K_{-,\ind{E}} + \partial_- K_{+,\ind{E}} = \eta \, \killing_{\ind{EC}} \big( R_\ind{A}{}^\ind{D} f_\ind{DB}{}^\ind{C} + R_\ind{B}{}^\ind{D} f_\ind{AD}{}^\ind{C} \big) \killing^{\ind{AF}} \killing^{\ind{BG}} K_{+,\ind{F}} K_{-,\ind{G}} ~,
\]
indicating Poisson--Lie symmetry, with the structure constants of the coduality group given by \eqref{eq:strucrel} up to an overall factor of $\eta$.
For $\eta = 0$, corresponding to the undeformed model, and $\eta \neq 0$, corresponding to the deformed model, the Drinfel'd doubles are different, with the former a contraction of the latter.
In the first case, the dual structure constants vanish, the dual group is abelian and the Drinfel'd double is the semi-abelian double.
In the second case, it is possible to rescale the dual generators to remove the dependence on $\eta$ in the structure constants, and show that the Drinfel'd double is indeed the complexification $\grp{D} \equiv \grp{G}^{\Complex}$.

This deformation can be extended to coset spaces \cite{Delduc:2013fga}.
The action of the $\eta$~deformed symmetric space sigma model is given by
\[
\Act_{\text{SSSM},\eta} = -\int \extder^2 \sigma \, \killing \big( g^{-1} \partial_+ g, \proj \frac{1}{\identity-\eta R_g \proj} g^{-1} \partial_- g \big) ~,
\]
where $\proj$ is the projector onto $\alg{p}$ as defined in the decomposition \eqref{eq:symspace}.
Up to an overall factor of $\eta$, this action has precisely the same form as \eqref{eq:actproj} provided that
\[
(\Bn+\Pi(g))^{\ind{AB}} = \killing^{\ind{AC}} (R_g)_\ind{C}{}^\ind{B} ~,
\]
which reduces to \eqref{eq:brrel} for $g$ equal to the identity.
Therefore, we have the relation
\[
\Pi(g)^{\ind{AB}}= \kappa^{\ind{AC}} (R_g - R)_\ind{C}{}^\ind{B} ~,
\]
between the operators $\Pi(g): \alg{b} \rightarrow \alg{g}$ and $R: \alg{g} \rightarrow \alg{g}$.

%%%%%%%%%%%%%%%%%%%%%%%%%%%%%%%%%%%%%%%%
\paragraph{Poisson--Lie duals of the $\eta$~deformation.}
Let us consider the Dynkin diagram of $\alg{g}$.
Any subalgebra $\alg{g}_0$ of $\alg{g}$ associated to a sub-Dynkin diagram has the property that the commutation relations \eqref{eq:commutdual} and \eqref{eq:commutdualcons} are satisfied and hence generates a subgroup with respect to which we can Poisson--Lie dualise.
To prove this let us recall the generators \eqref{eq:compactreal} of $\alg{g}$ and \eqref{eq:borel} of $\alg{b}$.
For a particular sub-Dynkin diagram we take the corresponding set of the Cartan generators and simple roots, which we label by the index $\bar\imath$.
The associated sets of positive and negative roots are labelled by the index $\bar m$.
The remaining Cartan elements and simple roots are labelled by the index $\hat\imath$, while the remaining roots are labelled by the index $\hat m$.
To summarise we have the following bases
\[\label{eq:setsetsetset}
\set{T_a}&=\set{i c_{\bar\imath\bar\jmath}h_{\bar\jmath}, i(e_{\bar m}+f_{\bar m}), -(e_{\bar m}-f_{\bar m})} ~, \qquad &
\set{\tilde{T}^a}&=\set{b_{\bar\imath j}h_j, e_{\bar m}, ie_{\bar m}} ~, \\
\set{T_\alpha}&=\set{i c_{\hat\imath j}h_j, i(e_{\hat m}+f_{\hat m}), -(e_{\hat m}-f_{\hat m})} ~, \qquad &
\set{\tilde{T}^\alpha}&=\set{b_{\hat\imath j}h_j, e_{\hat m}, ie_{\hat m}} ~,
\]
for the spaces $\alg{g}_0$, $\tilde{\alg{g}}_0$, $\alg{m}$ and $\tilde{\alg{m}}$ respectively.
It is important to note that we have set $c_{\bar\imath\hat\jmath} = 0$.
This is so that $\alg{g}_0$ indeed forms a subalgebra of $\alg{g}$ as required.

Let us now check the commutation relations \eqref{eq:commutdual}.
We begin by observing that in this setup $\tilde{\alg{g}}_0$ also forms a subalgebra by construction.
It transpires that the only non-trivial relation that needs to be considered is
\[
\com{b_{\hat\imath j} h_j}{e_{\bar k}}=b_{\hat\imath j} a_{j\bar k} e_{\bar k}=(c^{-t})_{\hat\imath\bar k} e_{\bar k} ~,
\]
where we have used relation \eqref{eq:bca} in the last equality.
This is the commutator of an element of $\tilde{\alg{g}}_0$ with an element of $\tilde{\alg{m}}$ that gives an element of $\tilde{\alg{g}}_0$.
Therefore, for consistency with \eqref{eq:commutdual}, we require that the right-hand side vanishes.
Indeed this is the case since $c_{\bar\imath\hat\jmath} = 0$ implies $(c^{-t})_{\hat\imath\bar\jmath} = 0$.
This in turn implies that
\[\label{eq:bbeemm}
\com{b_{\hat\imath j} h_j}{e_{\bar m}} = 0 ~.
\]
Finally, for the commutation relations of \eqref{eq:commutdual} to be realised, it is then sufficient to observe that
\[\label{eq:finalrels}
\com{h_i}{e_{\hat m}} \propto e_{\hat m} \in \tilde{\alg{m}} ~, \qquad
\com{e_{\bar m}}{e_{\hat n}} \in \set{e_{\hat p}} \subset \tilde{\alg{m}} ~, \qquad
\com{e_{\hat m}}{e_{\hat n}} \in \set{e_{\hat p}} \subset \tilde{\alg{m}} ~,
\]
are relations that are true by definition for a subalgebra associated to a sub-Dynkin diagram.
The relation \eqref{eq:commutdualcons} then follows from those in \eqref{eq:commutdual} by the ad-invariance of the inner product, which implies that $\alg{\tilde{k}} = \alg{g}_0 \dsum \tilde{\alg{m}}$ forms an algebra.

It is also possible to include, in the Lie group with respect to which we dualise, additional $\grp{U}_1$ factors built from the remaining Cartan generators.
To be precise, any subset of the generators $\set{i c_{\hat\imath j} h_j}$ can be moved from the set $\set{T_\alpha}$ to $\set{T_a}$ in \eqref{eq:setsetsetset}, while at the same time we move the corresponding elements of $\set{b_{\hat\imath j} h_j}$ from $\set{\tilde{T}^\alpha}$ to $\set{\tilde{T}^a}$.
As these are Cartan generators it follows that $\alg{g}_0$ and $\tilde{\alg{g}}_0$ will still form algebras.
Furthermore, as the Cartan generators are external in the Lie algebra $\tilde{\alg{m}}$, as seen from the first relation of \eqref{eq:finalrels}, and commute amongst themselves, $\tilde{\alg{m}}$ will still satisfy $\com{\tilde{\alg{m}}}{\tilde{\alg{m}}} \subset \tilde{\alg{m}}$.
The same relation also implies that the other condition of \eqref{eq:commutdual}, $\com{\tilde{\alg{g}}_0}{\tilde{\alg{m}}} \subset \tilde{\alg{m}}$, also remains true.
As we will see including such additional factors corresponds to taking the abelian dual.

\medskip

Finally, we note two dualities and the corresponding choices of $\tilde{\alg{k}}$ that are of particular interest in the context of the $\eta$~deformed model.
The first is the Poisson--Lie dual with respect to $\grp{G}$, which refer to as the $\lambdastar$~deformation.
In this case the Lie algebra whose degrees of freedom we integrate out is $\alg{\tilde{k}} = \alg{g}_0 = \alg{g}$, which is unimodular, that is $f_{ab}{}^b = 0$, as $\alg{g}$ is a simple Lie algebra.
As discussed in \secref{sec:intro}, this model is conjectured to be an analytic continuation of the corresponding $\lambda$~deformation of the symmetric space sigma model.

The second is the Poisson--Lie dual with respect to $\grp{U}_1{}^{\rank\grp{G}}$, which we refer to as the complete abelian dual.
In this case $\tilde{\alg{k}}$ is generated by
\[
\set{i c_{ij} h_j, e_\ind{M}, i e_\ind{M}} ~.
\]
As we are considering real Lie algebras, the factor of the imaginary unit in front of the Cartan generators implies that this is also a unimodular algebra.
In contrast, the Borel subalgebra \eqref{eq:borel} will in general not be unimodular.

%%%%%%%%%%%%%%%%%%%%%%%%%%%%%%%%%%%%%%%%%%%%%%%%%%%%%%%%%%%%%%%%%%%%%%%%%%%%%%%%
\subsection{The example of \texorpdfstring{$\Sp^2$}{S2}}\label{ssec:twosphere}

The first example we consider is the $\eta$~deformation of $\Sp^2$ and its Poisson--Lie duals.
The isometry algebra of $\Sp^2$ is $\alg{so}_3 \cong \alg{su}_2$.
From the analysis in \secref{ssec:etalambda} the Drinfel'd double associated to this $\eta$~deformed model is $\alg{sl}_2(\Complex)$, the complexification of $\alg{su}_2$.
The coduality algebra, generated by the Cartan subalgebra and the positive roots, is $\alg{b}_2$.

%%%%%%%%%%%%%%%%%%%%%%%%%%%%%%%%%%%%%%%%
\paragraph{$\grp{SU}_2$ and $\grp{B}_2$ embeddings in $\grp{SL}_2(\Complex)$.}
We consider the $2 \times 2$ matrix realisation of $\alg{sl}_2(\Complex)$.
Employing the basis of $\Mat(2;\Complex)$
\[
(N_\ind{ST})_\ind{UV} = \delta_\ind{SU}\delta_\ind{TV} ~, \qquad S,T,\ldots = 1,2 ~,
\]
we introduce the Cartan-Weyl basis
\[
h = N_{11} - N_{22} ~, \qquad e = N_{12} ~, \qquad f = N_{21} ~,
\]
with Cartan generator $h$, positive root $e$ and negative root $f$.
Together they satisfy the defining relations
\[
\com{h}{e} = 2 e ~, \qquad \com{h}{f} = - 2 f ~, \qquad \com{e}{f} = h ~.
\]

Let us now take the explicit basis for the Lie algebra $\alg{su}_2$
\[
T_1 = i h ~, \qquad T_2 = i (e + f) ~, \qquad T_3 = - (e - f) ~,
\]
and for $\alg{b}_2$
\[
\tilde{T}^1 = \half h ~, \qquad \tilde{T}^2 = e ~, \qquad \tilde{T}^3 = i e ~.
\]
The ad-invariant inner product on the double
\[
\< x, x' \> = \Im \Tr[x x'] , \qquad x,x' \in \alg{sl}_2(\Complex) ~,
\]
is normalised and maximally isotropic with respect to both $\alg{su}_2$ and $\alg{b}_2$
\[\label{eq:bilinsl2}
\< T_\ind{A}, \tilde{T}^\ind{B} \> = \delta_\ind{A}^\ind{B}, \qquad \< T_\ind{A}, T_\ind{B} \> = \< \tilde{T}^\ind{A}, \tilde{T}^\ind{B} \> = 0 ~.
\]
Finally let us note that the Killing form for the simple Lie algebra $\alg{su}_2$ in this basis is orthogonal and we normalise such that
\[
\killing_\ind{AB} \equiv \Tr[T_\ind{A} T_\ind{B}] = \sfrac{1}{4} f_\ind{AC}{}^\ind{D} f_\ind{BD}{}^\ind{C} = -2 \delta_\ind{AB} ~.
\]

\medskip

Starting from the model on the double \eqref{eq:action_double} with $l \in \grp{SL}_2(\Complex)$ we integrate out the degrees of freedom corresponding to different three-dimensional subalgebras to construct various Poisson--Lie dual sigma models.
In this example we shall consider three cases:
\begin{enumerate}[(i)]
\item integrating the degrees of freedom corresponding to $\alg{b}_2$, generated by $\set{\tilde{T}^1,\tilde{T}^2,\tilde{T}^3}$,
\item integrating the degrees of freedom corresponding to $\alg{su}_2$, generated by $\set{T_1,T_2,T_3}$,
\item integrating the degrees of freedom corresponding to the algebra generated by $\set{T_1,\tilde{T}^2,\tilde{T}^3}$.
\end{enumerate}
We can use the inner product \eqref{eq:bilinsl2} to identify the duals of these algebras as certain subspaces of the Drinfel'd double $\alg{sl}_2(\Complex)$: for (i) the dual space is spanned by $\set{T_1,T_2,T_3}$, for (ii) it is spanned by $\set{\tilde{T}^1,\tilde{T}^2,\tilde{T}^3}$, while for (iii) it is spanned by $\set{\tilde{T}^1,T_2,T_3}$.
For (i) and (ii) the dual spaces are subalgebras of $\alg{sl}_2(\Complex)$, however for (iii) this is not the case.
It is also worth noting that, in agreement with the discussion in \secref{ssec:etalambda}, the trace of the structure constants vanishes for (ii) and (iii) (corresponding to the $\lambdastar$~deformation and complete abelian dual respectively) and hence the algebras are unimodular.
This is in contrast to (i) (corresponding to the $\eta$~deformation), for which the algebra is not unimodular.

Let us consider each of these cases for the $\eta$~deformation of $\Sp^2$.
This model is a deformation of the symmetric space sigma model for the coset
\[
\Sp^2 \cong \frac{\grp{SU}_2}{\grp{U}_1} ~.
\]
Here we will follow \cite{Klimcik:1996np,Sfetsos:1999zm} and identify the generator of the $\grp{U}_1$ gauge group with $T_1$.
The backgrounds for cases (i) and (ii) were constructed in \cite{Klimcik:1996np,Sfetsos:1999zm} for a matrix $\En$ depending on two parameters.
In \cite{Hoare:2015gda} it was shown that in the special case
\[
\En = \frac{1}{2} \begin{pmatrix} \epsilon^{-1} & 0 & 0 \\ 0 & \eta^{-1} & 1 \\ 0 & -1 & \eta^{-1} \end{pmatrix} ~,
\]
in the $\epsilon \to 0$ limit, these backgrounds correspond to those of the $\eta$~ and $\lambdastar$~deformations respectively.
This matches the form of $\En$ in the action \eqref{eq:actproj} with $\Bn$ defined in \eqref{eq:brrel}.
We will first briefly review these results before turning to case (iii).

%%%%%%%%%%%%%%%%%%%%%%%%%%%%%%%%%%%%%%%%
\paragraph{Case (i).}
As described in \secref{ssec:etalambda}, to construct the $\eta$~deformation of $\Sp^2$, we start from the model on the double \eqref{eq:action_double} and decompose the group-valued field $l \in \grp{SL}_2(\Complex)$ as
\[
l = \tilde{g} g ~, \qquad g \in \grp{SU}_2 ~, \quad \tilde{g} \in \grp{B}_2 ~.
\]
We can then integrate out the degrees of freedom $\tilde{g}$ corresponding to case (i) above.
Using the $\grp{H}$ gauge freedom, $g \to g h$, to gauge-fix
\[
g = \exp[\phi \, T_1/2] \exp[\arccos r \, T_2/2] ~,
\]
we find the sigma model action for the following metric and $B$-field
\unskip\footnote{The $B$-field is defined with its normalisation factor, that is to say $B=\half B_\ind{IJ} \extder X^\ind{I} \wedge \extder X^\ind{J}$.}
\[\label{eq:etas2}
\eta^{-1} \extder s^2 & = \frac{1}{2} \frac{1}{1+\eta^2 r^2} \bigg( (1-r^2) \extder \phi^2 + \frac{\extder r^2}{(1-r^2)} \bigg) ~,
\\
\eta^{-1} B & = - \frac{1}{2} \frac{\eta r}{1+\eta^2 r^2} \extder \phi \wedge \extder r ~.
\]
This is indeed the $\eta$~deformation of $\Sp^2$ \cite{Arutyunov:2013ega,Hoare:2014pna}. 
Note that the $B$-field is a total derivative and hence can be set to zero by a gauge transformation.
The overall factor of $\eta$ comes from our definition of the Poisson--Lie symmetric model \eqref{eq:actproj} and hence also appears in the same way in the dual models.
Even though we have explicitly included this factor, we always think of it as being absorbed into an overall coupling constant.
In particular, we ignore it when taking the limit $\eta \rightarrow 0$. 

%%%%%%%%%%%%%%%%%%%%%%%%%%%%%%%%%%%%%%%%
\paragraph{Case (ii).}
To construct the Poisson--Lie dual model corresponding to case (ii), that is the $\lambdastar$~deformation, we instead parametrise the group-valued field $l \in \grp{SL}_2(\Complex)$ as
\[
l = g \tilde{g} ~, \qquad g \in \grp{SU}_2 ~, \quad \tilde{g} \in \grp{B}_2 ~.
\]
Integrating out the degrees of freedom $g$ and using the $\grp{H}$ gauge freedom, $\tilde{g} \to h^{-1} \tilde{g} h$, to gauge-fix \cite{Sfetsos:1999zm} 
\[
\tilde{g} = \exp[\rho \, \tilde{T}^2/2] \exp[\log(1 + \zeta) \, \tilde{T}^1/2] ~,
\]
we find the metric and $B$-field corresponding to the $\lambdastar$~deformed model for $\Sp^2$
\[\label{eq:s2pldsu2}
\eta^{-1} \extder s^2 & = \frac{1}{2(1+\eta^2)(1+\zeta)}\bigg(\frac{\extder \zeta^2}{\rho^2} + \frac{(1+\eta^2)^2}{4\eta^2} \bigg( \extder\rho + \Big(\frac{\zeta - \rho^2/4}{1 + \zeta} - \frac{2 \eta^2}{1+\eta^2}\Big) \frac{\extder \zeta}{\rho} \bigg)^2 \bigg) ~,
\\
\eta^{-1} B & = 0 ~.
\]
We also recall the field redefinition introduced in \cite{Hoare:2015gda}
\[
\rho = 2 \sqrt{p^2 - q^2 - 1} ~, \qquad \zeta = (p+q)^2 - 1 ~,
\]
for which the background takes the following particularly simple form
\[\label{eq:s2pldsu2alt}
\eta^{-1} \extder s^2 & = \frac{1}{2(p^2-q^2-1)} \bigg( \frac{\extder p^2}{\eta^2} + \extder q^2 \bigg) ~, \qquad \eta^{-1} B & = 0 ~.
\]
The undeformed limit, that is $\eta \to 0$, can be taken by first rescaling $\zeta \to \eta \zeta$ and $\rho \to \eta \rho$ in \eqref{eq:s2pldsu2} (or equivalently $p \to 1 + \eta^2 p$ and $q \to \eta q$ in \eqref{eq:s2pldsu2alt}).
The resulting background is, as expected, the non-abelian dual with respect to $\grp{SU}_2$ of $\Sp^2$.

%%%%%%%%%%%%%%%%%%%%%%%%%%%%%%%%%%%%%%%%
\paragraph{Case (iii).}
As a simple example of the setup described in \secref{ssec:firstorder} we consider case (iii), that is the Poisson--Lie dual with respect to $\grp{G}_0 \cong \grp{U}_1$ generated by $T_1$.
We write the field $k \in \grp{SL}_2(\Complex)$ of the action \eqref{eq:actionS1} as
\[
k = \tilde{m} \tilde{g}_0 g ~, \qquad \tilde{g}_0 \in \tilde{\grp{G}}_0 ~, \quad \tilde{m} \in \tilde{\grp{M}} ~, \quad g \in \grp{SU}_2 ~,
\]
where $\tilde{\grp{G}}_0$ and $\tilde{\grp{M}}$ are generated by $\set{\tilde{T}^1}$ and $\set{\tilde{T}^2,\tilde{T}^3}$ respectively.
The left-acting $\tilde{\grp{K}}$ gauge freedom, generated by $\set{T_1,\tilde{T}^2,\tilde{T}^3}$, can then be used to partially gauge-fix
\[
k = \tilde{g}_0 g^\prime ~.
\]
Using the remaining $\tilde{\grp{K}}$ and the right-acting $\grp{H}$ gauge freedoms to gauge-fix
\[
k = \exp[2 \tilde{\phi} \, \tilde{T}^1] \exp[\arccos r \, T_2/2] ~,
\]
the background of the dual model is given by
\[\label{eq:u1duals2}
\eta^{-1} \extder s^2 & = \frac{2}{\eta^2}\frac{1+\eta^2r^2}{1-r^2} \extder \tilde{\phi}^2 - 2 \frac{r}{1-r^2} \extder \tilde{\phi} \extder r + \frac{1}{2} \frac{\extder r^2}{1-r^2} ~,
\\
\eta^{-1} B & = 0 ~,
\]
where the field $\tilde{\phi}$ is actually associated to a non-compact direction in the Drinfel'd double.
After rescaling $\tilde{\phi} \to \eta \tilde{\phi}$, this background is precisely the abelian dual of \eqref{eq:etas2}.
The same rescaling also allows us to take the undeformed limit.
Note that redefining $\tilde{\phi} \to \eta \tilde{\phi} + \tfrac14 \log(1+\eta^2 r^2)$ diagonalises the metric of \eqref{eq:u1duals2} and the duality with \eqref{eq:etas2} becomes manifest, now up to a total derivative in the $B$-field.

\medskip

Starting from the $\lambda$~deformation of various low-dimensional spheres, the complete abelian dual of the corresponding $\eta$~deformation can be recovered via a combination of limits and analytic continuations \cite{Hoare:2015gda}.
Motivated by this we investigate a related limit of the $\lambdastar$~deformation that gives the abelian dual of the $\eta$~deformation of $\Sp^2$.
Considering the metric and $B$-field of the former given in \eqref{eq:s2pldsu2alt}, we redefine
\[
p \to \cosh 2\tilde{\phi} ~, \qquad q \to r \sinh 2\tilde{\phi} ~,
\]
which gives the background
\[\label{eq:newform}
\eta^{-1} \extder s^2 & = \frac{2}{\eta^2} \frac{1+\eta^2 r^2\coth^2 2\tilde{\phi}}{1-r^2} \extder \tilde{\phi}^2 + 2 \frac{r\coth2\tilde{\phi}}{1-r^2} \extder \tilde{\phi} \extder r + \frac{1}{2} \frac{\extder r^2}{1-r^2} ~,
\\
\eta^{-1} B & = 0 ~.
\]
This is equivalent to working with the action \eqref{eq:actionS1} and using the left-acting $\tilde{\grp{K}}$ and right-acting $\grp{U}_1$ gauge symmetries to gauge-fix
\[
k = \exp[2 \tilde{\phi} \, \tilde{T}^1] \exp[\arccos r \, T_2/2] ~.
\]
Here we emphasise that we are considering case (ii), and hence, in contrast to case (iii), the Lie algebra whose degrees of freedom are integrated out is $\tilde{\alg{k}} = \alg{su}_2$.
However, crucially in both cases we can use the (different) gauge symmetries to achieve the same parametrisation of $k \in \grp{SL}_2(\Complex)$.

Now sending either $\tilde{\phi} \to - \infty$ or $\tilde{\phi} \to \infty$ (and subsequently redefining $\tilde{\phi} \to - \tilde{\phi}$) in \eqref{eq:newform} we indeed recover the background of the abelian dual \eqref{eq:u1duals2}.
In order to explain this behaviour we focus on $\tilde{\phi} \to - \infty$.
In this limit the right action of $e^{2\tilde{\phi} \tilde{T}_1}$ on the algebras we integrate out in cases (ii) and (iii) coincides.
That is
\[
\set{T_1,T_2,T_3} e^{2\tilde{\phi} \tilde{T}_1} \quad \xrightarrow{\tilde{\phi} \to - \infty} \quad \set{T_1,\tilde{T}^3,-\tilde{T}^2} e^{2\tilde{\phi} \tilde{T}_1} ~.
\]
Consequently, there is no longer a distinction between the Poisson--Lie duals with respect to $\grp{U}_1$ and $\grp{SU}_2$.

%%%%%%%%%%%%%%%%%%%%%%%%%%%%%%%%%%%%%%%%%%%%%%%%%%%%%%%%%%%%%%%%%%%%%%%%%%%%%%%%
\subsection{The example of \texorpdfstring{$\Sp^5$}{S5}}\label{ssec:fivesphere}

Our second example is the $\eta$~deformation of $\Sp^5$ and its Poisson--Lie duals, a first step towards exploring the Poisson--Lie duals of the $\AdS_5 \times \Sp^5$ superstring sigma model.
The isometry algebra of $\Sp^5$ is $\alg{so}_6 \cong \alg{su}_4$.
From the analysis in \secref{ssec:etalambda} the Drinfel'd double associated to this $\eta$~deformed model is $\alg{sl}_4(\Complex)$, the complexification of $\alg{su}_4$.
The coduality algebra, generated by the Cartan subalgebra and the positive roots, is $\alg{b}_4$.
In this example the rank of the Drinfel'd double is 3 and hence there is a richer structure than for the $\eta$~deformation of $\Sp^2$ discussed in \secref{ssec:twosphere}.

%%%%%%%%%%%%%%%%%%%%%%%%%%%%%%%%%%%%%%%%
\paragraph{$\grp{SU}_4$ and $\grp{B}_4$ embeddings in $\grp{SL}_4(\Complex)$.}
We consider the $4 \times 4$ matrix representation of $\alg{sl}_4(\Complex)$ and define the basis of $\Mat(4;\Complex)$
\[
(N_\ind{ST})_\ind{UV} = \delta_\ind{SU} \delta_\ind{TV} , \qquad S,T,\ldots = 1,2,3,4 ~.
\]
The Cartan generators $h_j$, $j = 1,2,3$, the positive simple roots $e_j$ and the negative simple roots $f_j$ given by
\[
h_1 & = N_{11}-N_{22} ~, \qquad & h_2 & = N_{22} - N_{33} ~, \qquad & h_3 & = N_{33} - N_{44} ~, \\
e_1 & = N_{12} ~, \qquad & e_2 & = N_{23} ~, \qquad & e_3 & = N_{34} ~, \\
f_1 & = N_{21} ~, \qquad & f_2 & = N_{32} ~, \qquad & f_3 & = N_{43} ~,
\]
satisfy the defining relations
\[
\com{h_i}{h_j} = 0 ~, \qquad \com{h_i}{e_j} = a_{ij} e_j ~, \qquad \com{h_i}{f_j} = - a_{ij} f_j ~, \qquad \com{e_i}{f_j} = \delta_{ij} h_i ~,
\]
where $a_{ij}$ are elements of the Cartan matrix of $\alg{su}_4$.
The non simple roots follow from the commutators
\[
& e_4 = \com{e_1}{e_2} ~, \qquad e_5 = \com{e_2}{e_3} ~, \qquad e_6 = \com{e_1}{\com{e_2}{e_3}} ~, \\
& f_4 = \com{f_2}{f_1} ~, \qquad f_5 = \com{f_3}{f_2} ~, \qquad f_6 = \com{f_3}{\com{f_2}{f_1}} ~.
\]
Let us now consider the following explicit basis for the Lie algebra $\alg{su}_4$
\[
T_1 & = i c_{1i} h_i ~, \qquad & T_6 & = i c_{2i} h_i ~, \qquad & T_{11} & = i c_{3i} h_i ~, \\
T_2 & = i(e_1+f_1) ~, \qquad & T_7 & = i(e_2+f_2) ~, \qquad & T_{12} & = i(e_3+f_3) ~, \\
T_3 & = -(e_1-f_1) ~, \qquad & T_8 & = -(e_2-f_2) ~, \qquad & T_{13} & = -(e_3-f_3) ~, \\
T_4 & = i(e_4+f_4) ~, \qquad & T_9 & = i(e_5+f_5) ~, \qquad & T_{14} & = i(e_6+f_6) ~, \\
T_5 & = -(e_4-f_4) ~, \qquad & T_{10} & = -(e_5-f_5) ~, \qquad & T_{15} & = -(e_6-f_6) ~,
\]
and for $\alg{b}_4$
\[
\tilde{T}^1 & = b_{1i} h_i ~, \qquad & \tilde{T}^6 & = b_{2i} h_i ~, \qquad & \tilde{T}^{11} & = b_{3i} h_i ~, \\
\tilde{T}^2 & = e_1 ~, \qquad & \tilde{T}^7 & = e_2 ~, \qquad & \tilde{T}^{12} & = e_3 ~, \\
\tilde{T}^3 & = i e_1 ~, \qquad & \tilde{T}^8 & = i e_2 ~, \qquad & \tilde{T}^{13} & = i e_3 ~, \\
\tilde{T}^4 & = e_4 ~, \qquad & \tilde{T}^9 & = e_5 ~, \qquad & \tilde{T}^{14} & = e_6 ~, \\
\tilde{T}^5 & = i e_4 ~, \qquad & \tilde{T}^{10} & = i e_5 ~, \qquad & \tilde{T}^{15} & = i e_6 ~.
\]
The coefficients $c_{ij}$ and $b_{ij}$ should obey relation \eqref{eq:bca}. We also normalise the Killing form for the simple Lie algebra $\alg{su}_4$ such that
\[
\killing_\ind{AB} \equiv \Tr[T_\ind{A} T_\ind{B}] = \sfrac{1}{8} f_\ind{AC}{}^\ind{D} f_\ind{BD}{}^\ind{C} ~.
\]
As for the $S^2$ example, the ad-invariant inner product on the double
\[
\< x, x' \> = \Im \Tr[x x'] , \qquad x,x' \in \alg{sl}_4(\Complex) ~,
\]
is normalised and maximally isotropic with respect to both $\alg{su}_4$ and $\alg{b}_4$
\[\label{eq:bilinsl4}
\< T_\ind{A}, \tilde{T}^\ind{B} \> = \delta_\ind{A}^\ind{B}, \qquad \< T_\ind{A}, T_\ind{B} \> = \< \tilde{T}^\ind{A}, \tilde{T}^\ind{B} \> = 0 .
\]

%%%%%%%%%%%%%%%%%%%%%%%%%%%%%%%%%%%%%%%%
\paragraph{$\grp{SO}_5$ embeddings in $\grp{SL}_4(\Complex)$.}
The $\eta$~deformation of $\Sp^5$ is a deformation of the symmetric space sigma model for the coset
\[
\Sp^5 \cong \frac{\grp{SU}_4}{\grp{SO}_5} ~,
\]
and hence we need to specify the embedding of $\grp{H} = \grp{SO}_5$ in $\grp{SU}_4$.
To this end we introduce the following basis of $\alg{su}_4$
\[
n^{12} & = \ihalf(h_1+h_3) ~, \qquad & n^{34} & = \ihalf(h_1-h_3) ~, \qquad & n^{56} & = \ihalf(h_1+2h_2+h_3) ~, \\
n^{13} & = \half(T_3-T_{13}) ~, \qquad & n^{15} & = \half(T_8-T_{15}) ~, \qquad & n^{35} & = \half(T_5+T_{10}) ~, \\
n^{24} & = \half(-T_3-T_{13}) ~, \qquad & n^{26} & = \half(T_8+T_{15}) ~, \qquad & n^{46} & = \half(-T_5+T_{10}) ~, \\
n^{14} & = \half(T_2+T_{12}) ~, \qquad & n^{16} & = \half(T_7-T_{14}) ~, \qquad & n^{36} & = \half(T_4+T_9) ~, \\
n^{23} & = \half(T_2-T_{12}) ~, \qquad & n^{25} & = \half(-T_7-T_{14}) ~, \qquad & n^{45} & = \half(T_4-T_9) ~,
\]
satisfying the standard commutation relations of $\alg{so}_6 \cong \alg{su}_4$.
In what follows we take $\alg{h} = \alg{so}_5 = \vectorspan{n^{st} : s,t = 1,\ldots,5}$, while the five remaining generators $\{n^{s6} : s = 1,\ldots,5\}$ span $\alg{p}$.

As discussed in \secref{ssec:etalambda}, the requirement of $\grp{H}$ gauge symmetry relates the antisymmetric operator $\Bn$ to the structure constants of the dual algebra \eqref{eq:gc8text}.
This is condition solved if $\Bn$ and the R-matrix are related as in \eqref{eq:brrel}.
It is interesting to note that, if it is possible to choose a specific orientation of the R-matrix with respect to the $\grp{H}$ gauge group such that it maps $\alg{p}$ to $\alg{h}$, then the contribution of $\Bn$ in the action \eqref{eq:actproj} drops out and $\Pi(g)$ transforms covariantly under $\grp{H}$ gauge transformations.

%%%%%%%%%%%%%%%%%%%%%%%%%%%%%%%%%%%%%%%%
\paragraph{Dualisation with respect to subgroups $\grp{G}_0$.}
The Lie algebra $\alg{su}_4$ has a much richer subalgebra structure than the Lie algebra $\alg{su}_2$ considered in \secref{ssec:twosphere}.
Therefore, many more dual models can be constructed.
However, the commutation relations \eqref{eq:commutdual} impose constraints on the subgroups with respect to which we may Poisson--Lie dualise.
In \secref{ssec:etalambda} we proved that it is possible to Poisson--Lie dualise in subgroups associated to sub-Dynkin diagrams.
Considering $\grp{SU}_2 \subset \grp{SU}_4$ as an example, this leads to the following three choices: $\alg{su}_2 = \vectorspan{\ihalf h_j,i(e_j+f_j),-(e_j-f_j)}$ with $j = 1,2$ or $3$.
For $j = 1,3$, the embedding of the $\grp{SO}_5$ gauge group in $\grp{SU}_4$ is such that $\grp{G}_0 \subset \grp{H}$, while for $j = 2$ this is not the case.
Here we shall present various explicit results for $\grp{G}_0 = \{\identity, \grp{U}_1, \grp{SU}_2\}$ and conclude with a more general discussion of higher-dimensional subgroups.
To be precise, we will explore the following cases:
\begin{enumerate}[(i)]
\item integrating the degrees of freedom corresponding to $\alg{b}_4$, generated by $\set{\tilde{T}^\ind{A} : A = 1,\ldots,15}$,
\item integrating the degrees of freedom corresponding to the algebra generated by $\set{T_1,\tilde{T}^\ind{A} : A = 2,\ldots,15}$ with $T_1 = i h_1$,
\item integrating the degrees of freedom corresponding to the algebra generated by $\set{T_{1,2,3},\tilde{T}^\ind{A} : A \neq 1,2,3}$ with $T_1 = i h_1$,
\item integrating the degrees of freedom corresponding to the algebra generated by $\set{T_{6,7,8},\tilde{T}^\ind{A} : A \neq 6,7,8}$ with $T_6 = i h_2$.
\end{enumerate}
The backgrounds of the different models are computed using the action \eqref{eq:actionS1} with $\Gn^{-1} = P$, defined in \eqref{eq:projeps} and \eqref{eq:projeps2}, and $\Bn^\ind{AB} = \killing^\ind{AC} R_\ind{C}{}^\ind{B}$.
The various models differ in the choice of the Lie algebra $\tilde{\alg{k}}$ that is integrated out.
Of course, this implies that the field $k \in \tilde{\grp{K}}\backslash\grp{D}$ appearing in the action takes values in different spaces, and a convenient parametrisation has to be found in each case separately.
This is done by starting with the group-valued field $k \in \grp{SL}_4(\Complex)$ using the gauge symmetry
\[\label{eq:gthk}
k \to \tilde{k} k h ~, \qquad h \in \grp{H} ~, \quad \tilde{k} \in \tilde{\grp{K}} ~,
\]
to gauge-fix $n+\dim\grp{H}$ of the $2n$ degrees of freedom contained in $k$.
(Here $2n$ is the dimension of the Drinfel'd double: in this example we have $n = 15$).

%%%%%%%%%%%%%%%%%%%%%%%%%%%%%%%%%%%%%%%%
\paragraph{Case (i).}
The first case, in which the dual algebra $\alg{b}_4$ generated by $\set{\tilde{T}^\ind{A} : A = 1,\ldots,15}$ is integrated out, yields the original $\eta$~deformed model.
We set the constants $c_{ij}$ and $b_{ij}$ such that
\[\label{eq:basisi}
T_1 & = i h_1 ~, \qquad & T_6 & = \ihalf (h_1+2h_2+h_3) ~, \qquad & T_{11} & = i h_3 ~, \\
\tilde{T}^1 & = \half h_1 ~, \qquad & \tilde{T}^6 & = \half (h_1+2h_2+h_3) ~, \qquad & \tilde{T}^{11} & = \half h_3 ~.
\]
Parameterising $k = \tilde{g}g$ in the action \eqref{eq:actionS1}, with $g \in \grp{G} = \grp{SU}_4$ and $\tilde{g} \in \tilde{\grp{G}} = \grp{B}_4$, one simply takes $\tilde{k} = \tilde{g}^{-1}$ in the gauge transformation \eqref{eq:gthk} to gauge-fix 15 degrees of freedom, along with the following parametrisation of the coset $\grp{G}/\grp{H}$,
\[
g h = e^{\phi_1 T_1/2 + \phi_2 T_{11}/2 + \phi T_6} e^{\arccos{x} \, T_{13}} e^{\arcsin r \, (T_{10}-T_{5})/2} ~,
\]
fixing the remaining 10.
The corresponding metric and $B$-field are
\[\label{eq:etaorig}
\eta^{-1} \extder s^2 & = \frac{1}{1 + \eta^2 r^2} \bigg(\frac{\extder r^2}{1-r^2} + (1-r^2) \extder \phi^2 \bigg)
+ r^2 x^2 \Big(\frac{\extder \phi_1 - \extder \phi_2}{2} \Big)^2
\\
& \phantom{= \,} + \frac{r^2}{1 + \eta^2 r^4 x^2} \bigg(\frac{\extder x^2}{1-x^2} + (1-x^2)\Big(\frac{\extder \phi_1 + \extder \phi_2}{2} \Big)^2 \bigg) ~,
\\
\eta^{-1} B & = -\frac{\eta r^4 x}{1 + \eta^2 r^4 x^2} \Big(\frac{\extder \phi_1 + \extder \phi_2}{2}\Big) \wedge \extder x - \frac{\eta r}{1 + \eta^2 r^2} \extder \phi \wedge \extder r ~,
\]
which is indeed the $\eta$~deformation of $\Sp^5$ \cite{Arutyunov:2013ega}.
The parametrisation has been chosen in such a way that the three $\grp{U}_1$ isometries correspond to shifts in the angles $\phi$, $\phi_1$ and $\phi_2$.
This form of the background will be useful for comparing to the Poisson--Lie dual models that we consider, however it is worth noting that the reparametrisation $\half(\phi_1 + \phi_2) \to \phi_1$ and $\half(\phi_1 - \phi_2) \to \phi_2$ simplifies the metric and $B$-field.

%%%%%%%%%%%%%%%%%%%%%%%%%%%%%%%%%%%%%%%%
\paragraph{Case (ii).}
We begin our investigation with the dualisation in the abelian subgroup $\grp{G}_0 \cong \grp{U}_1$ generated by $T_1$.
In this case the commutation relations \eqref{eq:commutdual} are automatically satisfied, independent of the choice of $c_{ij}$ and $b_{ij}$.
We will use the same basis \eqref{eq:basisi} as in case (i) and hence the subgroup with respect to which we dualise lies completely within the gauge group.
Starting from the model on the double we integrate out the degrees of freedom corresponding to the algebra generated by $\set{T_1,\tilde{T}^\ind{A} : A = 2,\ldots,15}$.
The parametrisation
\[
k = e^{\tilde{\phi}_1 \tilde{T}^1/2 + \phi_2 T_{11}/2 + \phi T_6} e^{\arccos{x} \, T_{13}} e^{\arcsin r \, (T_{10}-T_{5})/2} ~,
\]
gives rise to the background
\[\label{eq:etau1dual}
\eta^{-1} \extder s^2 & = \frac{1}{1 + \eta^2 r^2} \bigg(\frac{\extder r^2}{1-r^2} + (1-r^2) \extder \phi^2 \bigg) +\frac{r^2}{1 + \eta^2 r^4 x^4} \bigg(\frac{\extder x^2}{1-x^2} + x^2(1-x^2) \extder \phi_2^2\bigg)
\\
& \phantom{= \,} + \frac{1}{4 \eta^2 r^2}\frac{1+\eta^2 r^4 x^2}{1+\eta^2 r^4 x^4} \extder \tilde{\phi}_1^2 - \frac{r^2 x }{1 + \eta^2 r^4 x^4} \extder \tilde{\phi}_1 \extder x ~,
\\
\eta^{-1} B & = \frac{1}{4\eta}\bigg(2 \frac{1-x^2}{1 + \eta^2 r^4 x^4} - 1\bigg) \extder \tilde{\phi}_1 \wedge \extder \phi_2 - \frac{\eta r^4 x^3}{1 + \eta^2 r^4 x^4} \extder \phi_2 \wedge \extder x - \frac{\eta r}{1 + \eta^2 r^2} \extder \phi \wedge \extder r ~,
\]
which is invariant under shifts in the angles $\phi$, $\tilde{\phi}_1$ and $\phi_2$ generating three $\grp{U}_1$ isometries.
Note that the field $\tilde{\phi}_1$ is again associated to a non-compact direction in the Drinfel'd double.
After rescaling $\tilde{\phi}_1 \to \eta \tilde{\phi}_1$, this background is nothing other than the abelian dual in $\phi_1$ of the $\eta$~deformed model \eqref{eq:etaorig}.
The same rescaling allows us to take the undeformed limit, in which we recover the corresponding abelian dual of $\Sp^5$.

%%%%%%%%%%%%%%%%%%%%%%%%%%%%%%%%%%%%%%%%
\paragraph{Case (iii).}
Let us now consider the dualisation with respect to the non-abelian group $\grp{SU}_2$ generated by $\alg{su}_2 = \vectorspan{T_1,T_2,T_3}$.
The commutation relations \eqref{eq:commutdual} are no longer automatically satisfied independent of the choice of $c_{ij}$ and $b_{ij}$.
Indeed, following the discussion in \secref{ssec:etalambda} we are required to set $c_{12} = c_{13} = 0$.
In fact, we will use the same basis \eqref{eq:basisi} as for cases (i) and (ii).
Again the subgroup with respect to which we dualise lies completely within the gauge group.

We start by constructing a suitable parametrisation for the field $k \in \tilde{\grp{K}}\backslash\grp{D}$ appearing in the action \eqref{eq:actionS1}.
The group-valued field $k \in \grp{SL}_4(\Complex)$ can be written as
\[
k = \tilde{g} g = \tilde{m} \tilde{g}_0 g ~, \qquad g \in \grp{SU}_4 ~, \quad \tilde{g} \in \grp{B}_4 ~, \quad \tilde{g}_0 \in \tilde{\grp{G}}_0 ~, \quad \tilde{m} \in \tilde{\grp{M}} ~,
\]
where $\tilde{\grp{G}}_0$ and $\tilde{\grp{M}}$ are generated by $\set{\tilde{T}^{1,2,3}}$ and $\set{\tilde{T}^\ind{A} : A = 4,\ldots,15}$ respectively.
We can then partially gauge-fix by taking $\tilde{k} = g_0 \tilde{m}^{-1}$ in the gauge transformation \eqref{eq:gthk}.
The resulting expression for $k$ is
\[
k = (g_0 \tilde{g}_0 \bar{g}_0^{-1})(\bar{g}_0 g h) ~, \qquad g_0, \bar{g}_0 \in \grp{SU}_2 ~, \quad h \in \grp{SO}_5 ~,
\]
where $g_0$, $\bar{g}_0$ and $h$ are understood as gauge degrees of freedom.
Using an explicit $2 \times 2$ matrix representation for $\grp{G}_0 = \grp{SU}_2$ and $\tilde{\grp{G}}_0 = \grp{B}_2$, one can show there exists a $\bar{g}_0 \in \grp{SU}_2$ such that $g_0 \tilde{g}_0 \bar{g}_0^{-1} \in \grp{B}_2$.
Furthermore, for $\alg{su}_2 = \vectorspan{T_1,T_2,T_3}$ we can choose $g_0$ and $\bar{g}_0$ such that
\[
g_0 \tilde{g}_0 \bar{g}_0^{-1} = \exp[\tilde{\phi}_1 \, \tilde{T}^1] ~.
\]
From now on we will take $g_0$ such that this relation is satisfied.
This however only partially fixes the degrees of freedom of $g_0$, with the remaining freedom given by $g_0 \to \exp[ \xi \, T_1]g_0$ and $\bar{g}_0 \to \exp[\xi \, T_1] \bar{g}_0$.
Therefore, $k$ is now given by
\[
k = \exp[\tilde{\phi}_1 \, \tilde{T}^1] \exp[\xi \, T_1] g^\prime h ~,
\]
where we have used that $\exp[\xi \, T_1]$ commutes with $\exp[\tilde{\phi}_1 \, \tilde{T}^1]$ and defined $g^\prime = \bar{g}_0 g \in \grp{SU}_4$ as a new field.
Using the parametrisation of the coset $\grp{G}/\grp{H}$
\[
g^\prime h = e^{\phi_1 T_1/2 + \phi_2 T_{11}/2 + \phi T_6} e^{\arccos{x} \, T_{13}} e^{\arcsin r \, (T_{10}-T_{5})/2} ~,
\]
it is then clear that we can choose the gauge degree of freedom $\xi$ to gauge-fix $\phi_1 = 0$.
Therefore it follows that we can use the same parametrisation as for the abelian dual in case (ii), namely
\[
k = e^{\tilde{\phi}_1 \tilde{T}^1/2 + \phi_2 T_{11}/2 + \phi T_6} e^{\arccos{x} \, T_{13}} e^{\arcsin r \,(T_{10}-T_{5})/2} ~.
\]
The metric and $B$-field are
\[
\eta^{-1} \extder s^2 & = \frac{1}{1 + \eta^2 r^2} \bigg(\frac{\extder r^2}{1-r^2} + (1-r^2) \extder \phi^2 \bigg) + \frac{1}{4 \eta^2 r^2} \frac{\extder z^2}{(1-z)^2}
\\
& \phantom{= \,} + \frac{r^2 x^2 (1-x^2)}{1+ \eta^2 r^4 (x^2+z^{-1}-1)^2} \bigg( \bigg(\frac{1}{2} \frac{\extder z}{1-z}+ \frac{\extder x}{x(1-x^2)}\bigg)^2 + \extder \phi_2^2 \bigg) ~,
\\
\eta^{-1} B & = \frac{1}{4 \eta}\bigg(\frac{2x^2}{z}\frac{z+ \eta^2 r^4 (x^2+z^{-1}-1)}{1+ \eta^2 r^4 (x^2+z^{-1}-1)^2} -1\bigg) \frac{\extder z}{(1-z)} \wedge \extder \phi_2
\\
& \phantom{= \,} - \frac{\eta r^4 x (x^2+z^{-1}-1)}{1+ \eta^2 r^4 (x^2+z^{-1}-1)^2}\extder \phi_2 \wedge \extder x - \frac{\eta r}{1 + \eta^2 r^2} \extder \phi \wedge \extder r ~,
\\
z & = 1 - e^{\tilde{\phi}_1} ~.
\]
While shifts in $\phi$ and $\phi_2$ remain isometries, shifts in $\tilde{\phi}_1$ no longer leave the background invariant.

Let us now consider two interesting limits of this background.
First, sending either $\tilde{\phi}_1 \to -\infty$ or $\tilde{\phi}_1 \to \infty$ (and subsequently redefining $x \to \sqrt{1-x^2}$ and $\tilde{\phi}_1 \to - \tilde{\phi}_1$) we recover the abelian dual \eqref{eq:etau1dual}.
This behaviour is explained by a similar logic to that used in the $\Sp^2$ example in \secref{ssec:twosphere}.
In the limit $\tilde{\phi}_1 \to - \infty$, the right action of $e^{\tilde{\phi}_1 \tilde{T}_1/2}$ on $\tilde{\alg{k}}$ coincides for cases (ii) and (iii).
Explicitly we have
\[
\set{T_1,T_2,T_3,\tilde{T}^\ind{A}} e^{\tilde{\phi}_1 \tilde{T}_1/2} \quad \xrightarrow{\tilde{\phi}_1 \to - \infty} \quad \set{T_1,\tilde{T}^3,-\tilde{T}^2,\tilde{T}^\ind{A}} e^{\tilde{\phi}_1 \tilde{T}_1/2} ~, \qquad A = 4,\ldots,15 ~.
\]
Therefore, there is again no longer any distinction between the Poisson--Lie duals with respect to $\grp{U}_1$ and $\grp{SU}_2$.

Second, after appropriately rescaling $\tilde{\phi}_1 \to \eta \tilde{\phi}_1$, we can take the undeformed limit.
As expected, the resulting background
\[\label{eq:nadsu2}
\eta^{-1} \extder s^2 & = \frac{\extder r^2}{1-r^2} + (1-r^2) \extder \phi^2 + \frac{r^2 \tilde{\phi}_1^2}{r^4 + \tilde{\phi}_1^2} \frac{\extder x^2}{1-x^2} + \frac{r^2 x^2 (1-x^2) \tilde{\phi}_1^2}{r^4 + \tilde{\phi}_1^2} \extder \phi_2^2 + \frac{1}{4 r^2}\extder \tilde{\phi}_1^2 ~,
\\
\eta^{-1} B & = \frac{1}{4} (1-2x^2) \extder \tilde{\phi}_1 \wedge \extder \phi_2 + \frac{ r^4 x \tilde{\phi}_1}{r^4 + \tilde{\phi}_1^2} \extder \phi_2 \wedge \extder x ~,
\]
is the non-abelian dual with respect to $\grp{SU}_2$ of $\Sp^5$.

%%%%%%%%%%%%%%%%%%%%%%%%%%%%%%%%%%%%%%%%
\paragraph{Case (iv).}
The last case we investigate in detail corresponds to the dualisation in the non-abelian group $\grp{SU}_2$ generated by $\alg{su}_2 = \vectorspan{T_6,T_7,T_8}$.
In order to satisfy the commutation relation \eqref{eq:commutdual} we need to take $c_{21} = c_{23} = 0$.
Therefore we set the constants $c_{ij}$ and $b_{ij}$ such that
\[\label{eq:basisiv}
T_1 & = i(h_1+h_2+h_3) ~, \qquad & T_6 & = i h_2 ~, \qquad & T_{11} & = \ihalf(h_1-h_3) ~, \\
\tilde{T}^1 & = \half(h_1+h_2+h_3) ~, \qquad & \tilde{T}^6 & = \half h_2 ~, \qquad & \tilde{T}^{11} & = \half(h_1-h_3) ~.
\]
One important difference with respect to case (ii) and (iii) regarding the technical computation is that the subgroup with respect to which we dualise does not lie completely inside the gauge group.
In order to find an appropriate parametrisation of $k \in \grp{SL}_4(\Complex)$ we proceed in the same spirit as in case (iii) and choose to work with
\[
k = e^{\tilde{\phi}_3 \tilde{T}^6/2 + \phi_1 (T_1-T_6)/2 + \phi_2 T_{11}} e^{\arccos x \, T_{15}} e^{\arccos r \,(T_{10}-T_{5})/2} ~.
\]
The resulting metric and $B$-field have a complicated structure.
Their expressions are not particularly enlightening and we refer the interested reader to \appref{app:metrics} for the details.

Rather let us briefly comment on the analogous limits to those discussed above in case (iii).
Shifting $\tilde{\phi}_3$ by a constant is not an isometry of the Poisson--Lie dual background \eqref{eq:appbfin}.
However, it becomes a symmetry if we send either $\tilde{\phi}_3 \to -\infty$ or $\tilde{\phi}_3 \to \infty$.
In both these limits we recover (up to a total derivative for the latter) the background of the associated abelian dual \eqref{eq:appbpen}.
Note that to see this for $\tilde{\phi}_3 \to \infty$ we also need to subsequently redefine $x \to \sqrt{1-x^2}$ and $\tilde{\phi}_3 \to - \tilde{\phi}_3$.

In the undeformed limit, after an appropriate rescaling of the dual field $\tilde{\phi}_3$, we again find the non-abelian dual with respect to $\grp{SU}_2$ \eqref{eq:nadsu2}.
While the two non-abelian duals recovered in the undeformed limits are the same for cases (iii) and (iv), this is not true for the two Poisson--Lie duals.
In particular, the integrated degrees of freedom are associated to algebras that are not isomorphic: the algebra in case (iv) has two one-dimensional ideals, while the algebra in case (iii) does not.

%%%%%%%%%%%%%%%%%%%%%%%%%%%%%%%%%%%%%%%%
\paragraph{Higher-dimensional subgroups.}
To conclude this section let us briefly summarise the subgroups of $\grp{SU}_4$ with respect to which we can Poisson--Lie dualise the $\eta$~deformed $\Sp^5$ symmetric space sigma model.
As discussed in \secref{ssec:etalambda} these subgroups are classified by sub-Dynkin diagrams.
For $\grp{SU}_4$ the possible choices are:
\begin{itemize}
\item $\grp{SU}_2$ type 1, generated by $\set{ih_j,i(e_j+f_j),-(e_j-f_j)}$ where $j = 1$ or $j = 3$.
\item $\grp{SU}_2$ type 2, generated by $\set{ih_j,i(e_j+f_j),-(e_j-f_j)}$ where $j = 2$.
\item $\grp{SU}_2 \times \grp{SU}_2$, generated by $\set{ih_j,i(e_j+f_j),-(e_j-f_j)}$ where $j = 1,3$.
\item $\grp{SU}_3$, generated by $\set{ih_j,i(e_j+f_j),-(e_j-f_j)}$ where $j = 1,2$ or $j = 2,3$.
\item $\grp{SU}_4$, generated by $\set{ih_j,i(e_j+f_j),-(e_j-f_j)}$ where $j = 1,2,3$.
\end{itemize}
For $\grp{SU}(2)$ type 1 and $\grp{SU}(3)$ we have two different options.
In these cases the Lie algebra whose degrees of freedom we integrate out, that is $\tilde{\alg{k}}$, are isomorphic and accordingly we expect the dual models to be the same.

It is only for the final case, which corresponds to the $\lambdastar$~deformation, that the algebra $\tilde{\alg{k}}$ is unimodular.
However, let us recall that we can include additional $\grp{U}_1$ factors, whose number is limited only by the fact that the total rank of the Lie group with respect to which we dualise should be less than or equal to that of $\grp{SU}_4$.
These factors can always be chosen such that $\tilde{\alg{k}}$ becomes unimodular.

%%%%%%%%%%%%%%%%%%%%%%%%%%%%%%%%%%%%%%%%%%%%%%%%%%%%%%%%%%%%%%%%%%%%%%%%%%%%%%%%
\section{Concluding remarks}\label{sec:conc}

In this paper we have initiated the investigation into the space of Poisson--Lie duals of the $\eta$~deformation of the symmetric space sigma model for the coset $\grp{G}/\grp{H}$ (with $\grp{G}$ a compact simple Lie group).
Starting from the model on the Drinfel'd double $\grp{D}$ \cite{Klimcik:1995dy,Klimcik:1996nq} with $\grp{D} \equiv \grp{G}^{\Complex}$ we identified a class of subgroups $\grp{G}_0 \subset \grp{G}$ with respect to which it is possible to dualise.
The corresponding Lie algebras $\alg{g}_0 = \Lie(\grp{G}_0)$ are constructed by picking a subset of the Cartan generators of $\grp{G}^{\Complex}$ along with the associated roots.
Such subalgebras are in correspondence with sub-Dynkin diagrams.
Additional $\grp{U}_1$ factors built from the remaining Cartan generators can also be included, and correspond to abelian dualities.
It would be interesting to investigate if there are further possible dualities, based on non-semisimple groups, when the Lie group $\grp{G}$ is taken to be non-compact.
This would be relevant, for example, for the $\eta$~deformation of $\AdS$ space.

In the undeformed limit the Poisson--Lie duals of the $\eta$~deformation reduce to the corresponding non-abelian duals of the symmetric space sigma model.
However, the restrictions outlined above suggest that not all non-abelian duals can be deformed.
Therefore the deformation restricts the space of allowed dualities.
On the other hand two non-abelian duals that are isomorphic before the deformation may no longer be so afterwards.

Applying this formalism to the specific cases of the $\eta$~deformations of $\Sp^2$ and $\Sp^5$ we have demonstrated these properties on a number of different examples.

\medskip

Curiously there is an alternative relation between the $\eta$ and $\lambda$~deformations \cite{Hoare:2015gda}.
Analytically continuing the latter and taking certain infinite limits one recovers the complete abelian dual of the former.
For the examples of $\Sp^2$ and $\Sp^5$ we have seen hints of the algebraic structure underlying the existence of these limits.
Starting from the Poisson--Lie dual with respect to an $\grp{SU}_2$ subgroup and taking an infinite limit in the field associated to its Cartan generator we recover the abelian dual of the $\eta$~deformation.
This should be interpreted as a certain contraction limit on the Lie algebra whose degrees of freedom are integrated out.
Understanding the interplay between the algebraic and geometric limits in detail may provide greater insight into the relationships between these models.

It may also prove interesting to explore the $\eta \to \infty$ limit of the Poisson--Lie duals.
Various ways of taking this limit in the $\eta$~deformed model have been studied \cite{Delduc:2013fga,Hoare:2014pna,Arutynov:2014ota,Arutyunov:2014cra,Pachol:2015mfa,Hoare:2016ibq}, with the limit of \cite{Arutynov:2014ota,Arutyunov:2014cra} having the particular algebraic interpretation as a contraction of the $q$~deformed algebra \cite{Pachol:2015mfa}.

\medskip

Our initial motivation for exploring the space of Poisson--Lie duals was to gain a better understanding of the Weyl anomaly of the $\eta$~deformation of the $\AdS_5 \times \Sp^5$ superstring.
The preservation of Weyl invariance under non-abelian duality \cite{Alvarez:1994np,Elitzur:1994ri} and Poisson--Lie duality \cite{Tyurin:1995bu,Bossard:2001au,VonUnge:2002xjf} has been discussed extensively in the literature.
In such expositions a Weyl anomaly is typically associated to integrating out the degrees of freedom corresponding to a non-unimodular algebra, that is whose structure constants satisfy $f_{ab}{}^b \neq 0$.

While the models discussed in this paper are not conformal, it is interesting to note that, starting from the model on the Drinfel'd double, the $\eta$~deformation is associated to integrating out a non-unimodular algebra (the Borel subalgebra $\alg{b}$), while its complete abelian dual and the $\lambdastar$~deformation are associated to integrating out unimodular algebras (an alternative real form of the Borel subalgebra and the simple Lie algebra $\alg{g}$ respectively).
Lifting this to the $\AdS_5 \times \Sp^5$ superstring may provide an explanation of why the $\eta$~deformation does not define a supergravity background, yet its complete abelian dual and the $\lambda$, or $\lambdastar$, deformation do \cite{Arutyunov:2015qva,Hoare:2015gda,Hoare:2015wia,Borsato:2016zcf,Borsato:2016ose}.
\unskip\footnote{BH would like to thank Arkady~Tseytlin for a number of discussions on these and related issues.}
Indeed, we have checked that the aforementioned unimodularity properties also hold for the isometry algebra of the $\AdS_2 \times \Sp^2$ superstring, $\alg{psu}_{1,1|2}$, as well as for the dual Borel subalgebra of the Drinfel'd double $\alg{psl}_{2|2}(\Complex)$.

There are number of open questions that would need to be addressed to formalise this proposal.
In this paper we have assumed that the $\eta$ and $\lambdastar$ deformations are Poisson--Lie dual, a claim that remains to be proven for the deformed symmetric space and superstring sigma models.
Furthermore, it would be necessary to understand the model on the Drinfel'd double in the path integral formalism together with its quantum properties, crucially including conformal invariance.
There has been a substantial amount of work already in this direction \cite{Alekseev:1995ym,Tyurin:1995bu,Bossard:2001au,VonUnge:2002xjf,Valent:2009nv,Sfetsos:2009dj,Avramis:2009xi,Sfetsos:2009vt,Hlavaty:2012sg,Hassler:2017yza,Jurco:2017gii}, however, a systematic approach that could be applied to the $\AdS_5 \times \Sp^5$ superstring remains lacking.

Were the Weyl anomaly to be understood in this way, it would identify those Poisson--Lie duals of the $\eta$~deformed $\AdS_5 \times \Sp^5$ superstring that define string theories and those that do not.
One particularly interesting Weyl-invariant example would be the Poisson--Lie dual with respect to the bosonic subgroup of $\grp{PSU}_{2,2|4}$.
This would have the same metric and $B$-field as the $\lambdastar$~deformation, but with different, possibly simpler, Ramond-Ramond fluxes and dilaton.
In fact there already exist candidate supergravity backgrounds for this model and its integrable cousins \cite{Sfetsos:2014cea,Demulder:2015lva}.

Finally, tracking the fate of the conserved charges, $q$~deformed symmetry and classical integrability under Poisson--Lie duality would open the door to solving these string theories.

\paragraph{Note added.} After this paper was posted to arXiv, the interesting article \cite{Severa:2017kcs} appeared.
In this paper the author demonstrates classical integrability, via the construction of a Lax pair, for a class of Poisson--Lie symmetric models.
This class includes the Poisson--Lie duals of the $\eta$~deformation considered in this paper, thereby confirming their classical integrability.

%%%%%%%%%%%%%%%%%%%%%%%%%%%%%%%%%%%%%%%%%%%%%%%%%%%%%%%%%%%%%%%%%%%%%%%%%%%%%%%%
\pdfbookmark[1]{Acknowledgements}{ack}
\section*{Acknowledgements}

We would like to thank Arkady~Tseytlin for many interesting and insightful discussions, and Niklas Beisert and Arkady Tseytlin for comments on the draft.
This work is supported by grant no.~615203 from the European Research Council under the FP7.

%%%%%%%%%%%%%%%%%%%%%%%%%%%%%%%%%%%%%%%%%%%%%%%%%%%%%%%%%%%%%%%%%%%%%%%%%%%%%%%%
\appendix

%%%%%%%%%%%%%%%%%%%%%%%%%%%%%%%%%%%%%%%%%%%%%%%%%%%%%%%%%%%%%%%%%%%%%%%%%%%%%%%%
\section{Gauge invariance of the \texorpdfstring{$\eta$}{eta}~deformed model}\label{app:gauge}

In this appendix we show that if the components of the operator $\Bn$ satisfy \eqref{eq:gc8text} then the gauge invariance condition \eqref{eq:requirement} is fulfilled.
In the following derivations we recall that the operator $\Bn$ maps $\alg{b} \to \alg{g}$ while $P$ maps $\alg{g} \to \alg{b}$.
We are therefore able to make the replacements
\[
& \Bn \to \Proj{\alg{g}} \Bn \Proj{\alg{b}} ~, & \qquad & \Bn \to \Bn \Proj{\alg{b}} ~, & \qquad & \Bn \to \Proj{\alg{g}} \Bn ~,
\\ & P \to \Proj{\alg{b}} P \Proj{\alg{g}} ~, & \qquad & P \to P \Proj{\alg{g}} ~, & \qquad & P \to \Proj{\alg{b}} P ~, \]
where it is useful to do so.
The reverse replacement can be made in the final case of each line, while for the first two cases it can be made if the operator is already acting on an object taking values in the relevant space.
Similarly one may make the replacements
\[
& \ad_{\theta}^n \Proj{\alg{g}} \to \Proj{\alg{g}} \ad_{\theta}^n \Proj{\alg{g}} ~, & \qquad & \theta \in \alg{g} ~,
\\ & \Ad_g \Proj{\alg{g}} \to \Proj{\alg{g}} \Ad_g \Proj{\alg{g}} ~, & \qquad & g \in \grp{G} ~,
\]
along with the reverse replacements.

We start by recalling the definition of $\Pi(g)$ \eqref{eq:pimap} and expanding \eqref{eq:requirement}
\[\label{eq:gc1}
P \Ad_h \Proj{\alg{g}} ( \Bn + \Ad_h^{-1} \Ad_g^{-1} \Proj{\alg{b}} \Ad_g \Ad_h ) \Proj{\alg{b}} \Ad_h^{-1} P = P \Proj{\alg{g}} (\Bn + \Ad_g^{-1} \Proj{\alg{b}} \Ad_g) \Proj{\alg{b}} P ~.
\]
This condition should be understood as an operator equation acting on $\alg{g}$, with each term in \eqref{eq:gc1} mapping from $\alg{g} \to \alg{b}$.

Let us analyse the condition \eqref{eq:gc1}.
In the second term on the left-hand side we insert the identity operator $\Proj{\alg{g}} + \Proj{\alg{b}}$ between the $\Ad_g$ and $\Ad_h$ operators.
The term arising from the $\Proj{\alg{g}}$ insertion now contains the combination $\Proj{\alg{b}} \Ad_g \Proj{\alg{g}}$, which is the zero operator.
Using \eqref{eq:rel2} to replace $P_{\alg{b}} \Ad_h P_{\alg{b}} \Ad_h^{-1}$ with $\Proj{\alg{b}}$ in the remaining term, \eqref{eq:gc1} then simplifies to
\[\label{eq:gc3}
P ( \Proj{\alg{g}} \Bn \Proj{\alg{b}} - \Ad_h \Proj{\alg{g}} \Bn \Proj{\alg{b}} \Ad_h^{-1} ) P = P (\Ad_h \Proj{\alg{g}} \Ad_h^{-1} - \Proj{\alg{g}}) \Ad_g^{-1} \Proj{\alg{b}} \Ad_g \Proj{\alg{b}} P ~.
\]
We now observe that the operator $(\Ad_h \Proj{\alg{g}} \Ad_h^{-1} - \Proj{\alg{g}})$ is the zero operator when acting on $\alg{g}$ as $h \in \grp{H} \subset \grp{G}$.
Therefore we are free to insert $\Proj{\alg{b}}$ after this operator in \eqref{eq:gc3}.
Again using \eqref{eq:rel2} to replace $P_{\alg{b}} \Ad_g P_{\alg{b}} \Ad_g^{-1}$ with $\Proj{\alg{b}}$ the dependence on $g$ drops out and we are left with
\[\label{eq:gc4}
P ( \Proj{\alg{g}} \Bn \Proj{\alg{b}} - \Ad_h \Proj{\alg{g}} \Bn \Proj{\alg{b}} \Ad_h^{-1} ) P = P \Ad_h \Proj{\alg{g}} \Ad_h^{-1} \Proj{\alg{b}} P ~,
\]
where we have also used that $\Proj{\alg{g}} \Proj{\alg{b}} = 0$.

It is still the case that each term in \eqref{eq:gc4} is an operator that maps from $\alg{g}$ to $\alg{b}$.
Therefore we act on \eqref{eq:gc4} with the operator $\Proj{\alg{b}} \Ad_h^{-1} \Proj{\alg{b}}$, which is invertible on $\alg{b}$, from the left to give
\[\label{eq:gc5}
P ( \Ad_h^{-1} \Proj{\alg{g}} \Bn \Proj{\alg{b}} - \Proj{\alg{g}} \Bn \Proj{\alg{b}} \Ad_h^{-1} ) P = P \Proj{\alg{g}} \Ad_h^{-1} \Proj{\alg{b}} P ~.
\]

Let us now assume $h$ takes the form $\exp[\xi]$, $\xi \in \alg{h}$ and demand that \eqref{eq:gc5} is satisfied at each order in $\xi$.
The $\Order(\xi^n)$ term in the expansion (up to a numerical factor) is
\[\label{eq:gc6}
P (\ad_\xi^n \Proj{\alg{g}} \Bn \Proj{\alg{b}} - \Proj{\alg{g}} \Bn \Proj{\alg{b}} \ad_\xi^n ) P = P \Proj{\alg{g}} \ad_\xi^n \Proj{\alg{b}} P ~.
\]
This relation is trivially satisfied for $n = 0$.
Assuming that \eqref{eq:gc6} is true for $n = 1$ and $n = k-1$, for $n = k$ we have
\[
P \ad_\xi^k \Proj{\alg{g}} \Bn \Proj{\alg{b}} P & = P \ad_\xi \ad_\xi^{k-1} \Proj{\alg{g}} \Bn \Proj{\alg{b}} P
\\
& = \Proj{\alg{b}} \ad_\xi P \Proj{\alg{g}} \Bn \Proj{\alg{b}} \ad_\xi^{k-1} P + \Proj{\alg{b}} \ad_\xi P \Proj{\alg{g}} \ad_{\xi}^{k-1} \Proj{\alg{b}} P
\\
& = P \Proj{\alg{g}} \Bn \Proj{\alg{b}} \ad_\xi P \ad_\xi^{k-1} + P \Proj{\alg{g}} \ad_\xi \Proj{\alg{b}} P \ad_\xi^{k-1} + P \ad_\xi \Proj{\alg{g}} \ad_{\xi}^{k-1} \Proj{\alg{b}} P
\\
& = P \Proj{\alg{g}} \Bn \Proj{\alg{b}} \ad_\xi^k P + P \Proj{\alg{g}} \ad_\xi ( \Proj{\alg{b}} + \Proj{\alg{g}}) \ad_\xi^{k-1} \Proj{\alg{b}} P
\\
& = P \Proj{\alg{g}} \Bn \Proj{\alg{b}} \ad_\xi^k P + P \Proj{\alg{g}} \ad_\xi^k \Proj{\alg{b}} P ~,
\]
where we have used the identity \eqref{eq:projadcom}.
It then follows that \eqref{eq:gc6} is true for $n = k$.
Therefore if we find a solution to \eqref{eq:gc6} for $n = 1$ by induction this will also be a solution for all $n$.

The final step is to find a solution to \eqref{eq:gc6} for $n = 1$.
Let us take $\xi = U_{\bar c}$ and act with both sides of \eqref{eq:gc6} with $n = 1$ on $U_{\bar\alpha}$.
Doing so we find
\[\label{eq:gc7}
\Bn^{\bar\alpha\ind{D}} f_{\bar c\ind{D}}{}^{\bar\beta} - f_{\ind{D}\bar c}{}^{\bar\alpha} \Bn^{\ind{D}\bar\beta} = \tilde{f}^{\bar\alpha\bar\beta}{}_{\bar c} ~,
\]
recovering \eqref{eq:gc8text} as claimed.

%%%%%%%%%%%%%%%%%%%%%%%%%%%%%%%%%%%%%%%%%%%%%%%%%%%%%%%%%%%%%%%%%%%%%%%%%%%%%%%%
\section{List of parametrisations and backgrounds} \label{app:metrics}
In this appendix we present the metrics and $B$-fields for case (iv) of \secref{ssec:fivesphere}.
Recalling that the relevant basis for the Cartan generators, $T_{1,6,11}$ and $\tilde{T}^{1,6,11}$, is given in \eqref{eq:basisiv}, we use the parametrisation
\[
g & = e^{\phi_3 T_6 + \phi_1 (T_1-T_6)/2 + \phi_2 T_{11}} e^{\arccos x \, T_{15}} e^{\arccos r \,(T_{10}-T_{5})/2} ~,
\]
for the $\Sp^5$ background and its $\eta$~deformation, while for the various dual models we take
\[
k & = e^{\tilde{\phi}_3 \tilde{T}^6/2 + \phi_1 (T_1-T_6)/2 + \phi_2 T_{11}} e^{\arccos x \, T_{15}} e^{\arccos r \,(T_{10}-T_{5})/2} ~.
\]
With these parametrisations we find the following metrics and $B$-fields:
\begin{itemize}
\item $\Sp^5$ background:
\[
\extder s^2 & = r^2 (1-x^2) (\extder \phi_1-\extder \phi_3)^2 + (1-r^2) \extder \phi_2^2 + r^2 x^2 \extder \phi_3^2 + \frac{\extder r^2}{1-r^2} + r^2 \frac{\extder x^2}{1-x^2} ~, \\
B & = 0 ~.
\]
\item Abelian dual in $\set{ih_2}$:
\[
\extder s^2 & = r^2 x^2 (1-x^2) \extder \phi_1^2 + (1-r^2) \extder \phi_2^2 + \frac{1}{4r^2} \extder \tilde{\phi}_3^2 + \frac{\extder r^2}{1-r^2} + r^2 \frac{\extder x^2}{1-x^2} ~,
\\
B & = \half (1-x^2) \extder \phi_1 \wedge \extder \tilde{\phi}_3 ~.
\]
\item Non-abelian dual in $\set{ih_2,i(e_2+f_2),-(e_2-f_2)}$:
\[
\extder s^2 & = \frac{r^2 x^2 (1-x^2) \tilde{\phi}_3^2}{\tilde{\phi}_3^2 + r^4} \extder \phi_1^2 + (1-r^2) \extder \phi_2^2 + \frac{1}{4r^2} \extder \tilde{\phi}_3^2 + \frac{\extder r^2}{1-r^2} + \frac{r^2\tilde{\phi}_3^2}{\tilde{\phi}_3^2+r^4} \frac{\extder x^2}{1-x^2} ~,
\\
B & = \half (1-x^2) \extder \phi_1 \wedge \extder \tilde{\phi}_3 - \frac{r^4 x \tilde{\phi}_3}{\tilde{\phi}_3^2 + r^4} \extder \phi_1 \wedge \extder x ~.
\]
\item $\eta$~deformation of $\Sp^5$:
%%%%%%%%%%%%%%%%%%%%%%%%%%%%%%%%%%%%%%%%
\def\denoma{1+\eta^2 (1-r^2)(1-r^2 x^2)}
\def\denomb{1+\eta^2 (1-r^2 x^2)}
%%%%%%%%%%%%%%%%%%%%%%%%%%%%%%%%%%%%%%%%
\begingroup\makeatletter\def\f@size{9}\check@mathfonts
\def\maketag@@@#1{\hbox{\m@th\normalsize\normalfont#1}}
\[
\eta^{-1} \extder s^2 & = \frac{r^2(1-x^2)}{\denoma} (\extder \phi_1 - \extder \phi_3)^2 + (1-r^2) \extder \phi_2^2 + \frac{r^2 x^2}{\denomb} \extder \phi_3^2 \\
& \phantom{= \,} + \frac{1}{\denoma} \frac{((1-x^2) \extder r - r (1-r^2) x \extder x)^2}{(1-r^2)(1-x^2)(1-r^2x^2)} \\
& \phantom{= \,} + \frac{1}{\denomb} \frac{(x \extder r + r \extder x)^2}{1-r^2 x^2} ~,
\\
\eta^{-1} B & = \frac{\eta r}{\denoma} (\extder \phi_1 - \extder \phi_3) \wedge ((1-x^2) \extder r - r (1-r^2) x \extder x) \\
& \phantom{= \,} + \frac{\eta r x}{\denomb} \extder \phi_3 \wedge (x \extder r + r \extder x) ~.
\]
\endgroup
\item Poisson--Lie dual in $\set{ih_2}$:
\begingroup\makeatletter\def\f@size{9}\check@mathfonts
\def\maketag@@@#1{\hbox{\m@th\normalsize\normalfont#1}}
\[\label{eq:appbpen}
\eta^{-1} \extder s^2 & = \frac{r^2 x^2 (1-x^2)}{N_0} \extder \phi_1^2 + (1-r^2) \extder \phi_2^2 + \frac{(1+\eta^2 \beta_0)(1+\eta^2 (1-r^2) \beta_0)}{4 \eta^2 r^2 N_0} \extder \tilde{\phi}_3^2 \\
& \phantom{= \,} + \frac{1}{N_0} \Big( \frac{\extder r^2}{1-r^2} +r^2 \frac{\extder x^2}{1-x^2} + \eta^2(1-r^2) \Big( \alpha_0 \frac{\extder r}{1-r^2} -2rx \extder x \Big)^2 \Big) \\
& \phantom{= \,} + \frac{x}{N_0} \Big(2-r^2+2\eta^2(1-r^2)\beta_0\Big) \extder \tilde{\phi}_3 \extder x - \frac{1}{r N_0} \Big(1-2x^2+\eta^2 \alpha_0 \beta_0 \Big) \extder \tilde{\phi}_3 \extder r ~,
\\
\eta^{-1} B & = \frac{(1-x^2)(1+\eta^2 \beta_0)}{2 \eta N_0} \extder \phi_1 \wedge \extder \tilde{\phi}_3 + \frac{\eta r^2 x \alpha_0}{N_0} \extder \phi_1 \wedge \extder x + \frac{2 \eta r x^2 (1-x^2)}{N_0} \extder \phi_1 \wedge \extder r ~,
\\
N_0 & = 1 + \eta^2 (1-r^2x^2)^2 ~, \qquad
\alpha_0 = 1-(2-r^2)x^2 ~, \qquad
\beta_0 = 1-r^2x^2 ~.
\]
\endgroup
\item Poisson--Lie dual in $\set{ih_2,i(e_2+f_2),-(e_2-f_2)}$:
\begingroup
\makeatletter\def\f@size{9}\check@mathfonts
\def\maketag@@@#1{\hbox{\m@th\normalsize\normalfont#1}}
\[\label{eq:appbfin}
\eta^{-1} \extder s^2 & = \frac{r^2 x^2(1-x^2)}{N} \extder \phi_1^2 + (1-r^2) \extder \phi_2^2 \\
& \phantom{= \,} + \frac{(1+\eta^2\beta)(1+\eta^2(1-r^2)\beta) + \eta^2r^4z^{-2}(1-z)}{4 \eta^2 r^2 N} \extder \tilde{\phi}_3^2 \\
& \phantom{= \,} + \frac{1}{N} \Big( \frac{\extder r^2}{1-r^2} + r^2 \frac{\extder x^2}{1-x^2} + \eta^2(1-r^2)\Big(\alpha \frac{\extder r}{1-r^2} - 2rx \extder x\Big)^2 \Big) \\
& \phantom{= \,} + \frac{x}{N} \Big( 2-r^2+2\eta^2(1-r^2)\beta \Big) \extder \tilde{\phi}_3 \extder x -\frac{1}{rN} \Big((1-2x^2) + \eta^2 \alpha \beta \Big) \extder \tilde{\phi}_3 \extder r ~,
\\
\eta^{-1} B & = \frac{(1-x^2)(1+\eta^2\beta+\eta^2r^2(\alpha+2x^2)z^{-1}(1-z))}{2 \eta N} \extder \phi_1 \wedge \extder \tilde{\phi}_3 \\
& \phantom{= \,} + \frac{\eta r^2 x \alpha}{N} \extder \phi_1 \wedge \extder x + \frac{2 \eta r x^2(1-x^2)}{N} \extder \phi_1 \wedge \extder r ~,
\\
z & = 1-e^{\tilde{\phi}_3} ~, \qquad
N = 1 + \eta^2 \big( (r^2(x^2+z^{-1}-1)+1)^2-4r^2x^2z^{-1} \big) ~,
\\
\alpha & = 1-2x^2+r^2(x^2+z^{-1}-1) ~, \qquad
\beta = 1+r^2(x^2+z^{-1}(1-2x^2)-1) ~.
\]
\endgroup
\end{itemize}

%%%%%%%%%%%%%%%%%%%%%%%%%%%%%%%%%%%%%%%%%%%%%%%%%%%%%%%%%%%%%%%%%%%%%%%%%%%%%%%%
\begin{bibtex}[\jobname]

@article{Delduc:2013qra,
author         = "Delduc, Francois and Magro, Marc and Vicedo, Benoit",
title          = "{An integrable deformation of the $AdS_5 \times S^5$ superstring action}",
journal        = "Phys. Rev. Lett.",
volume         = "112",
year           = "2014",
number         = "5",
pages          = "051601",
doi            = "10.1103/PhysRevLett.112.051601",
eprint         = "1309.5850",
archivePrefix  = "arXiv",
primaryClass   = "hep-th",
SLACcitation   = "%%CITATION = ARXIV:1309.5850;%%"
}

@article{Delduc:2014kha,
author         = "Delduc, Francois and Magro, Marc and Vicedo, Benoit",
title          = "{Derivation of the action and symmetries of the $q$-deformed $AdS_5 \times S^5$ superstring}",
journal        = "JHEP",
volume         = "10",
year           = "2014",
pages          = "132",
doi            = "10.1007/JHEP10(2014)132",
eprint         = "1406.6286",
archivePrefix  = "arXiv",
primaryClass   = "hep-th",
SLACcitation   = "%%CITATION = ARXIV:1406.6286;%%"
}

@article{Hollowood:2014qma,
author         = "Hollowood, Timothy J. and Miramontes, J. Luis and Schmidtt, David M.",
title          = "{An integrable deformation of the $AdS_5 \times S^5$ superstring}",
journal        = "J. Phys.",
volume         = "A47",
year           = "2014",
number         = "49",
pages          = "495402",
doi            = "10.1088/1751-8113/47/49/495402",
eprint         = "1409.1538",
archivePrefix  = "arXiv",
primaryClass   = "hep-th",
SLACcitation   = "%%CITATION = ARXIV:1409.1538;%%"
}

@article{Green:1983wt,
author         = "Green, Michael B. and Schwarz, John H.",
title          = "{Covariant description of superstrings}",
journal        = "Phys. Lett.",
volume         = "136B",
year           = "1984",
pages          = "367-370",
doi            = "10.1016/0370-2693(84)92021-5",
reportNumber   = "QMC-83-7",
SLACcitation   = "%%CITATION = PHLTA,136B,367;%%"
}

@article{Green:1983sg,
author         = "Green, Michael B. and Schwarz, John H.",
title          = "{Properties of the covariant formulation of superstring theories}",
journal        = "Nucl. Phys.",
volume         = "B243",
year           = "1984",
pages          = "285-306",
doi            = "10.1016/0550-3213(84)90030-0",
reportNumber   = "Print-84-0264 (QUEEN MARY COLL.)",
SLACcitation   = "%%CITATION = NUPHA,B243,285;%%"
}

@article{Witten:1985nt,
author         = "Witten, Edward",
title          = "{Twistor-like transform in ten dimensions}",
journal        = "Nucl. Phys.",
volume         = "B266",
year           = "1986",
pages          = "245-264",
doi            = "10.1016/0550-3213(86)90090-8",
reportNumber   = "PRINT-85-0458 (PRINCETON)",
SLACcitation   = "%%CITATION = NUPHA,B266,245;%%"
}

@article{Grisaru:1985fv,
author         = "Grisaru, Marcus T. and Howe, Paul S. and Mezincescu, L. and Nilsson, B. and Townsend, P. K.",
title          = "{$N=2$ superstrings in a supergravity background}",
journal        = "Phys. Lett.",
volume         = "162B",
year           = "1985",
pages          = "116-120",
doi            = "10.1016/0370-2693(85)91071-8",
reportNumber   = "Print-85-0603 (CAMBRIDGE)",
SLACcitation   = "%%CITATION = PHLTA,162B,116;%%"
}

@article{Metsaev:1998it,
author         = "Metsaev, R. R. and Tseytlin, Arkady A.",
title          = "{Type IIB superstring action in $AdS_5 \times S^5$ background}",
journal        = "Nucl. Phys.",
volume         = "B533",
year           = "1998",
pages          = "109-126",
doi            = "10.1016/S0550-3213(98)00570-7",
eprint         = "hep-th/9805028",
archivePrefix  = "arXiv",
primaryClass   = "hep-th",
reportNumber   = "FIAN-TD-98-21, IMPERIAL-TP-97-98-44, NSF-ITP-98-055",
SLACcitation   = "%%CITATION = HEP-TH/9805028;%%"
}

@article{Berkovits:1999zq,
author         = "Berkovits, N. and Bershadsky, M. and Hauer, T. and Zhukov, S. and Zwiebach, B.",
title          = "{Superstring theory on $AdS_2 \times S^2$ as a coset supermanifold}",
journal        = "Nucl. Phys.",
volume         = "B567",
year           = "2000",
pages          = "61-86",
doi            = "10.1016/S0550-3213(99)00683-5",
eprint         = "hep-th/9907200",
archivePrefix  = "arXiv",
primaryClass   = "hep-th",
reportNumber   = "IFT-P-060-99, HUTP-99-A044, MIT-CTP-2878, CTP-MIT-2878",
SLACcitation   = "%%CITATION = HEP-TH/9907200;%%"
}

@article{Klimcik:2002zj,
author         = "Klim\v{c}\'{i}k, Ctirad",
title          = "{Yang--Baxter $\sigma$-models and dS/AdS T-duality}",
journal        = "JHEP",
volume         = "12",
year           = "2002",
pages          = "051",
doi            = "10.1088/1126-6708/2002/12/051",
eprint         = "hep-th/0210095",
archivePrefix  = "arXiv",
primaryClass   = "hep-th",
reportNumber   = "IML-02-XY",
SLACcitation   = "%%CITATION = HEP-TH/0210095;%%"
}

@article{Klimcik:2008eq,
author         = "Klim\v{c}\'{i}k, Ctirad",
title          = "{On integrability of the Yang--Baxter $\sigma$-model}",
journal        = "J. Math. Phys.",
volume         = "50",
year           = "2009",
pages          = "043508",
doi            = "10.1063/1.3116242",
eprint         = "0802.3518",
archivePrefix  = "arXiv",
primaryClass   = "hep-th",
SLACcitation   = "%%CITATION = ARXIV:0802.3518;%%"
}

@article{Delduc:2013fga,
author         = "Delduc, Francois and Magro, Marc and Vicedo, Benoit",
title          = "{On classical $q$-deformations of integrable $\sigma$-models}",
journal        = "JHEP",
volume         = "11",
year           = "2013",
pages          = "192",
doi            = "10.1007/JHEP11(2013)192",
eprint         = "1308.3581",
archivePrefix  = "arXiv",
primaryClass   = "hep-th",
SLACcitation   = "%%CITATION = ARXIV:1308.3581;%%"
}

@article{Delduc:2016ihq,
author         = "Delduc, Francois and Lacroix, Sylvain and Magro, Marc and Vicedo, Benoit",
title          = "{On $q$-deformed symmetries as Poisson--Lie symmetries and application to Yang--Baxter type models}",
journal        = "J. Phys.",
volume         = "A49",
year           = "2016",
number         = "41",
pages          = "415402",
doi            = "10.1088/1751-8113/49/41/415402",
eprint         = "1606.01712",
archivePrefix  = "arXiv",
primaryClass   = "hep-th",
SLACcitation   = "%%CITATION = ARXIV:1606.01712;%%"
}

@article{Arutyunov:2013ega,
author         = "Arutyunov, Gleb and Borsato, Riccardo and Frolov, Sergey",
title          = "{S-matrix for strings on $\eta$-deformed $AdS_5 \times S^5$}",
journal        = "JHEP",
volume         = "04",
year           = "2014",
pages          = "002",
doi            = "10.1007/JHEP04(2014)002",
eprint         = "1312.3542",
archivePrefix  = "arXiv",
primaryClass   = "hep-th",
reportNumber   = "ITP-UU-13-31, SPIN-13-23, HU-MATHEMATIK-2013-24, TCD-MATH-13-16",
SLACcitation   = "%%CITATION = ARXIV:1312.3542;%%"
}

@article{Arutyunov:2015qva,
author         = "Arutyunov, Gleb and Borsato, Riccardo and Frolov, Sergey",
title          = "{Puzzles of $\eta$-deformed $AdS_5 \times S^5$}",
journal        = "JHEP",
volume         = "12",
year           = "2015",
pages          = "049",
doi            = "10.1007/JHEP12(2015)049",
eprint         = "1507.04239",
archivePrefix  = "arXiv",
primaryClass   = "hep-th",
reportNumber   = "ITP-UU-15-10, TCD-MATH-15-05, ZMP-HH-15-19",
SLACcitation   = "%%CITATION = ARXIV:1507.04239;%%"
}

@article{Hoare:2015gda,
author         = "Hoare, B. and Tseytlin, A. A.",
title          = "{On integrable deformations of superstring sigma models related to $AdS_n \times S^n$ supercosets}",
journal        = "Nucl. Phys.",
volume         = "B897",
year           = "2015",
pages          = "448-478",
doi            = "10.1016/j.nuclphysb.2015.06.001",
eprint         = "1504.07213",
archivePrefix  = "arXiv",
primaryClass   = "hep-th",
reportNumber   = "IMPERIAL-TP-AT-2015-02, HU-EP-15-21",
SLACcitation   = "%%CITATION = ARXIV:1504.07213;%%"
}

@article{Hoare:2015wia,
author         = "Hoare, B. and Tseytlin, A. A.",
title          = "{Type IIB supergravity solution for the T-dual of the $\eta$-deformed $AdS_5 \times S^5$ superstring}",
journal        = "JHEP",
volume         = "10",
year           = "2015",
pages          = "060",
doi            = "10.1007/JHEP10(2015)060",
eprint         = "1508.01150",
archivePrefix  = "arXiv",
primaryClass   = "hep-th",
reportNumber   = "HU-EP-15-34, IMPERIAL-TP-AT-2015-05",
SLACcitation   = "%%CITATION = ARXIV:1508.01150;%%"
}

@article{Arutyunov:2015mqj,
author         = "Arutyunov, G. and Frolov, S. and Hoare, B. and Roiban, R. and Tseytlin, A. A.",
title          = "{Scale invariance of the $\eta$-deformed $AdS_5 \times S^5$ superstring, T-duality and modified type II equations}",
journal        = "Nucl. Phys.",
volume         = "B903",
year           = "2016",
pages          = "262-303",
doi            = "10.1016/j.nuclphysb.2015.12.012",
eprint         = "1511.05795",
archivePrefix  = "arXiv",
primaryClass   = "hep-th",
reportNumber   = "ZMP-HH-15-27, TCDMATH-15-12, IMPERIAL-TP-AT-2015-08",
SLACcitation   = "%%CITATION = ARXIV:1511.05795;%%"
}

@article{Wulff:2016tju,
author         = "Wulff, L. and Tseytlin, A. A.",
title          = "{Kappa-symmetry of superstring sigma model and generalized 10d supergravity equations}",
journal        = "JHEP",
volume         = "06",
year           = "2016",
pages          = "174",
doi            = "10.1007/JHEP06(2016)174",
eprint         = "1605.04884",
archivePrefix  = "arXiv",
primaryClass   = "hep-th",
reportNumber   = "IMPERIAL-TP-LW-2016-02",
SLACcitation   = "%%CITATION = ARXIV:1605.04884;%%"
}

@article{Sakatani:2016fvh,
author         = "Sakatani, Yuho and Uehara, Shozo and Yoshida, Kentaroh",
title          = "{Generalized gravity from modified DFT}",
journal        = "JHEP",
volume         = "04",
year           = "2017",
pages          = "123",
doi            = "10.1007/JHEP04(2017)123",
eprint         = "1611.05856",
archivePrefix  = "arXiv",
primaryClass   = "hep-th",
reportNumber   = "KUNS-2653",
SLACcitation   = "%%CITATION = ARXIV:1611.05856;%%"
}

@article{Baguet:2016prz,
author         = "Baguet, Arnaud and Magro, Marc and Samtleben, Henning",
title          = "{Generalized IIB supergravity from exceptional field theory}",
journal        = "JHEP",
volume         = "03",
year           = "2017",
pages          = "100",
doi            = "10.1007/JHEP03(2017)100",
eprint         = "1612.07210",
archivePrefix  = "arXiv",
primaryClass   = "hep-th",
SLACcitation   = "%%CITATION = ARXIV:1612.07210;%%"
}

@article{Sakamoto:2017wor,
author         = "Sakamoto, Jun-ichi and Sakatani, Yuho and Yoshida, Kentaroh",
title          = "{Weyl invariance for generalized supergravity backgrounds from the doubled formalism}",
journal        = "PTEP",
volume         = "2017",
year           = "2017",
number         = "5",
pages          = "053B07",
doi            = "10.1093/ptep/ptx067",
eprint         = "1703.09213",
archivePrefix  = "arXiv",
primaryClass   = "hep-th",
reportNumber   = "KUNS-2668",
SLACcitation   = "%%CITATION = ARXIV:1703.09213;%%"
}

@article{Hoare:2016ibq,
author         = "Hoare, Ben and van Tongeren, Stijn J.",
title          = "{Non-split and split deformations of $AdS_5$}",
journal        = "J. Phys.",
volume         = "A49",
year           = "2016",
number         = "48",
pages          = "484003",
doi            = "10.1088/1751-8113/49/48/484003",
eprint         = "1605.03552",
archivePrefix  = "arXiv",
primaryClass   = "hep-th",
SLACcitation   = "%%CITATION = ARXIV:1605.03552;%%"
}

@article{Araujo:2017enj,
author         = "Araujo, T. and Colg\'{a}in, E. \'{O} and Sakamoto, J. and Sheikh-Jabbari, M. M. and Yoshida, K.",
title          = "{$I$ in generalized supergravity}",
year           = "2017",
eprint         = "1708.03163",
archivePrefix  = "arXiv",
primaryClass   = "hep-th",
reportNumber   = "APCTP-PRE2017---015, KUNS-2696, IPM-P-2017-024, IPM-P-2017-028",
SLACcitation   = "%%CITATION = ARXIV:1708.03163;%%"
}

@article{Sfetsos:2013wia,
author         = "Sfetsos, Konstadinos",
title          = "{Integrable interpolations: from exact CFTs to non-abelian T-duals}",
journal        = "Nucl. Phys.",
volume         = "B880",
year           = "2014",
pages          = "225-246",
doi            = "10.1016/j.nuclphysb.2014.01.004",
eprint         = "1312.4560",
archivePrefix  = "arXiv",
primaryClass   = "hep-th",
reportNumber   = "DMUS-MP-13-23, DMUS--MP--13-23",
SLACcitation   = "%%CITATION = ARXIV:1312.4560;%%"
}

@article{Hollowood:2014rla,
author         = "Hollowood, Timothy J. and Miramontes, J. Luis and Schmidtt, David M.",
title          = "{Integrable deformations of strings on symmetric spaces}",
journal        = "JHEP",
volume         = "11",
year           = "2014",
pages          = "009",
doi            = "10.1007/JHEP11(2014)009",
eprint         = "1407.2840",
archivePrefix  = "arXiv",
primaryClass   = "hep-th",
SLACcitation   = "%%CITATION = ARXIV:1407.2840;%%"
}

@article{Borsato:2016zcf,
author         = "Borsato, R. and Tseytlin, A. A. and Wulff, L.",
title          = "{Supergravity background of $\lambda$-deformed model for $AdS_2 \times S^2$ supercoset}",
journal        = "Nucl. Phys.",
volume         = "B905",
year           = "2016",
pages          = "264-292",
doi            = "10.1016/j.nuclphysb.2016.02.018",
eprint         = "1601.08192",
archivePrefix  = "arXiv",
primaryClass   = "hep-th",
reportNumber   = "IMPERIAL-TP-RB-2016-01",
SLACcitation   = "%%CITATION = ARXIV:1601.08192;%%"
}

@article{Chervonyi:2016bfl,
author         = "Chervonyi, Yuri and Lunin, Oleg",
title          = "{Generalized $\lambda$-deformations of $AdS_p \times S^p$}",
journal        = "Nucl. Phys.",
volume         = "B913",
year           = "2016",
pages          = "912-941",
doi            = "10.1016/j.nuclphysb.2016.10.014",
eprint         = "1608.06641",
archivePrefix  = "arXiv",
primaryClass   = "hep-th",
SLACcitation   = "%%CITATION = ARXIV:1608.06641;%%"
}

@article{Borsato:2016ose,
author         = "Borsato, Riccardo and Wulff, Linus",
title          = "{Target space supergeometry of $\eta$ and $\lambda$-deformed strings}",
journal        = "JHEP",
volume         = "10",
year           = "2016",
pages          = "045",
doi            = "10.1007/JHEP10(2016)045",
eprint         = "1608.03570",
archivePrefix  = "arXiv",
primaryClass   = "hep-th",
reportNumber   = "IMPERIAL-TP-LW-2016-03",
SLACcitation   = "%%CITATION = ARXIV:1608.03570;%%"
}

@article{Appadu:2017xku,
author         = "Appadu, Calan and Hollowood, Timothy J. and Miramontes, J. Luis and Price, Dafydd and Schmidtt, David M.",
title          = "{Giant magnons of string theory in the lambda background}",
year           = "2017",
eprint         = "1704.05437",
archivePrefix  = "arXiv",
primaryClass   = "hep-th",
SLACcitation   = "%%CITATION = ARXIV:1704.05437;%%"
}

@article{Klimcik:1995ux,
author         = "Klim\v{c}\'{i}k, C. and \v{S}evera, P.",
title          = "{Dual non-abelian duality and the Drinfel'd double}",
journal        = "Phys. Lett.",
volume         = "B351",
year           = "1995",
pages          = "455-462",
doi            = "10.1016/0370-2693(95)00451-P",
eprint         = "hep-th/9502122",
archivePrefix  = "arXiv",
primaryClass   = "hep-th",
reportNumber   = "CERN-TH-95-39, CERN-TH-95-039",
SLACcitation   = "%%CITATION = HEP-TH/9502122;%%"
}

@article{Klimcik:1995jn,
author         = "Klim\v{c}\'{i}k, C.",
title          = "{Poisson--Lie T-duality}",
journal        = "Nucl. Phys. Proc. Suppl.",
volume         = "46",
year           = "1996",
pages          = "116-121",
doi            = "10.1016/0920-5632(96)00013-8",
eprint         = "hep-th/9509095",
archivePrefix  = "arXiv",
primaryClass   = "hep-th",
reportNumber   = "CERN-TH-95-248",
SLACcitation   = "%%CITATION = HEP-TH/9509095;%%"
}

@article{Vicedo:2015pna,
author         = "Vicedo, Benoit",
title          = "{Deformed integrable $\sigma$-models, classical R-matrices and classical exchange algebra on Drinfel'd doubles}",
journal        = "J. Phys.",
volume         = "A48",
year           = "2015",
number         = "35",
pages          = "355203",
doi            = "10.1088/1751-8113/48/35/355203",
eprint         = "1504.06303",
archivePrefix  = "arXiv",
primaryClass   = "hep-th",
SLACcitation   = "%%CITATION = ARXIV:1504.06303;%%"
}

@article{Hoare:2014pna,
author         = "Hoare, B. and Roiban, R. and Tseytlin, A. A.",
title          = "{On deformations of $AdS_n \times S^n$ supercosets}",
journal        = "JHEP",
volume         = "06",
year           = "2014",
pages          = "002",
doi            = "10.1007/JHEP06(2014)002",
eprint         = "1403.5517",
archivePrefix  = "arXiv",
primaryClass   = "hep-th",
reportNumber   = "IMPERIAL-TP-AT-2014-02, HU-EP-14-10",
SLACcitation   = "%%CITATION = ARXIV:1403.5517;%%"
}

@article{Sfetsos:2015nya,
author         = "Sfetsos, Konstantinos and Siampos, Konstantinos and Thompson, Daniel C.",
title          = "{Generalised integrable $\lambda$- and $\eta$-deformations and their relation}",
journal        = "Nucl. Phys.",
volume         = "B899",
year           = "2015",
pages          = "489-512",
doi            = "10.1016/j.nuclphysb.2015.08.015",
eprint         = "1506.05784",
archivePrefix  = "arXiv",
primaryClass   = "hep-th",
SLACcitation   = "%%CITATION = ARXIV:1506.05784;%%"
}

@article{Sfetsos:1999zm,
author         = "Sfetsos, Konstadinos",
title          = "{Duality-invariant class of two-dimensional field theories}",
journal        = "Nucl. Phys.",
volume         = "B561",
year           = "1999",
pages          = "316-340",
doi            = "10.1016/S0550-3213(99)00485-X",
eprint         = "hep-th/9904188",
archivePrefix  = "arXiv",
primaryClass   = "hep-th",
reportNumber   = "CERN-TH-99-112",
SLACcitation   = "%%CITATION = HEP-TH/9904188;%%"
}

@article{Klimcik:1995dy,
author         = "Klim\v{c}\'{i}k, C. and \v{S}evera, P.",
title          = "{Poisson--Lie T-duality and loop groups of Drinfel'd doubles}",
journal        = "Phys. Lett.",
volume         = "B372",
year           = "1996",
pages          = "65-71",
doi            = "10.1016/0370-2693(96)00025-1",
eprint         = "hep-th/9512040",
archivePrefix  = "arXiv",
primaryClass   = "hep-th",
reportNumber   = "CERN-TH-95-330",
SLACcitation   = "%%CITATION = HEP-TH/9512040;%%"
}

@article{Klimcik:1996nq,
author         = "Klim\v{c}\'{i}k, C. and \v{S}evera, P.",
title          = "{Non-abelian momentum-winding exchange}",
journal        = "Phys. Lett.",
volume         = "B383",
year           = "1996",
pages          = "281-286",
doi            = "10.1016/0370-2693(96)00755-1",
eprint         = "hep-th/9605212",
archivePrefix  = "arXiv",
primaryClass   = "hep-th",
reportNumber   = "CERN-TH-96-142",
SLACcitation   = "%%CITATION = HEP-TH/9605212;%%"
}

@article{Tseytlin:1990nb,
author         = "Tseytlin, Arkady A.",
title          = "{Duality symmetric formulation of string world sheet dynamics}",
journal        = "Phys. Lett.",
volume         = "B242",
year           = "1990",
pages          = "163-174",
doi            = "10.1016/0370-2693(90)91454-J",
reportNumber   = "KCL-TP-1990-2",
SLACcitation   = "%%CITATION = PHLTA,B242,163;%%"
}

@article{Tseytlin:1990va,
author         = "Tseytlin, Arkady A.",
title          = "{Duality symmetric closed string theory and interacting chiral scalars}",
journal        = "Nucl. Phys.",
volume         = "B350",
year           = "1991",
pages          = "395-440",
doi            = "10.1016/0550-3213(91)90266-Z",
reportNumber   = "KCL-TP-1990-3",
SLACcitation   = "%%CITATION = NUPHA,B350,395;%%"
}

@article{Klimcik:2015gba,
author         = "Klim\v{c}\'{i}k, Ctirad",
title          = "{$\eta$ and $\lambda$ deformations as $\mathcal{E}$-models}",
journal        = "Nucl. Phys.",
volume         = "B900",
year           = "2015",
pages          = "259-272",
doi            = "10.1016/j.nuclphysb.2015.09.011",
eprint         = "1508.05832",
archivePrefix  = "arXiv",
primaryClass   = "hep-th",
SLACcitation   = "%%CITATION = ARXIV:1508.05832;%%"
}

@article{Klimcik:1996np,
author         = "Klim\v{c}\'{i}k, C. and \v{S}evera, P.",
title          = "{Dressing cosets}",
journal        = "Phys. Lett.",
volume         = "B381",
year           = "1996",
pages          = "56-61",
doi            = "10.1016/0370-2693(96)00669-7",
eprint         = "hep-th/9602162",
archivePrefix  = "arXiv",
primaryClass   = "hep-th",
reportNumber   = "CERN-TH-96-43",
SLACcitation   = "%%CITATION = HEP-TH/9602162;%%"
}

@article{Squellari:2011dg,
author         = "Squellari, R.",
title          = "{Dressing cosets revisited}",
journal        = "Nucl. Phys.",
volume         = "B853",
year           = "2011",
pages          = "379-403",
doi            = "10.1016/j.nuclphysb.2011.07.025",
eprint         = "1105.0162",
archivePrefix  = "arXiv",
primaryClass   = "hep-th",
SLACcitation   = "%%CITATION = ARXIV:1105.0162;%%"
}

@article{Zarembo:2010sg,
author         = "Zarembo, K.",
title          = "{Strings on semisymmetric superspaces}",
journal        = "JHEP",
volume         = "05",
year           = "2010",
pages          = "002",
doi            = "10.1007/JHEP05(2010)002",
eprint         = "1003.0465",
archivePrefix  = "arXiv",
primaryClass   = "hep-th",
reportNumber   = "ITEP-TH-12-10, LPTENS-10-12, UUITP-05-10",
SLACcitation   = "%%CITATION = ARXIV:1003.0465;%%"
}

@article{Wulff:2014kja,
author         = "Wulff, Linus",
title          = "{Superisometries and integrability of superstrings}",
journal        = "JHEP",
volume         = "05",
year           = "2014",
pages          = "115",
doi            = "10.1007/JHEP05(2014)115",
eprint         = "1402.3122",
archivePrefix  = "arXiv",
primaryClass   = "hep-th",
reportNumber   = "IMPERIAL-TP-LW-2014-01",
SLACcitation   = "%%CITATION = ARXIV:1402.3122;%%"
}

@article{Wulff:2015mwa,
author         = "Wulff, Linus",
title          = "{On integrability of strings on symmetric spaces}",
journal        = "JHEP",
volume         = "09",
year           = "2015",
pages          = "115",
doi            = "10.1007/JHEP09(2015)115",
eprint         = "1505.03525",
archivePrefix  = "arXiv",
primaryClass   = "hep-th",
reportNumber   = "IMPERIAL-TP-LW-2015-01",
SLACcitation   = "%%CITATION = ARXIV:1505.03525;%%"
}

@article{Wulff:2017hzy,
author         = "Wulff, Linus",
title          = "{Integrability of the superstring in $AdS_3 \times S^2 \times S^2 \times T^3$}",
journal        = "J. Phys.",
volume         = "A50",
year           = "2017",
number         = "23",
pages          = "23LT01",
doi            = "10.1088/1751-8121/aa70b5",
eprint         = "1702.08788",
archivePrefix  = "arXiv",
primaryClass   = "hep-th",
SLACcitation   = "%%CITATION = ARXIV:1702.08788;%%"
}

@article{Wulff:2017zbl,
author         = "Wulff, Linus",
title          = "{All symmetric $AdS_{n>2}$ solutions of type II supergravity}",
year           = "2017",
eprint         = "1706.02118",
archivePrefix  = "arXiv",
primaryClass   = "hep-th",
SLACcitation   = "%%CITATION = ARXIV:1706.02118;%%"
}

@article{Wulff:2017lxh,
author         = "Wulff, Linus",
title          = "{Condition on Ramond-Ramond fluxes for factorization of worldsheet scattering in anti-de Sitter space}",
year           = "2017",
eprint         = "1708.09673",
archivePrefix  = "arXiv",
primaryClass   = "hep-th",
SLACcitation   = "%%CITATION = ARXIV:1708.09673;%%"
}

@article{Arutynov:2014ota,
author         = "Arutyunov, Gleb and de Leeuw, Marius and van Tongeren, Stijn J.",
title          = "{The exact spectrum and mirror duality of the $(AdS_5 \times S^5)_\eta$ superstring}",
journal        = "Theor. Math. Phys.",
volume         = "182",
year           = "2015",
number         = "1",
pages          = "23-51",
doi            = "10.1007/s11232-015-0243-9",
note           = "[Teor. Mat. Fiz. 182, 28 (2014)]",
eprint         = "1403.6104",
archivePrefix  = "arXiv",
primaryClass   = "hep-th",
SLACcitation   = "%%CITATION = ARXIV:1403.6104;%%"
}

@article{Arutyunov:2014cra,
author         = "Arutyunov, Gleb and van Tongeren, Stijn J.",
title          = "{$AdS_5 \times S^5$ mirror model as a string sigma model}",
journal        = "Phys. Rev. Lett.",
volume         = "113",
year           = "2014",
pages          = "261605",
doi            = "10.1103/PhysRevLett.113.261605",
eprint         = "1406.2304",
archivePrefix  = "arXiv",
primaryClass   = "hep-th",
reportNumber   = "HU-EP-14-21, HU-MATH-14-12, ITP-UU-14-18, SPIN-14-16",
SLACcitation   = "%%CITATION = ARXIV:1406.2304;%%"
}

@article{Pachol:2015mfa,
author         = "Pacho\l{}, Anna and van Tongeren, Stijn J.",
title          = "{Quantum deformations of the flat space superstring}",
journal        = "Phys. Rev.",
volume         = "D93",
year           = "2016",
pages          = "026008",
doi            = "10.1103/PhysRevD.93.026008",
eprint         = "1510.02389",
archivePrefix  = "arXiv",
primaryClass   = "hep-th",
reportNumber   = "HU-EP-15-48, HU-MATH-15-13",
SLACcitation   = "%%CITATION = ARXIV:1510.02389;%%"
}

@article{Sfetsos:2014cea,
author         = "Sfetsos, Konstantinos and Thompson, Daniel C.",
title          = "{Spacetimes for $\lambda$-deformations}",
journal        = "JHEP",
volume         = "12",
year           = "2014",
pages          = "164",
doi            = "10.1007/JHEP12(2014)164",
eprint         = "1410.1886",
archivePrefix  = "arXiv",
primaryClass   = "hep-th",
SLACcitation   = "%%CITATION = ARXIV:1410.1886;%%"
}

@article{Demulder:2015lva,
author         = "Demulder, Saskia and Sfetsos, Konstantinos and Thompson, Daniel C.",
title          = "{Integrable $\lambda$-deformations: squashing coset CFTs and $AdS_5 \times S^5$}",
journal        = "JHEP",
volume         = "07",
year           = "2015",
pages          = "019",
doi            = "10.1007/JHEP07(2015)019",
eprint         = "1504.02781",
archivePrefix  = "arXiv",
primaryClass   = "hep-th",
SLACcitation   = "%%CITATION = ARXIV:1504.02781;%%"
}

@article{Alvarez:1994np,
author         = "\'{A}lvarez, Enrique and \'{A}lvarez-Gaum\'{e}, Luis and Lozano, Yolanda",
title          = "{On non-abelian duality}",
journal        = "Nucl. Phys.",
volume         = "B424",
year           = "1994",
pages          = "155-183",
doi            = "10.1016/0550-3213(94)90093-0",
eprint         = "hep-th/9403155",
archivePrefix  = "arXiv",
primaryClass   = "hep-th",
reportNumber   = "CERN-TH-7204-94",
SLACcitation   = "%%CITATION = HEP-TH/9403155;%%"
}

@article{Elitzur:1994ri,
author         = "Elitzur, S. and Giveon, A. and Rabinovici, E. and Schwimmer, A. and Veneziano, G.",
title          = "{Remarks on non-abelian duality}",
journal        = "Nucl. Phys.",
volume         = "B435",
year           = "1995",
pages          = "147-171",
doi            = "10.1016/0550-3213(94)00426-F",
eprint         = "hep-th/9409011",
archivePrefix  = "arXiv",
primaryClass   = "hep-th",
reportNumber   = "CERN-TH-7414-94, RI-9-94, WIS-7-94",
SLACcitation   = "%%CITATION = HEP-TH/9409011;%%"
}

@article{Alekseev:1995ym,
author         = "Alekseev, A. {\relax Yu}. and Klim\v{c}\'{i}k, C. and Tseytlin, Arkady A.",
title          = "{Quantum Poisson--Lie T-duality and WZNW model}",
journal        = "Nucl. Phys.",
volume         = "B458",
year           = "1996",
pages          = "430-444",
doi            = "10.1016/0550-3213(95)00575-7",
eprint         = "hep-th/9509123",
archivePrefix  = "arXiv",
primaryClass   = "hep-th",
reportNumber   = "CERN-TH-95-251, ETH-TH-95-26, IMPERIAL-TP-94-95-61, UUITP-16-95",
SLACcitation   = "%%CITATION = HEP-TH/9509123;%%"
}

@article{Tyurin:1995bu,
author         = "Tyurin, Eugene and von Unge, Rikard",
title          = "{Poisson--Lie T-duality: the path-integral derivation}",
journal        = "Phys. Lett.",
volume         = "B382",
year           = "1996",
pages          = "233-240",
doi            = "10.1016/0370-2693(96)00680-6",
eprint         = "hep-th/9512025",
archivePrefix  = "arXiv",
primaryClass   = "hep-th",
reportNumber   = "ITP-SB-95-50, USITP-95-11",
SLACcitation   = "%%CITATION = HEP-TH/9512025;%%"
}

@article{Bossard:2001au,
author         = "Bossard, A. and Mohammedi, N.",
title          = "{Poisson--Lie duality in the string effective action}",
journal        = "Nucl. Phys.",
volume         = "B619",
year           = "2001",
pages          = "128-154",
doi            = "10.1016/S0550-3213(01)00541-7",
eprint         = "hep-th/0106211",
archivePrefix  = "arXiv",
primaryClass   = "hep-th",
SLACcitation   = "%%CITATION = HEP-TH/0106211;%%"
}

@article{VonUnge:2002xjf,
author         = "von Unge, Rikard",
title          = "{Poisson--Lie T-plurality}",
journal        = "JHEP",
volume         = "07",
year           = "2002",
pages          = "014",
doi            = "10.1088/1126-6708/2002/07/014",
eprint         = "hep-th/0205245",
archivePrefix  = "arXiv",
primaryClass   = "hep-th",
SLACcitation   = "%%CITATION = HEP-TH/0205245;%%"
}

@article{Valent:2009nv,
author         = "Valent, Galliano and Klim\v{c}\'{i}k, Ctirad and Squellari, Romain",
title          = "{One loop renormalizability of the Poisson--Lie sigma models}",
journal        = "Phys. Lett.",
volume         = "B678",
year           = "2009",
pages          = "143-148",
doi            = "10.1016/j.physletb.2009.06.001",
eprint         = "0902.1459",
archivePrefix  = "arXiv",
primaryClass   = "hep-th",
SLACcitation   = "%%CITATION = ARXIV:0902.1459;%%"
}

@article{Sfetsos:2009dj,
author         = "Sfetsos, Konstadinos and Siampos, Konstadinos",
title          = "{Quantum equivalence in Poisson--Lie T-duality}",
journal        = "JHEP",
volume         = "06",
year           = "2009",
pages          = "082",
doi            = "10.1088/1126-6708/2009/06/082",
eprint         = "0904.4248",
archivePrefix  = "arXiv",
primaryClass   = "hep-th",
SLACcitation   = "%%CITATION = ARXIV:0904.4248;%%"
}

@article{Avramis:2009xi,
author         = "Avramis, Spyros D. and Derendinger, Jean-Pierre and Prezas, Nikolaos",
title          = "{Conformal chiral boson models on twisted doubled tori and non-geometric string vacua}",
journal        = "Nucl. Phys.",
volume         = "B827",
year           = "2010",
pages          = "281-310",
doi            = "10.1016/j.nuclphysb.2009.11.003",
eprint         = "0910.0431",
archivePrefix  = "arXiv",
primaryClass   = "hep-th",
SLACcitation   = "%%CITATION = ARXIV:0910.0431;%%"
}

@article{Sfetsos:2009vt,
author         = "Sfetsos, K. and Siampos, K. and Thompson, Daniel C.",
title          = "{Renormalization of Lorentz non-invariant actions and manifest T-duality}",
journal        = "Nucl. Phys.",
volume         = "B827",
year           = "2010",
pages          = "545-564",
doi            = "10.1016/j.nuclphysb.2009.11.001",
eprint         = "0910.1345",
archivePrefix  = "arXiv",
primaryClass   = "hep-th",
reportNumber   = "QMUL-PH-09-22",
SLACcitation   = "%%CITATION = ARXIV:0910.1345;%%"
}

@article{Hlavaty:2012sg,
author         = "Hlavat\'{y}, Ladislav and Navr\'{a}til, Josef and \v{S}nobl, Libor",
title          = "{On renormalization of Poisson--Lie T-plural sigma models}",
journal        = "Acta Polytech.",
volume         = "53",
year           = "2013",
number         = "5",
pages          = "433-437",
doi            = "10.14311/AP.2013.53.0433",
eprint         = "1212.5936",
archivePrefix  = "arXiv",
primaryClass   = "hep-th",
SLACcitation   = "%%CITATION = ARXIV:1212.5936;%%"
}

@article{Hassler:2017yza,
author         = "Hassler, Falk",
title          = "{Poisson--Lie T-duality in double field theory}",
year           = "2017",
eprint         = "1707.08624",
archivePrefix  = "arXiv",
primaryClass   = "hep-th",
SLACcitation   = "%%CITATION = ARXIV:1707.08624;%%"
}

@article{Jurco:2017gii,
author         = "Jur\v{c}o, Branislav and Vysok\'{y}, Jan",
title          = "{Poisson--Lie T-duality of string effective actions: a new approach to the dilaton puzzle}",
year           = "2017",
eprint         = "1708.04079",
archivePrefix  = "arXiv",
primaryClass   = "hep-th",
SLACcitation   = "%%CITATION = ARXIV:1708.04079;%%"
}

@article{Severa:2017kcs,
author         = "\v{S}evera, Pavol",
title          = "{On integrability of 2-dimensional $\sigma$-models of Poisson--Lie type}",
year           = "2017",
eprint         = "1709.02213",
archivePrefix  = "arXiv",
primaryClass   = "hep-th",
SLACcitation   = "%%CITATION = ARXIV:1709.02213;%%"
}

\end{bibtex}

\bibliographystyle{nb}
\bibliography{\jobname}

%bibliography generated by nb.bst v1.06 (C) 2003-2011 Niklas Beisert
\begin{thebibliography}{10}
\providecommand{\href}[2]{#2}
\providecommand{\arxivref}[2]{\href{http://arxiv.org/abs/#1}{#2}}
\providecommand{\doiref}[2]{\href{http://dx.doi.org/#1}{#2}}
\providecommand{\nbbstauthor}[1]{#1}
\providecommand{\nbbstjournal}[1]{\textsf{#1}}
\providecommand{\nbbsttitle}[1]{\textit{#1}}
\providecommand{\nbbsturl}[1]{\texttt{#1}}
\providecommand{\nbbsteprint}[1]{\texttt{#1}}
\providecommand{\nbbststyle}{\raggedright\small\parskip0pt}
\nbbststyle

\bibitem{Delduc:2013qra}
\nbbstauthor{F.~Delduc, M.~Magro and B.~Vicedo},
\nbbsttitle{``{An integrable deformation of the $AdS_5 \times S^5$ superstring
  action}''},
\nbbstjournal{\doiref{10.1103/PhysRevLett.112.051601}{Phys.~Rev.~Lett.~112,~051601~(2014)}},
\nbbsteprint{\arxivref{1309.5850}{arxiv:1309.5850}}.
%%CITATION = ARXIV:1309.5850;%%

\bibitem{Delduc:2014kha}
\nbbstauthor{F.~Delduc, M.~Magro and B.~Vicedo},
\nbbsttitle{``{Derivation of the action and symmetries of the $q$-deformed
  $AdS_5 \times S^5$ superstring}''},
\nbbstjournal{\doiref{10.1007/JHEP10(2014)132}{JHEP~1410,~132~(2014)}},
\nbbsteprint{\arxivref{1406.6286}{arxiv:1406.6286}}.
%%CITATION = ARXIV:1406.6286;%%

\bibitem{Hollowood:2014qma}
\nbbstauthor{T.~J.~Hollowood, J.~L.~Miramontes and D.~M.~Schmidtt},
\nbbsttitle{``{An integrable deformation of the $AdS_5 \times S^5$
  superstring}''},
\nbbstjournal{\doiref{10.1088/1751-8113/47/49/495402}{J.~Phys.~A47,~495402~(2014)}},
\nbbsteprint{\arxivref{1409.1538}{arxiv:1409.1538}}.
%%CITATION = ARXIV:1409.1538;%%

\bibitem{Green:1983wt}
\nbbstauthor{M.~B.~Green and J.~H.~Schwarz},
\nbbsttitle{``{Covariant description of superstrings}''},
\nbbstjournal{\doiref{10.1016/0370-2693(84)92021-5}{Phys.~Lett.~136B,~367~(1984)}}.
%%CITATION = PHLTA,136B,367;%%

\bibitem{Green:1983sg}
\nbbstauthor{M.~B.~Green and J.~H.~Schwarz},
\nbbsttitle{``{Properties of the covariant formulation of superstring
  theories}''},
\nbbstjournal{\doiref{10.1016/0550-3213(84)90030-0}{Nucl.~Phys.~B243,~285~(1984)}}.
%%CITATION = NUPHA,B243,285;%%

\bibitem{Witten:1985nt}
\nbbstauthor{E.~Witten},
\nbbsttitle{``{Twistor-like transform in ten dimensions}''},
\nbbstjournal{\doiref{10.1016/0550-3213(86)90090-8}{Nucl.~Phys.~B266,~245~(1986)}}.
%%CITATION = NUPHA,B266,245;%%

\bibitem{Grisaru:1985fv}
\nbbstauthor{M.~T.~Grisaru, P.~S.~Howe, L.~Mezincescu, B.~Nilsson and
  P.~K.~Townsend},
\nbbsttitle{``{$N=2$ superstrings in a supergravity background}''},
\nbbstjournal{\doiref{10.1016/0370-2693(85)91071-8}{Phys.~Lett.~162B,~116~(1985)}}.
%%CITATION = PHLTA,162B,116;%%

\bibitem{Metsaev:1998it}
\nbbstauthor{R.~R.~Metsaev and A.~A.~Tseytlin},
\nbbsttitle{``{Type IIB superstring action in $AdS_5 \times S^5$
  background}''},
\nbbstjournal{\doiref{10.1016/S0550-3213(98)00570-7}{Nucl.~Phys.~B533,~109~(1998)}},
\nbbsteprint{\arxivref{hep-th/9805028}{hep-th/9805028}}.
%%CITATION = HEP-TH/9805028;%%

\bibitem{Berkovits:1999zq}
\nbbstauthor{N.~Berkovits, M.~Bershadsky, T.~Hauer, S.~Zhukov and B.~Zwiebach},
\nbbsttitle{``{Superstring theory on $AdS_2 \times S^2$ as a coset
  supermanifold}''},
\nbbstjournal{\doiref{10.1016/S0550-3213(99)00683-5}{Nucl.~Phys.~B567,~61~(2000)}},
\nbbsteprint{\arxivref{hep-th/9907200}{hep-th/9907200}}.
%%CITATION = HEP-TH/9907200;%%

\bibitem{Klimcik:2002zj}
\nbbstauthor{C.~Klim\v{c}\'{i}k},
\nbbsttitle{``{Yang--Baxter $\sigma$-models and dS/AdS T-duality}''},
\nbbstjournal{\doiref{10.1088/1126-6708/2002/12/051}{JHEP~0212,~051~(2002)}},
\nbbsteprint{\arxivref{hep-th/0210095}{hep-th/0210095}}.
%%CITATION = HEP-TH/0210095;%%

\bibitem{Klimcik:2008eq}
\nbbstauthor{C.~Klim\v{c}\'{i}k},
\nbbsttitle{``{On integrability of the Yang--Baxter $\sigma$-model}''},
\nbbstjournal{\doiref{10.1063/1.3116242}{J.~Math.~Phys.~50,~043508~(2009)}},
\nbbsteprint{\arxivref{0802.3518}{arxiv:0802.3518}}.
%%CITATION = ARXIV:0802.3518;%%

\bibitem{Delduc:2013fga}
\nbbstauthor{F.~Delduc, M.~Magro and B.~Vicedo},
\nbbsttitle{``{On classical $q$-deformations of integrable $\sigma$-models}''},
\nbbstjournal{\doiref{10.1007/JHEP11(2013)192}{JHEP~1311,~192~(2013)}},
\nbbsteprint{\arxivref{1308.3581}{arxiv:1308.3581}}.
%%CITATION = ARXIV:1308.3581;%%

\bibitem{Arutyunov:2013ega}
\nbbstauthor{G.~Arutyunov, R.~Borsato and S.~Frolov},
\nbbsttitle{``{S-matrix for strings on $\eta$-deformed $AdS_5 \times S^5$}''},
\nbbstjournal{\doiref{10.1007/JHEP04(2014)002}{JHEP~1404,~002~(2014)}},
\nbbsteprint{\arxivref{1312.3542}{arxiv:1312.3542}}.
%%CITATION = ARXIV:1312.3542;%%

\bibitem{Delduc:2016ihq}
\nbbstauthor{F.~Delduc, S.~Lacroix, M.~Magro and B.~Vicedo},
\nbbsttitle{``{On $q$-deformed symmetries as Poisson--Lie symmetries and
  application to Yang--Baxter type models}''},
\nbbstjournal{\doiref{10.1088/1751-8113/49/41/415402}{J.~Phys.~A49,~415402~(2016)}},
\nbbsteprint{\arxivref{1606.01712}{arxiv:1606.01712}}.
%%CITATION = ARXIV:1606.01712;%%

\bibitem{Arutyunov:2015qva}
\nbbstauthor{G.~Arutyunov, R.~Borsato and S.~Frolov},
\nbbsttitle{``{Puzzles of $\eta$-deformed $AdS_5 \times S^5$}''},
\nbbstjournal{\doiref{10.1007/JHEP12(2015)049}{JHEP~1512,~049~(2015)}},
\nbbsteprint{\arxivref{1507.04239}{arxiv:1507.04239}}.
%%CITATION = ARXIV:1507.04239;%%

\bibitem{Hoare:2015gda}
\nbbstauthor{B.~Hoare and A.~A.~Tseytlin},
\nbbsttitle{``{On integrable deformations of superstring sigma models related
  to $AdS_n \times S^n$ supercosets}''},
\nbbstjournal{\doiref{10.1016/j.nuclphysb.2015.06.001}{Nucl.~Phys.~B897,~448~(2015)}},
\nbbsteprint{\arxivref{1504.07213}{arxiv:1504.07213}}.
%%CITATION = ARXIV:1504.07213;%%

\bibitem{Hoare:2015wia}
\nbbstauthor{B.~Hoare and A.~A.~Tseytlin},
\nbbsttitle{``{Type IIB supergravity solution for the T-dual of the
  $\eta$-deformed $AdS_5 \times S^5$ superstring}''},
\nbbstjournal{\doiref{10.1007/JHEP10(2015)060}{JHEP~1510,~060~(2015)}},
\nbbsteprint{\arxivref{1508.01150}{arxiv:1508.01150}}.
%%CITATION = ARXIV:1508.01150;%%

\bibitem{Arutyunov:2015mqj}
\nbbstauthor{G.~Arutyunov, S.~Frolov, B.~Hoare, R.~Roiban and A.~A.~Tseytlin},
\nbbsttitle{``{Scale invariance of the $\eta$-deformed $AdS_5 \times S^5$
  superstring, T-duality and modified type II equations}''},
\nbbstjournal{\doiref{10.1016/j.nuclphysb.2015.12.012}{Nucl.~Phys.~B903,~262~(2016)}},
\nbbsteprint{\arxivref{1511.05795}{arxiv:1511.05795}}.
%%CITATION = ARXIV:1511.05795;%%

\bibitem{Wulff:2016tju}
\nbbstauthor{L.~Wulff and A.~A.~Tseytlin},
\nbbsttitle{``{Kappa-symmetry of superstring sigma model and generalized 10d
  supergravity equations}''},
\nbbstjournal{\doiref{10.1007/JHEP06(2016)174}{JHEP~1606,~174~(2016)}},
\nbbsteprint{\arxivref{1605.04884}{arxiv:1605.04884}}.
%%CITATION = ARXIV:1605.04884;%%

\bibitem{Hoare:2016ibq}
\nbbstauthor{B.~Hoare and S.~J.~van~Tongeren},
\nbbsttitle{``{Non-split and split deformations of $AdS_5$}''},
\nbbstjournal{\doiref{10.1088/1751-8113/49/48/484003}{J.~Phys.~A49,~484003~(2016)}},
\nbbsteprint{\arxivref{1605.03552}{arxiv:1605.03552}}.
%%CITATION = ARXIV:1605.03552;%%

\bibitem{Borsato:2016ose}
\nbbstauthor{R.~Borsato and L.~Wulff},
\nbbsttitle{``{Target space supergeometry of $\eta$ and $\lambda$-deformed
  strings}''},
\nbbstjournal{\doiref{10.1007/JHEP10(2016)045}{JHEP~1610,~045~(2016)}},
\nbbsteprint{\arxivref{1608.03570}{arxiv:1608.03570}}.
%%CITATION = ARXIV:1608.03570;%%

\bibitem{Araujo:2017enj}
\nbbstauthor{T.~Araujo, E.~O.~Colg\'{a}in, J.~Sakamoto, M.~M.~Sheikh-Jabbari
  and K.~Yoshida},
\nbbsttitle{``{$I$ in generalized supergravity}''},
\nbbsteprint{\arxivref{1708.03163}{arxiv:1708.03163}}.
%%CITATION = ARXIV:1708.03163;%%

\bibitem{Sakatani:2016fvh}
\nbbstauthor{Y.~Sakatani, S.~Uehara and K.~Yoshida},
\nbbsttitle{``{Generalized gravity from modified DFT}''},
\nbbstjournal{\doiref{10.1007/JHEP04(2017)123}{JHEP~1704,~123~(2017)}},
\nbbsteprint{\arxivref{1611.05856}{arxiv:1611.05856}}.
%%CITATION = ARXIV:1611.05856;%%

\bibitem{Baguet:2016prz}
\nbbstauthor{A.~Baguet, M.~Magro and H.~Samtleben},
\nbbsttitle{``{Generalized IIB supergravity from exceptional field theory}''},
\nbbstjournal{\doiref{10.1007/JHEP03(2017)100}{JHEP~1703,~100~(2017)}},
\nbbsteprint{\arxivref{1612.07210}{arxiv:1612.07210}}.
%%CITATION = ARXIV:1612.07210;%%

\bibitem{Sakamoto:2017wor}
\nbbstauthor{J.-i.~Sakamoto, Y.~Sakatani and K.~Yoshida},
\nbbsttitle{``{Weyl invariance for generalized supergravity backgrounds from
  the doubled formalism}''},
\nbbstjournal{\doiref{10.1093/ptep/ptx067}{PTEP~2017,~053B07~(2017)}},
\nbbsteprint{\arxivref{1703.09213}{arxiv:1703.09213}}.
%%CITATION = ARXIV:1703.09213;%%

\bibitem{Sfetsos:2013wia}
\nbbstauthor{K.~Sfetsos},
\nbbsttitle{``{Integrable interpolations: from exact CFTs to non-abelian
  T-duals}''},
\nbbstjournal{\doiref{10.1016/j.nuclphysb.2014.01.004}{Nucl.~Phys.~B880,~225~(2014)}},
\nbbsteprint{\arxivref{1312.4560}{arxiv:1312.4560}}.
%%CITATION = ARXIV:1312.4560;%%

\bibitem{Hollowood:2014rla}
\nbbstauthor{T.~J.~Hollowood, J.~L.~Miramontes and D.~M.~Schmidtt},
\nbbsttitle{``{Integrable deformations of strings on symmetric spaces}''},
\nbbstjournal{\doiref{10.1007/JHEP11(2014)009}{JHEP~1411,~009~(2014)}},
\nbbsteprint{\arxivref{1407.2840}{arxiv:1407.2840}}.
%%CITATION = ARXIV:1407.2840;%%

\bibitem{Borsato:2016zcf}
\nbbstauthor{R.~Borsato, A.~A.~Tseytlin and L.~Wulff},
\nbbsttitle{``{Supergravity background of $\lambda$-deformed model for $AdS_2
  \times S^2$ supercoset}''},
\nbbstjournal{\doiref{10.1016/j.nuclphysb.2016.02.018}{Nucl.~Phys.~B905,~264~(2016)}},
\nbbsteprint{\arxivref{1601.08192}{arxiv:1601.08192}}.
%%CITATION = ARXIV:1601.08192;%%

\bibitem{Chervonyi:2016bfl}
\nbbstauthor{Y.~Chervonyi and O.~Lunin},
\nbbsttitle{``{Generalized $\lambda$-deformations of $AdS_p \times S^p$}''},
\nbbstjournal{\doiref{10.1016/j.nuclphysb.2016.10.014}{Nucl.~Phys.~B913,~912~(2016)}},
\nbbsteprint{\arxivref{1608.06641}{arxiv:1608.06641}}.
%%CITATION = ARXIV:1608.06641;%%

\bibitem{Appadu:2017xku}
\nbbstauthor{C.~Appadu, T.~J.~Hollowood, J.~L.~Miramontes, D.~Price and
  D.~M.~Schmidtt},
\nbbsttitle{``{Giant magnons of string theory in the lambda background}''},
\nbbsteprint{\arxivref{1704.05437}{arxiv:1704.05437}}.
%%CITATION = ARXIV:1704.05437;%%

\bibitem{Klimcik:1995ux}
\nbbstauthor{C.~Klim\v{c}\'{i}k and P.~\v{S}evera},
\nbbsttitle{``{Dual non-abelian duality and the Drinfel'd double}''},
\nbbstjournal{\doiref{10.1016/0370-2693(95)00451-P}{Phys.~Lett.~B351,~455~(1995)}},
\nbbsteprint{\arxivref{hep-th/9502122}{hep-th/9502122}}.
%%CITATION = HEP-TH/9502122;%%

\bibitem{Klimcik:1995jn}
\nbbstauthor{C.~Klim\v{c}\'{i}k},
\nbbsttitle{``{Poisson--Lie T-duality}''},
\nbbstjournal{\doiref{10.1016/0920-5632(96)00013-8}{Nucl.~Phys.~Proc.~Suppl.~46,~116~(1996)}},
\nbbsteprint{\arxivref{hep-th/9509095}{hep-th/9509095}}.
%%CITATION = HEP-TH/9509095;%%

\bibitem{Vicedo:2015pna}
\nbbstauthor{B.~Vicedo},
\nbbsttitle{``{Deformed integrable $\sigma$-models, classical R-matrices and
  classical exchange algebra on Drinfel'd doubles}''},
\nbbstjournal{\doiref{10.1088/1751-8113/48/35/355203}{J.~Phys.~A48,~355203~(2015)}},
\nbbsteprint{\arxivref{1504.06303}{arxiv:1504.06303}}.
%%CITATION = ARXIV:1504.06303;%%

\bibitem{Hoare:2014pna}
\nbbstauthor{B.~Hoare, R.~Roiban and A.~A.~Tseytlin},
\nbbsttitle{``{On deformations of $AdS_n \times S^n$ supercosets}''},
\nbbstjournal{\doiref{10.1007/JHEP06(2014)002}{JHEP~1406,~002~(2014)}},
\nbbsteprint{\arxivref{1403.5517}{arxiv:1403.5517}}.
%%CITATION = ARXIV:1403.5517;%%

\bibitem{Sfetsos:2015nya}
\nbbstauthor{K.~Sfetsos, K.~Siampos and D.~C.~Thompson},
\nbbsttitle{``{Generalised integrable $\lambda$- and $\eta$-deformations and
  their relation}''},
\nbbstjournal{\doiref{10.1016/j.nuclphysb.2015.08.015}{Nucl.~Phys.~B899,~489~(2015)}},
\nbbsteprint{\arxivref{1506.05784}{arxiv:1506.05784}}.
%%CITATION = ARXIV:1506.05784;%%

\bibitem{Sfetsos:1999zm}
\nbbstauthor{K.~Sfetsos},
\nbbsttitle{``{Duality-invariant class of two-dimensional field theories}''},
\nbbstjournal{\doiref{10.1016/S0550-3213(99)00485-X}{Nucl.~Phys.~B561,~316~(1999)}},
\nbbsteprint{\arxivref{hep-th/9904188}{hep-th/9904188}}.
%%CITATION = HEP-TH/9904188;%%

\bibitem{Klimcik:1995dy}
\nbbstauthor{C.~Klim\v{c}\'{i}k and P.~\v{S}evera},
\nbbsttitle{``{Poisson--Lie T-duality and loop groups of Drinfel'd doubles}''},
\nbbstjournal{\doiref{10.1016/0370-2693(96)00025-1}{Phys.~Lett.~B372,~65~(1996)}},
\nbbsteprint{\arxivref{hep-th/9512040}{hep-th/9512040}}.
%%CITATION = HEP-TH/9512040;%%

\bibitem{Klimcik:1996nq}
\nbbstauthor{C.~Klim\v{c}\'{i}k and P.~\v{S}evera},
\nbbsttitle{``{Non-abelian momentum-winding exchange}''},
\nbbstjournal{\doiref{10.1016/0370-2693(96)00755-1}{Phys.~Lett.~B383,~281~(1996)}},
\nbbsteprint{\arxivref{hep-th/9605212}{hep-th/9605212}}.
%%CITATION = HEP-TH/9605212;%%

\bibitem{Tseytlin:1990nb}
\nbbstauthor{A.~A.~Tseytlin},
\nbbsttitle{``{Duality symmetric formulation of string world sheet
  dynamics}''},
\nbbstjournal{\doiref{10.1016/0370-2693(90)91454-J}{Phys.~Lett.~B242,~163~(1990)}}.
%%CITATION = PHLTA,B242,163;%%

\bibitem{Tseytlin:1990va}
\nbbstauthor{A.~A.~Tseytlin},
\nbbsttitle{``{Duality symmetric closed string theory and interacting chiral
  scalars}''},
\nbbstjournal{\doiref{10.1016/0550-3213(91)90266-Z}{Nucl.~Phys.~B350,~395~(1991)}}.
%%CITATION = NUPHA,B350,395;%%

\bibitem{Klimcik:2015gba}
\nbbstauthor{C.~Klim\v{c}\'{i}k},
\nbbsttitle{``{$\eta$ and $\lambda$ deformations as $\mathcal{E}$-models}''},
\nbbstjournal{\doiref{10.1016/j.nuclphysb.2015.09.011}{Nucl.~Phys.~B900,~259~(2015)}},
\nbbsteprint{\arxivref{1508.05832}{arxiv:1508.05832}}.
%%CITATION = ARXIV:1508.05832;%%

\bibitem{Klimcik:1996np}
\nbbstauthor{C.~Klim\v{c}\'{i}k and P.~\v{S}evera},
\nbbsttitle{``{Dressing cosets}''},
\nbbstjournal{\doiref{10.1016/0370-2693(96)00669-7}{Phys.~Lett.~B381,~56~(1996)}},
\nbbsteprint{\arxivref{hep-th/9602162}{hep-th/9602162}}.
%%CITATION = HEP-TH/9602162;%%

\bibitem{Squellari:2011dg}
\nbbstauthor{R.~Squellari},
\nbbsttitle{``{Dressing cosets revisited}''},
\nbbstjournal{\doiref{10.1016/j.nuclphysb.2011.07.025}{Nucl.~Phys.~B853,~379~(2011)}},
\nbbsteprint{\arxivref{1105.0162}{arxiv:1105.0162}}.
%%CITATION = ARXIV:1105.0162;%%

\bibitem{Zarembo:2010sg}
\nbbstauthor{K.~Zarembo},
\nbbsttitle{``{Strings on semisymmetric superspaces}''},
\nbbstjournal{\doiref{10.1007/JHEP05(2010)002}{JHEP~1005,~002~(2010)}},
\nbbsteprint{\arxivref{1003.0465}{arxiv:1003.0465}}.
%%CITATION = ARXIV:1003.0465;%%

\bibitem{Wulff:2014kja}
\nbbstauthor{L.~Wulff},
\nbbsttitle{``{Superisometries and integrability of superstrings}''},
\nbbstjournal{\doiref{10.1007/JHEP05(2014)115}{JHEP~1405,~115~(2014)}},
\nbbsteprint{\arxivref{1402.3122}{arxiv:1402.3122}}.
%%CITATION = ARXIV:1402.3122;%%

\bibitem{Wulff:2015mwa}
\nbbstauthor{L.~Wulff},
\nbbsttitle{``{On integrability of strings on symmetric spaces}''},
\nbbstjournal{\doiref{10.1007/JHEP09(2015)115}{JHEP~1509,~115~(2015)}},
\nbbsteprint{\arxivref{1505.03525}{arxiv:1505.03525}}.
%%CITATION = ARXIV:1505.03525;%%

\bibitem{Wulff:2017hzy}
\nbbstauthor{L.~Wulff},
\nbbsttitle{``{Integrability of the superstring in $AdS_3 \times S^2 \times S^2
  \times T^3$}''},
\nbbstjournal{\doiref{10.1088/1751-8121/aa70b5}{J.~Phys.~A50,~23LT01~(2017)}},
\nbbsteprint{\arxivref{1702.08788}{arxiv:1702.08788}}.
%%CITATION = ARXIV:1702.08788;%%

\bibitem{Wulff:2017zbl}
\nbbstauthor{L.~Wulff},
\nbbsttitle{``{All symmetric $AdS_{n>2}$ solutions of type II supergravity}''},
\nbbsteprint{\arxivref{1706.02118}{arxiv:1706.02118}}.
%%CITATION = ARXIV:1706.02118;%%

\bibitem{Wulff:2017lxh}
\nbbstauthor{L.~Wulff},
\nbbsttitle{``{Condition on Ramond-Ramond fluxes for factorization of
  worldsheet scattering in anti-de Sitter space}''},
\nbbsteprint{\arxivref{1708.09673}{arxiv:1708.09673}}.
%%CITATION = ARXIV:1708.09673;%%

\bibitem{Tyurin:1995bu}
\nbbstauthor{E.~Tyurin and R.~von~Unge},
\nbbsttitle{``{Poisson--Lie T-duality: the path-integral derivation}''},
\nbbstjournal{\doiref{10.1016/0370-2693(96)00680-6}{Phys.~Lett.~B382,~233~(1996)}},
\nbbsteprint{\arxivref{hep-th/9512025}{hep-th/9512025}}.
%%CITATION = HEP-TH/9512025;%%

\bibitem{Bossard:2001au}
\nbbstauthor{A.~Bossard and N.~Mohammedi},
\nbbsttitle{``{Poisson--Lie duality in the string effective action}''},
\nbbstjournal{\doiref{10.1016/S0550-3213(01)00541-7}{Nucl.~Phys.~B619,~128~(2001)}},
\nbbsteprint{\arxivref{hep-th/0106211}{hep-th/0106211}}.
%%CITATION = HEP-TH/0106211;%%

\bibitem{Arutynov:2014ota}
\nbbstauthor{G.~Arutyunov, M.~de~Leeuw and S.~J.~van~Tongeren},
\nbbsttitle{``{The exact spectrum and mirror duality of the $(AdS_5 \times
  S^5)_\eta$ superstring}''},
\nbbstjournal{\doiref{10.1007/s11232-015-0243-9}{Theor.~Math.~Phys.~182,~23~(2015)}},
\nbbsteprint{\arxivref{1403.6104}{arxiv:1403.6104}},
[Teor. Mat. Fiz. 182, 28 (2014)].
%%CITATION = ARXIV:1403.6104;%%

\bibitem{Arutyunov:2014cra}
\nbbstauthor{G.~Arutyunov and S.~J.~van~Tongeren},
\nbbsttitle{``{$AdS_5 \times S^5$ mirror model as a string sigma model}''},
\nbbstjournal{\doiref{10.1103/PhysRevLett.113.261605}{Phys.~Rev.~Lett.~113,~261605~(2014)}},
\nbbsteprint{\arxivref{1406.2304}{arxiv:1406.2304}}.
%%CITATION = ARXIV:1406.2304;%%

\bibitem{Pachol:2015mfa}
\nbbstauthor{A.~Pacho\l{} and S.~J.~van~Tongeren},
\nbbsttitle{``{Quantum deformations of the flat space superstring}''},
\nbbstjournal{\doiref{10.1103/PhysRevD.93.026008}{Phys.~Rev.~D93,~026008~(2016)}},
\nbbsteprint{\arxivref{1510.02389}{arxiv:1510.02389}}.
%%CITATION = ARXIV:1510.02389;%%

\bibitem{Alvarez:1994np}
\nbbstauthor{E.~\'{A}lvarez, L.~\'{A}lvarez~Gaum\'{e} and Y.~Lozano},
\nbbsttitle{``{On non-abelian duality}''},
\nbbstjournal{\doiref{10.1016/0550-3213(94)90093-0}{Nucl.~Phys.~B424,~155~(1994)}},
\nbbsteprint{\arxivref{hep-th/9403155}{hep-th/9403155}}.
%%CITATION = HEP-TH/9403155;%%

\bibitem{Elitzur:1994ri}
\nbbstauthor{S.~Elitzur, A.~Giveon, E.~Rabinovici, A.~Schwimmer and
  G.~Veneziano},
\nbbsttitle{``{Remarks on non-abelian duality}''},
\nbbstjournal{\doiref{10.1016/0550-3213(94)00426-F}{Nucl.~Phys.~B435,~147~(1995)}},
\nbbsteprint{\arxivref{hep-th/9409011}{hep-th/9409011}}.
%%CITATION = HEP-TH/9409011;%%

\bibitem{VonUnge:2002xjf}
\nbbstauthor{R.~von~Unge},
\nbbsttitle{``{Poisson--Lie T-plurality}''},
\nbbstjournal{\doiref{10.1088/1126-6708/2002/07/014}{JHEP~0207,~014~(2002)}},
\nbbsteprint{\arxivref{hep-th/0205245}{hep-th/0205245}}.
%%CITATION = HEP-TH/0205245;%%

\bibitem{Alekseev:1995ym}
\nbbstauthor{A.~{\relax Yu}.~Alekseev, C.~Klim\v{c}\'{i}k and A.~A.~Tseytlin},
\nbbsttitle{``{Quantum Poisson--Lie T-duality and WZNW model}''},
\nbbstjournal{\doiref{10.1016/0550-3213(95)00575-7}{Nucl.~Phys.~B458,~430~(1996)}},
\nbbsteprint{\arxivref{hep-th/9509123}{hep-th/9509123}}.
%%CITATION = HEP-TH/9509123;%%

\bibitem{Valent:2009nv}
\nbbstauthor{G.~Valent, C.~Klim\v{c}\'{i}k and R.~Squellari},
\nbbsttitle{``{One loop renormalizability of the Poisson--Lie sigma models}''},
\nbbstjournal{\doiref{10.1016/j.physletb.2009.06.001}{Phys.~Lett.~B678,~143~(2009)}},
\nbbsteprint{\arxivref{0902.1459}{arxiv:0902.1459}}.
%%CITATION = ARXIV:0902.1459;%%

\bibitem{Sfetsos:2009dj}
\nbbstauthor{K.~Sfetsos and K.~Siampos},
\nbbsttitle{``{Quantum equivalence in Poisson--Lie T-duality}''},
\nbbstjournal{\doiref{10.1088/1126-6708/2009/06/082}{JHEP~0906,~082~(2009)}},
\nbbsteprint{\arxivref{0904.4248}{arxiv:0904.4248}}.
%%CITATION = ARXIV:0904.4248;%%

\bibitem{Avramis:2009xi}
\nbbstauthor{S.~D.~Avramis, J.-P.~Derendinger and N.~Prezas},
\nbbsttitle{``{Conformal chiral boson models on twisted doubled tori and
  non-geometric string vacua}''},
\nbbstjournal{\doiref{10.1016/j.nuclphysb.2009.11.003}{Nucl.~Phys.~B827,~281~(2010)}},
\nbbsteprint{\arxivref{0910.0431}{arxiv:0910.0431}}.
%%CITATION = ARXIV:0910.0431;%%

\bibitem{Sfetsos:2009vt}
\nbbstauthor{K.~Sfetsos, K.~Siampos and D.~C.~Thompson},
\nbbsttitle{``{Renormalization of Lorentz non-invariant actions and manifest
  T-duality}''},
\nbbstjournal{\doiref{10.1016/j.nuclphysb.2009.11.001}{Nucl.~Phys.~B827,~545~(2010)}},
\nbbsteprint{\arxivref{0910.1345}{arxiv:0910.1345}}.
%%CITATION = ARXIV:0910.1345;%%

\bibitem{Hlavaty:2012sg}
\nbbstauthor{L.~Hlavat\'{y}, J.~Navr\'{a}til and L.~\v{S}nobl},
\nbbsttitle{``{On renormalization of Poisson--Lie T-plural sigma models}''},
\nbbstjournal{\doiref{10.14311/AP.2013.53.0433}{Acta~Polytech.~53,~433~(2013)}},
\nbbsteprint{\arxivref{1212.5936}{arxiv:1212.5936}}.
%%CITATION = ARXIV:1212.5936;%%

\bibitem{Hassler:2017yza}
\nbbstauthor{F.~Hassler},
\nbbsttitle{``{Poisson--Lie T-duality in double field theory}''},
\nbbsteprint{\arxivref{1707.08624}{arxiv:1707.08624}}.
%%CITATION = ARXIV:1707.08624;%%

\bibitem{Jurco:2017gii}
\nbbstauthor{B.~Jur\v{c}o and J.~Vysok\'{y}},
\nbbsttitle{``{Poisson--Lie T-duality of string effective actions: a new
  approach to the dilaton puzzle}''},
\nbbsteprint{\arxivref{1708.04079}{arxiv:1708.04079}}.
%%CITATION = ARXIV:1708.04079;%%

\bibitem{Sfetsos:2014cea}
\nbbstauthor{K.~Sfetsos and D.~C.~Thompson},
\nbbsttitle{``{Spacetimes for $\lambda$-deformations}''},
\nbbstjournal{\doiref{10.1007/JHEP12(2014)164}{JHEP~1412,~164~(2014)}},
\nbbsteprint{\arxivref{1410.1886}{arxiv:1410.1886}}.
%%CITATION = ARXIV:1410.1886;%%

\bibitem{Demulder:2015lva}
\nbbstauthor{S.~Demulder, K.~Sfetsos and D.~C.~Thompson},
\nbbsttitle{``{Integrable $\lambda$-deformations: squashing coset CFTs and
  $AdS_5 \times S^5$}''},
\nbbstjournal{\doiref{10.1007/JHEP07(2015)019}{JHEP~1507,~019~(2015)}},
\nbbsteprint{\arxivref{1504.02781}{arxiv:1504.02781}}.
%%CITATION = ARXIV:1504.02781;%%

\bibitem{Severa:2017kcs}
\nbbstauthor{P.~\v{S}evera},
\nbbsttitle{``{On integrability of 2-dimensional $\sigma$-models of
  Poisson--Lie type}''},
\nbbsteprint{\arxivref{1709.02213}{arxiv:1709.02213}}.
%%CITATION = ARXIV:1709.02213;%%

\end{thebibliography}

\end{document}